\documentclass{aa}  
\usepackage{xcolor}

\newcommand{\lya}{Ly$\alpha$\ }
\newcommand{\SBlim}{\ensuremath{\mathrm{SB}_{\mathrm{lim,\, Ly}\alpha}}}
\newcommand{\rsh}{\ensuremath{r_\mathrm{s,H}}}

\usepackage{natbib,twoopt}
\usepackage[breaklinks=true]{hyperref}
\bibpunct{(}{)}{;}{a}{}{,}
\makeatletter
  \newcommandtwoopt{\citeads}[3][][]{\href{http://adsabs.harvard.edu/abs/#3}
    {\def\hyper@linkstart##1##2{}
     \let\hyper@linkend\@empty\citealp[#1][#2]{#3}}}
  \newcommandtwoopt{\citepads}[3][][]{\href{http://adsabs.harvard.edu/abs/#3}
    {\def\hyper@linkstart##1##2{}
     \let\hyper@linkend\@empty\citep[#1][#2]{#3}}}
  \newcommandtwoopt{\citetads}[3][][]{\href{http://adsabs.harvard.edu/abs/#3}
    {\def\hyper@linkstart##1##2{}
     \let\hyper@linkend\@empty\citet[#1][#2]{#3}}}
  \newcommandtwoopt{\citeyearads}[3][][]
    {\href{http://adsabs.harvard.edu/abs/#3}
    {\def\hyper@linkstart##1##2{}
     \let\hyper@linkend\@empty\citeyear[#1][#2]{#3}}}
\makeatother

\addto\extrasenglish{

}

\defcitealias{2017A&A...608A...8L}{FL17}
\defcitealias{2016A&A...587A..98W}{LW16}

\usepackage{graphicx}
\usepackage{txfonts}
\usepackage{amsmath}
\begin{document}

   \title{Lyman-alpha haloes in the aftermath of reionisation}

   \author{Daniil Smirnov\inst{1}\corrauth{dsmirnov@aip.de}                   \and
          Lutz Wisotzki\inst{1}\email{lwisotzki@aip.de}                       \and
          Tanya Urrutia\inst{1}\email{turrutia@aip.de}                        \and
          John Pharo\inst{1}\email{jpharo@aip.de}                             \and
          Ramona Augustin\inst{1}\email{raugustin@aip.de}                     \and
          Yucheng Guo\inst{2}\email{yuchengg@asu.edu}                         \and
          Daria Kozlova\inst{1}\email{dkozlova@aip.de}                        \and
          Haruka Kusakabe\inst{3}\email{haruka.kusakabe.takeishi@gmail.com}   \and
          Jorryt Matthee\inst{4}\email{jorryt.matthee@ist.ac.at}              \and
          Ismael Pessa\inst{1}\email{ipessa@aip.de}                           \and
          Joop Schaye\inst{5}\email{schaye@strw.leidenuniv.nl}                
          }

   \institute{Leibniz-Institut für Astrophysik Potsdam (AIP), An der Sternwarte 16, 14482 Potsdam, Germany
             \and
             School of Earth \& Space Exploration, Arizona State University, 781 Terrace Mall, Tempe, AZ 85287, USA
             \and
             Department of General Systems Studies, Graduate School of Arts and Sciences, The University of Tokyo, 3-8-1 Komaba, Meguro-ku, Tokyo, 153-8902, Japan
             \and
             Institute of Science and Technology Austria (ISTA), Am Campus 1, 3400 Klosterneuburg, Austria
             \and
             Leiden Observatory, Leiden University, PO Box 9513, 2300 RA Leiden, the Netherlands
             }

   \date{Received XXX; accepted YYY}

   \abstract{We present a comparative study of \lya haloes (LAHs) around low-luminosity (L$_{\mathrm{Ly}\alpha}\lesssim 10^{42}$\,erg\,s$^{-1}$) Ly$\alpha$-emitting galaxies (LAEs) at very high redshifts $z\geq6$ and a reference sample at $z\sim 3$ covering a similar \lya luminosity and host galaxy stellar mass range. Using data from the Multi-Unit Spectroscopic Explorer (MUSE) at the ESO VLT, we extracted the samples such that at the different redshifts we obtain the same \emph{intrinsic} surface brightness sensitivity, accounting for cosmological dimming. We detect extended \lya emission around 6 out of 18 high-$z$ LAEs in the MUSE eXtremely Deep Field (MXDF), more than doubling the number of known such objects at $z\geq6$. We obtain an only slightly higher individual LAH detection fraction of 40\% among the lower redshift comparison sample. Yet the typical exponential scale lengths at $z\geq6$ are three times smaller than those at $z\sim3$. Stacking the LAEs with undetected haloes gives again drastically different results for the two samples, with a highly significant halo detection at $z\sim 3$ but no trace of extended \lya emission at $z\geq6$. We also find the \lya spectral line widths of the high-$z$ sample to be $\sim$1.7 smaller
   in comparison to the lower redshift objects. We discuss the potential mechanisms driving such strong changes. In a reionisation-driven scenario the higher neutral fraction in the intergalactic and circumgalactic media might lead to substantial scattering losses of escaping \lya radiation, leaving detectable only emission from the vicinity of the star-forming regions. In an alternative scenario the LAH properties might be linked more closely to the evolution of their host galaxies than previously thought.
   }


   \maketitle
   \nolinenumbers

\section{Introduction}

The \lya emission line is one of the principal tracers of star-forming galaxies at high redshifts\citepads{1967ApJ...147..868P}.
It is now well established that this emission is not confined to the sites of actual star formation, but often extends far beyond the stellar components of galaxies. Typically more than half, and sometimes even more than 90\% of the total \lya luminosity of a galaxy emerges from this  circumgalactic region (\citeads{2016A&A...587A..98W}, \citeads{2017A&A...608A...8L} hereafter cited as LW16 and FL17, respectively). The physical mechanisms to produce these \lya haloes (LAHs) can be diverse, but the dominant process is probably recombination radiation followed by resonant scattering at \ion{H}{i} clouds in the circumgalactic medium (CGM) (e.g., \citeads{2006A&A...460..397V}, \citeads{2019PASJ...71...55K}, \citeads{2021MNRAS.501.5757M}, \citeads{2021MNRAS.506.5129B}). LAHs are therefore valuable tracers of the cool and partly neutral CGM at high redshifts.

Recent years have seen a tremendous push of the observational frontier towards earlier cosmic epochs and into the epoch of reionisation. The visibility of \lya in galaxies at very high redshifts is now one of the key probes of the rapidly evolving \ion{H}{i} content of the Universe (e.g., \citeads{2020ARA&A..58..617O}, \citeads{2024A&A...688A.106N}, \citeads{2025ApJ...984...95R}). While the internal production of \lya radiation depends mainly on the star formation rate, its escape and propagation along the line of sight can be greatly reduced through scattering. Predicting the shape and the spatial extent of \lya emission during reionisation is challenging (e.g., \citeads{2021ApJ...914...44A}, \citeads{2024A&A...689A..10M}, \citeads{2025ApJS..278...33K} and references therein), but qualitatively, an immediate consequence of a more neutral medium surrounding a \lya-bright galaxy would be to disperse its line emission, both spatially and spectrally. A large neutral fraction in the CGM or in the adjacent intergalactic medium (IGM) would therefore cause the solid angle of \lya emission to spread out and render the emission essentially invisible (unless the system resides in a substantial ionized bubble; e.g., \citeads{2025Natur.639..897W}). On the other hand, an only moderately increased abundance of neutral gas in the outskirts of a galaxy at the end of reionisation might lead to just a more extended LAH, still observable albeit with lower surface brightness.

Measurements of the prominence and properties of \lya haloes could thus provide additional constraints on the reionisation process and topology, especially its late stages at $z \sim 6$--7. Observational data on LAHs at these redshifts are however scarce, with currently only very few known haloes at $z\geq6$ (\citeads{2009ApJ...696.1164O}, \citetalias{2017A&A...608A...8L}, \citeads{2018PASJ...70S..15S}, \citeads{2020MNRAS.492.1778M}, \citeads{2020ApJ...891..177Z}). Some of these are extremely luminous (L$_{\mathrm{Ly}\alpha} > 10^{43}\,$erg\,s$^{-1}$) and fall into the category of \lya ``blobs'', with counterparts that show complex clumpy merger UV morphology and tentative AGN signatures (\citeads{2025A&A...699A.154M}, \citeads{2025ApJ...995..150K}), making them more akin to \lya nebulae around QSOs (e.g., \citeads{2016ApJ...831...39B}, \citeads{2017ApJ...848...78F}, \citeads{2019MNRAS.482.3162A}). Statistical studies of haloes around star-forming galaxies at this redshift are also limited. \citetads{2014MNRAS.442..110M} found an extended \lya halo in a stack of 119 narrowband images of Lyman alpha emitters (LAEs) at $z=6.6$ with a rather large scale length of $\sim12\,$kpc. The other detection comes from \citetads{2021ApJ...916...22K} and \citetads{2022ApJ...931...97K} at $2\sigma$ significance at $z=6.6$ also based on the stacking of narrowband imaging data. And while the James Webb Space Telescope (JWST) has recently opened the discovery space for \lya observations towards $z\ga 7$, it is not particularly sensitive to low surface brightness emission. This explains the absence of JWST-detected LAHs, with only one tentative sighting for the $z=7$ galaxy DP7 \citepads{2021ApJ...908L..30P}. 

An obvious reason for the large absence of meaningful data of \lya haloes at $z\geq6$ is that such measurements are extremely difficult, due to the unfavourable combination of intrinsically very low surface brightness emission and cosmological surface brightness dimming. In this study we aim to improve on this situation by taking advantage of the MUSE eXtremely Deep Field (MXDF; \citeads{2023A&A...670A...4B}), arguably the most sensitive observational dataset for such purposes in existence. We conduct a systematic search to detect extended \lya emission around faint LAEs at $z\geq6$ and measure the properties of such emission. We then compare these properties with those of LAHs at $z\sim 3$, taking great care to eliminate possible biases arising from redshift-dependent differences in the sensitivity to \emph{intrinsic} surface brightness. 

The paper is organized as follows: in \autoref{sec:samp} we introduce the data and the derived samples. Section~\ref{sec:big_det} describes the construction of \lya and UV continuum images as well as our LAH detection procedure. The parameters of individually detected LAHs are determined and characterised in \autoref{sec:pars}. Section~\ref{sec:spec} is focused on spectral line properties of studied LAEs. We discuss the implications of our findings in \autoref{sec:disc} and conclude in \autoref{sec:conc}. We use AB magnitudes throughout and express all distances in physical kpc. We assume a $\Lambda$CDM cosmology with $H_0$ = 70 km s$^{-1}$ Mpc$^{-1}$, $\Omega_m$ = 0.3, and $\Omega_\Lambda$ = 0.7.
    
\section{Data and sample construction}\label{sec:samp}

Since our goal is to compare the properties of \lya haloes at two very different redshift ranges, cosmological surface-brightness dimming is a major issue, with a factor $\sim$10 loss in effective sensitivity between $z\sim 3$ and $z\sim 6$. The typical exponential scale length of a LAH with luminosity $L_{\mathrm{Ly}\alpha} \sim 10^{42}$\,erg\,s$^{-1}$ at $z=3$ is only $\sim$4~kpc  \citepalias{2017A&A...608A...8L}, comparable to the size of the ground-based point spread function (PSF) even under good conditions. Observing a halo of the same intrinsic properties but at $z\geq6$, using the same instrumental setup and depth, would lead to a $2\times$ smaller isophotal radius and complete noise dominance over any real extended emission at $\ga 1$--2 scale lengths. These effects taken together could make all the difference for the detectability of \lya haloes and must therefore be taken into account.

In the following we describe our approach to address this issue by comparing two samples at different redshifts that were extracted from observational data matched in intrinsic (i.e., corrected for cosmological dimming) surface brightness sensitivity. For both samples we use observations conducted with the Multi-Unit Spectroscopic Explorer (MUSE; \citeads{2010SPIE.7735E..08B}) mounted on the ESO Very Large Telescope (VLT)). MUSE is a panoramic integral field spectrograph with an instantaneous field of view of $1'\times 1'$ at $0\farcs2\times0\farcs2$ sampling and a wavelength range from 4700\,\AA\ to 9350\,\AA. The corresponding redshift range in \lya is $z=2.87$ to 6.67. One MUSE spectral pixel is 1.25\,\AA\ and the spectral resolution is $\sim2.6$\,\AA, depending weakly on wavelength. All MUSE datasets used in this study  benefit from improved image quality from Ground-Layer Adaptive Optics (GLAO), resulting in a PSF full width at half maximum of typically better than 0\farcs5.

\subsection{High-redshift (\texorpdfstring{$z\geq6$}{z ≥ 6}) sample}
\label{sec:z6sample}

We selected our high-redshift sample from the MUSE eXtremely Deep Field (MXDF, \citeads{2023A&A...670A...4B}), a blind spectroscopic survey consisting of many exposures in a single pointing inside the Hubble Ultra Deep Field (HUDF). The MXDF has a coadded exposure time of 141 hours over a circular field of view of $1'$ in diameter and reaches a 3$\sigma$ surface brightness limit of $\sim 1.0\times10^{-19}$ ergs$^{-1}$ cm$^{-2}$ arcsec$^{-2}$ in a $1\arcsec\times 1\arcsec$ aperture for an unresolved emission line around 7300\,\AA. Our sample was drawn from the AMUSED catalogue and database by \citetads{2023A&A...670A...4B}\footnote{https://amused.univ-lyon1.fr/project/UDF/}, comprising altogether 2221 objects in the HUDF with spectroscopically measured redshifts. The location of the MXDF inside the HUDF implies that extensive space-based imaging data is also available (see \autoref{sec:uvdata}).

\begin{figure}
    \centering
    \includegraphics[width=\linewidth]{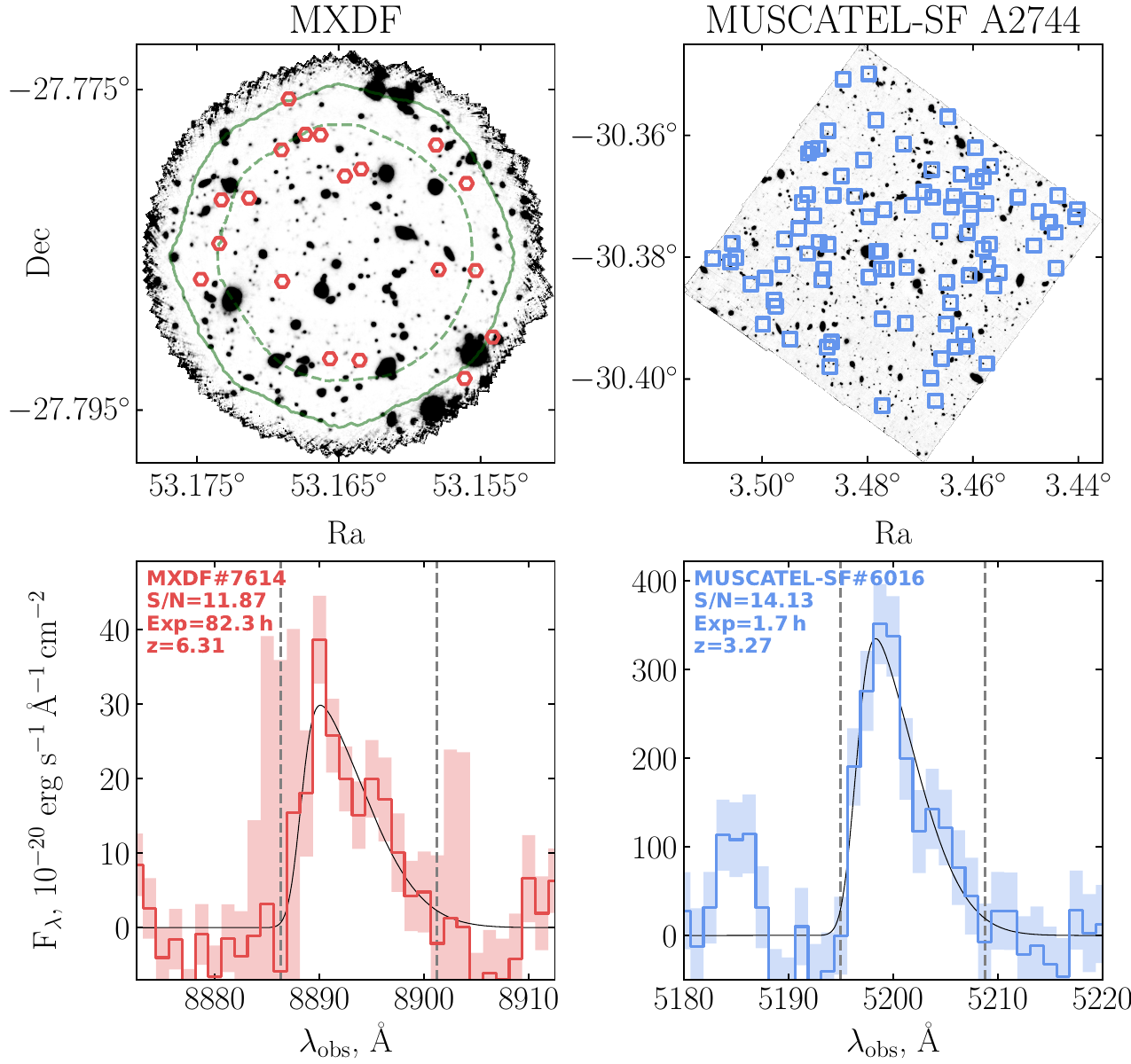}
    \caption{\textit{Top:} Spatial distribution of LAEs from high and lower redshift samples in their respective fields. The dashed and solid lines in the top left panel show the MXDF 30- and 140-hour exposure time isocontours. \textit{Bottom:} Examples of \lya spectra from both samples. The dashed vertical lines indicate the spectral bandwidth used to extract the pseudo-narrowband images used in the analysis. The thin black lines show skewed Gaussian fits to the observed line profiles.}
    \label{FoV}
\end{figure}

We selected all galaxies in AMUSED with $z\geq6$ falling within the MXDF footprint. We discarded one of these (AMUSED ID 8003) that was detected only in the shallower UDF-MOSAIC data, but not recovered in the MXDF datacube, implying it might be a noise spike rather than a real object. This left us with a set of 18 confirmed \lya emitters. Their main properties are summarised in \autoref{table:1}. Most of the sample lies in the high-exposure ($t_{\mathrm{exp}} > 100^\mathrm{h}$) area of the MXDF, with only 4 galaxies outside that region ($t_\mathrm{exp}\approx30^\mathrm{h}$). The top left panel of \autoref{FoV} shows the spatial locations of these objects superimposed on the MXDF white-light image. 

Since all galaxies in this sample are single-line detections, we have to ensure that their classification as high-$z$ LAEs is robust and the sample is not contaminated by low-redshift interlopers. This is supported by several arguments: (i) As discussed in \autoref{sec:spec}, the \lya line is resolved for all objects and generally displays the characteristic skewed spectral profile, excluding the [\ion{O}{ii}] $\lambda\lambda$3727, 3729 doublet as a possible contaminant and also arguing strongly against [\ion{O}{iii}] $\lambda$5007 which for such faint galaxies is generally unresolved in MUSE. (ii) Most objects show detectable continuum counterparts in JWST F115W images but are undetected in the extremely deep HST F775W data, leading to photometric redshifts consistent with their spectroscopic $z$. (iii) Based on published luminosity functions of the relevant lines \citepads{2016MNRAS.461.1076C}, we expect that more than 90\% of the single line emitters at wavelengths $\lambda > 8500$\,\AA\ and close to the MXDF flux limit are LAEs.

The redshifts of the objects are indicated by the red dashes in the bottom right corner of\ \ \autoref{SB_sens}. Due to the strong impact of the terrestrial night sky background on the achievable sensitivity at wavelengths $\ga 7000$\,\AA, the selected LAEs are typically found in narrow spectral windows in-between the atmospheric OH lines clearly visible in the variance spectrum. This leads to several of the \lya lines being adjacent to high-variance regions which impacts the propagation of formal uncertainties and makes the estimation of realistic error bars a crucial part of the analysis. 

To investigate the \lya line spectral profiles, we re-extracted all LAE spectra from a continuum-subtracted datacube, with the continuum estimated by median filtering the original cube with a window of 151 pixels in dispersion direction, ensuring the removal of the stellar continuum of the galaxy itself as well as fore- and background objects. Spectra were extracted by summing the data over an aperture of $1''$ in radius. Overlapping foreground sources were masked to eliminate possible contamination from any intervening emission lines. In the bottom left panel of \autoref{FoV} we show an example \lya spectral line profile displaying the characteristic skewed line shape.

        \begin{table}
            \def\arraystretch{1.15}
            \caption{Properties of the high-$z$ LAE sample. }
            \label{table:1}
            \centering
            \begin{tabular}{c c c c c c}
              \hline\hline
              ID & $z$ & $\log_{10} (\mathrm{L}^\mathrm{CoG}_{\mathrm{Ly}\alpha} / [\mathrm{erg\,s^{-1}}])$ & M$_\mathrm{UV}$  \\
              \hline
              802  & 6.11 & $41.82\pm 0.19$ & $-16.83\pm 0.15$\\
              852  & 6.64 & $42.05\pm 0.05$ & $-18.69\pm 0.08$\\
              6332 & 6.33 & $41.90\pm 0.28$ & $-16.41\pm 0.21$\\
              7610 & 6.10 & $42.11\pm 0.06$ & $-17.71\pm 0.13$\\
              7614 & 6.31 & $42.08\pm 0.15$ & $-18.96\pm 0.07$\\
              7692 & 6.31 & $41.68\pm 0.22$ & $-18.09\pm 0.09$\\
              7699 & 6.64 & $42.06\pm 0.03$ & $-19.46\pm 0.10$\\
              8145 & 6.04 & $41.48\pm 0.14$ & $-$ \\
              8172 & 6.00 & $41.47\pm 0.15$ & $-15.33\pm 1.07$\\
              8179 & 6.60 & $41.47\pm 0.39$ & $-$ \\
              8209 & 6.57 & $40.93\pm 0.13$ & $-$ \\
              8211 & 6.00 & $41.36\pm 0.13$ & $-$ \\
              8214 & 6.63 & $41.70\pm 0.08$ & $-18.63\pm 0.09$\\
              8367 & 6.58 & $41.15\pm 0.16$ & $-$ \\
              8379 & 6.63 & $41.44\pm 0.14$  & $-16.63\pm 0.3$\\
              8384 & 6.05 & $41.13\pm 0.15$ & $-$ \\
              8433 & 6.55 & $41.17\pm 0.16$ & $-$ \\
              8461 & 6.49 & $41.46\pm 0.19$ & $-$ \\
              \hline
            \end{tabular}
            \tablefoot{
                ID: source identifier in the AMUSED database. $z$: redshift determined from the \lya line peak position. $\log_{10} \mathrm{L}$: \lya luminosities derived with the curve of growth method on the NB images. M$_\mathrm{UV}$: rest-frame UV magnitude in the F115W filter of the associated UV counterpart from JADES if present.
            }
        \end{table}

\subsection{Comparison sample at \texorpdfstring{$z\sim 3.2$}{z∼3.2}}
\label{sec:z3sample}

We employed two main criteria to select our lower-redshift comparison sample. (i) It should feature a large difference in mean redshift from our $z\geq6$ dataset, preferably in the range $3\la z \la 4$ where LAH properties have been extensively studied. (ii) The observational data used for the comparison sample should have a similar \emph{intrinsic} \lya SB sensitivity as the MXDF data at $z\geq6$, after accounting for cosmological dimming as well as for the (mild) variations of the MUSE instrument throughput with wavelength. Instead of degrading deeper exposures with artificial noise, which always introduces the danger of being too idealised, we searched our portfolio of MUSE surveys for a dataset with the appropriately matched quality. Besides the MXDF we considered the following surveys (numbers in parentheses indicating the mean exposure times): 
UDF-MOSAIC ($10^\mathrm{h}$) and UDF-10 ($31^\mathrm{h}$, \citeads{2017A&A...608A...1B}, \citeads{2023A&A...670A...4B}); MUSE-Wide ($1^\mathrm{h}$, \citeads{2019A&A...624A.141U}); and the recent MUSCATEL MUSE Cosmic Assembly survey Targeting Extragalactic Legacy fields) survey (Urrutia et al., in preparation; see also below) with three different depths, MUSCATEL-DF (25$^\mathrm{h}$), -MF (5$^\mathrm{h}$), and -SF (1.7$^\mathrm{h}$).

To quantify the SB sensitivity at any given wavelength, we assumed it to be proportional to the so-called effective noise $\sigma_\mathrm{eff}(\lambda)$ \citepads{2019A&A...624A.141U} which we routinely determine for each datacube in the postprocessing following the data reduction (see also \citetads{2020A&A...641A..28W} for a comprehensive discussion of the noise properties of MUSE data). The effective noise is given in calibrated flux units per pixel and includes both the wavelength-dependent MUSE throughput and the co-added exposure level of the corresponding datacube, multiplied by an approximate correction for covariance losses due to correlated noise. To this we applied a $(1+z)^4$ modulation factor to account also for cosmological surface brightness dimming. We thus define the \emph{intrinsic} \lya SB sensitivity within a standard aperture of 1\arcsec\ in one spectral layer as  
\begin{equation}
    \mathrm{SB}_{\mathrm{lim,\, Ly}\alpha}(\lambda) = \sigma_\mathrm{eff}(\lambda)\, \left[z_{\mathrm{Ly}\alpha}(\lambda)+1\right]^4\times1.25\,\AA\label{eq:SB_lim}.
 \end{equation}

We note that this definition of $\SBlim(\lambda)$ is not a prediction of the final SB limit of the constructed pseudo-narrowband images, but a measure of a cosmologically corrected depth of MUSE deep surveys.

\begin{figure}
\centering
\includegraphics[width=\linewidth]{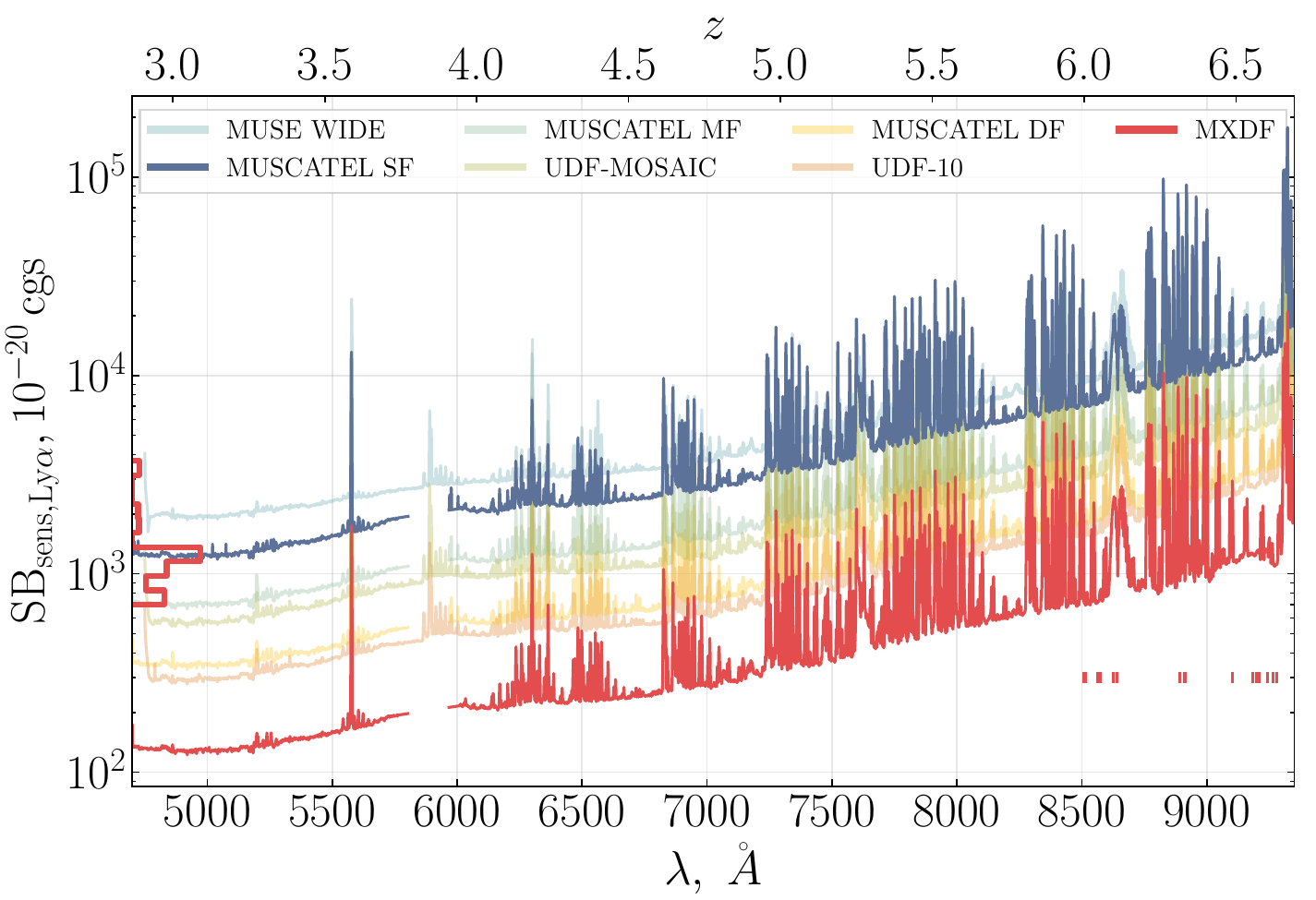}
\caption{Intrinsic (redshift-corrected) surface brightness sensitivity limit $\SBlim$ as a function of wavelength (\lya redshift) for MUSE datacubes from different surveys as indicated in the legend (see text for references). The red horizontal histogram on the left shows the distribution of $\SBlim$ of our $z\geq6$ sample, and the red dashes mark the observed \lya wavelengths of the objects in that sample.}
\label{SB_sens}
\end{figure}
        
In \autoref{SB_sens} we compare the values of $\SBlim$ of typical MUSE cubes obtained in several of our `deep field' spectroscopic surveys. As the abundance of skylines makes it difficult to easily estimate the sensitivity levels for $\lambda \ga 8000$\,\AA, a histogram of $\SBlim$ values for the high-redshift sample evaluated at the location of each object is plotted along the vertical axis on the left. 

This figure suggests that LAEs at $z\sim 3$--3.5 from MUSCATEL-SF would provide the best match in $\SBlim$ to our ultradeep LAE sample at $z\geq6$. The MUSCATEL project (PIs: L.~Wisotzki, R.~Bacon, T.~Contini) is a recently concluded blind MUSE survey in the  parallel fields of the Hubble Frontier Fields (HFF, \citeads{2021ApJS..256...27P}, \citeads{2017ApJ...837...97L}), focussing on regions near the clusters A2744, M0416, AS1063, and A370 -- but sufficiently far from the cluster centres so that magnification effects due to cluster gravitational lensing are very small and for the present purpose negligible.

MUSCATEL encompasses three different levels of coadded exposure time: ``Shallow fields'' (SF) with $t_{\mathrm{exp}} = 100^\mathrm{m}$, ``medium-deep'' (MF) with $t_{\mathrm{exp}} = 5^\mathrm{h}$, and ``deep'' (DF) with $t_{\mathrm{exp}} = 25^\mathrm{h}$. A full data release of datacubes and catalogues and an accompanying survey publication are in preparation. The MUSCATEL data set has also been described in \citeads{2026A&A...708A.214P}, where the authors drew a sample of galaxies from the MUSCATEL survey to study their galactic outflow properties. In broad terms, the strategy for MUSCATEL has been similar to that adopted in previous MUSE surveys of the HUDF and surrounding regions. In particular, LAEs were selected as significant emission line peaks in the 3D datacubes using the matched-filter algorithm LSDCat (\citeads{2017A&A...602A.111H}, \citeads{2023AN....34420091H}), followed by redshift (and redshift confidence) assignment by three independent human classifiers. For the present study we restricted ourselves to the ``SF'' subset in the parallel field to A2744, which provides data with the best seeing conditions among the MUSCATEL-SF observations. We initially selected all LAEs at $3 \le z \le 3.4$, of which we then removed two due to their close proximity to a bright galaxy or a star. This left us with 90 objects for the $z\sim3$ comparison sample. We expect minimal contamination of the sample by the low-redshift line emitters (primarily [\ion{O}{II}] and [\ion{O}{III}]) as they would have distinct spectral shapes, well-resolved stellar counterparts and are too few, given the survey volume at the relevant redshifts. See \citetads{2015A&A...575A..75B} and \citetads{2023A&A...670A...4B} for an extended discussion on redshift classification and potential low-$z$ interlopers.

The spatial distribution of these sources over the MUSCATEL-SF A2744 footprint is shown in \autoref{FoV} alongside an example \lya spectrum. The extraction of their spectra for this study was done in the same way as for the $z\geq6$ MXDF sample, see \autoref{sec:z6sample}.

\subsection{Rest-frame UV data}
\label{sec:uvdata}

In addition to the MUSE observations we used space-based imaging at rest-frame UV wavelengths to estimate the stellar continuum properties of the galaxies in our study. For the high-redshift sample we exploited the JWST Advanced Deep Extragalactic Survey (JADES \citeads{2026ApJS..283....6E}, \citeads{2024A&A...690A.288B}, \citeads{2025ApJS..277....4D}), which covers the entire footprint of the MXDF with 11 filters ranging from F090W to F444W. Among the available filters we chose the NIRCam F115W filter as it is free from \lya contamination and provides the deepest observations in the rest-frame UV at $z\sim6$. For the $z\sim 3$ sample in the MUSCATEL-SF A2744 field we employed HST/ACS observations from the HFF project \citepads{2018ApJS..235...14S}, selecting the ultradeep F814W band as the main reference for similar reasons. For each dataset we also consulted the associated source catalogue with photometric redshift estimates.
        
For each \lya emitter in both samples we identified by visual inspection the most likely UV counterpart (in 11/18 objects of the MXDF and all of the MUSCATEL-SF sample), or decided that no such counterpart could be detected (8/18 and 0/90, respectively). For 8 objects in the lower-redshift sample and for 2 objects in the $z\geq6$ sample we found a clear UV source (albeit rather faint, $m_\mathrm{F814W}^\mathrm{lim}=29.1\,$mag and $m_\mathrm{F115W}^\mathrm{lim}=30.5\,$mag) spatially coinciding with the \lya emission, which however had no record in the corresponding published catalogue. In all those cases we assumed the uncatalogued UV source to be the counterpart. For 7 of the 18 high-redshift LAEs we deemed the angular and/or photometric redshift distances between LAE centroid and the closest UV source as too large (more than 5.5 kpc in all cases), so that no counterpart was assigned. For the $z\sim3$ sample we encountered no such case. We found a mean offset of $0\farcs25$ between \lya and UV centroids for both samples, only slightly more than one MUSE spaxel and in excellent agreement with previous measurements \citepads{2022A&A...666A..78C}.

We used the available broadband photometric data also to estimate stellar masses ($M_\star$) and star formation rates (SFRs) by fitting the spectral energy distributions (SEDs) using \texttt{Prospector} (\citeads{2017ApJ...837..170L}, \citeads{2021ApJS..254...22J}). \texttt{Prospector} is a Bayesian forward-modelling code developed to derive the physical properties of galaxies from photometric or spectroscopic data using Markov chain Monte Carlo sampling. It uses FSPS stellar population models (\citeads{2009ApJ...699..486C}, \citeads{2010ApJ...712..833C}) with Chabrier IMF \citepads{2003PASP..115..763C}, self-consistent nebular emission generated with \texttt{Cloudy} as described in \citetads{2017ApJ...840...44B}, and dust attenuation law with \citetads{2000ApJ...533..682C} extinction curve as a default. The main advantage of \texttt{Prospector} is a very flexible treatment of star formation histories (SFHs): the built-in functions allow for various parametric models (such as $\tau$ and delayed-$\tau$) and five non-parametric ones (\texttt{continuity}, \texttt{continuity\char`_flex}, \texttt{continuity\char`_psb}, \texttt{dirichlet}, and \texttt{stochastic}), which mainly vary in the resulting smoothness of the reconstructed SFH. For all galaxies with detected UV counterparts we estimated their masses and SFRs with the delayed-$\tau$ SFH and a log-uniform mass prior.

The choice of a specific star formation history can significantly influence SED derived stellar mass and SFR by altering the mass-to-light ratio (e.g. \citeads{2020ApJ...904...33L}). To explore the effect of the adopted SFH on the estimated galaxy properties, we conducted the following experiment: at both redshifts we selected 4 galaxies that uniformly probe the M$_\text{UV}$ distribution in each sample. Then, we again estimated stellar masses and SFRs, now using \texttt{Prospector} assuming six different star formation histories: $\tau$ and delayed-$\tau$, \texttt{continuity}, \texttt{continuity\char`_flex}, \texttt{dirichlet}, and \texttt{stochastic}\footnote{We excluded the \texttt{continuity\char`_psb} SFH prescription as it is tailored to galaxies in the post-starburst phase.}. We report that at $z\geq6$ the stellar masses inferred assuming different star formation histories are generally consistent with each other within $\sim0.3\,$dex. At $z\sim3$ agreement between different codes and SFH treatments gets weaker and the difference increases to $\sim0.6\,$dex. This happens because we lack the photometric coverage redward of the Balmer break or even miss the break completely in $\sim35\%$ of the objects due to the smaller F160W footprint. At both redshifts we recover rising SFHs for the nonparametric prescriptions and very low ages for the $\tau$ models, indicating active star formation in these galaxies at the time of observation. From this experiment we conclude that derived stellar masses and SFRs are fairly robust against variations in the star formation histories given the available data.

For objects where no UV-counterpart is detected with measured broadband photometry we get constraints on the stellar mass and SFR from the photometric limits of the corresponding catalogues. Following \texttt{Prospector} user guidelines, we treat these non-detections as having a flux of 0 in each band with an associated uncertainty derived from the catalogue $5\sigma$ flux upper limits. For 8 $z\geq6$ UV non-detections the resulting stellar mass and SFR upper limits (quoted as a 99th percentile of the posterior) are essentially the same and amount to $10^{7.8}\,$M$_\odot$ and $0.09\,$M$_\odot$\,yr$^{-1}$. For $z\sim3$ the same exercise gives $10^{8.5}\,$M$_\odot$ and $0.3\,$M$_\odot$\,yr$^{-1}$.

\begin{figure*}
\centering
\includegraphics[width=\linewidth]{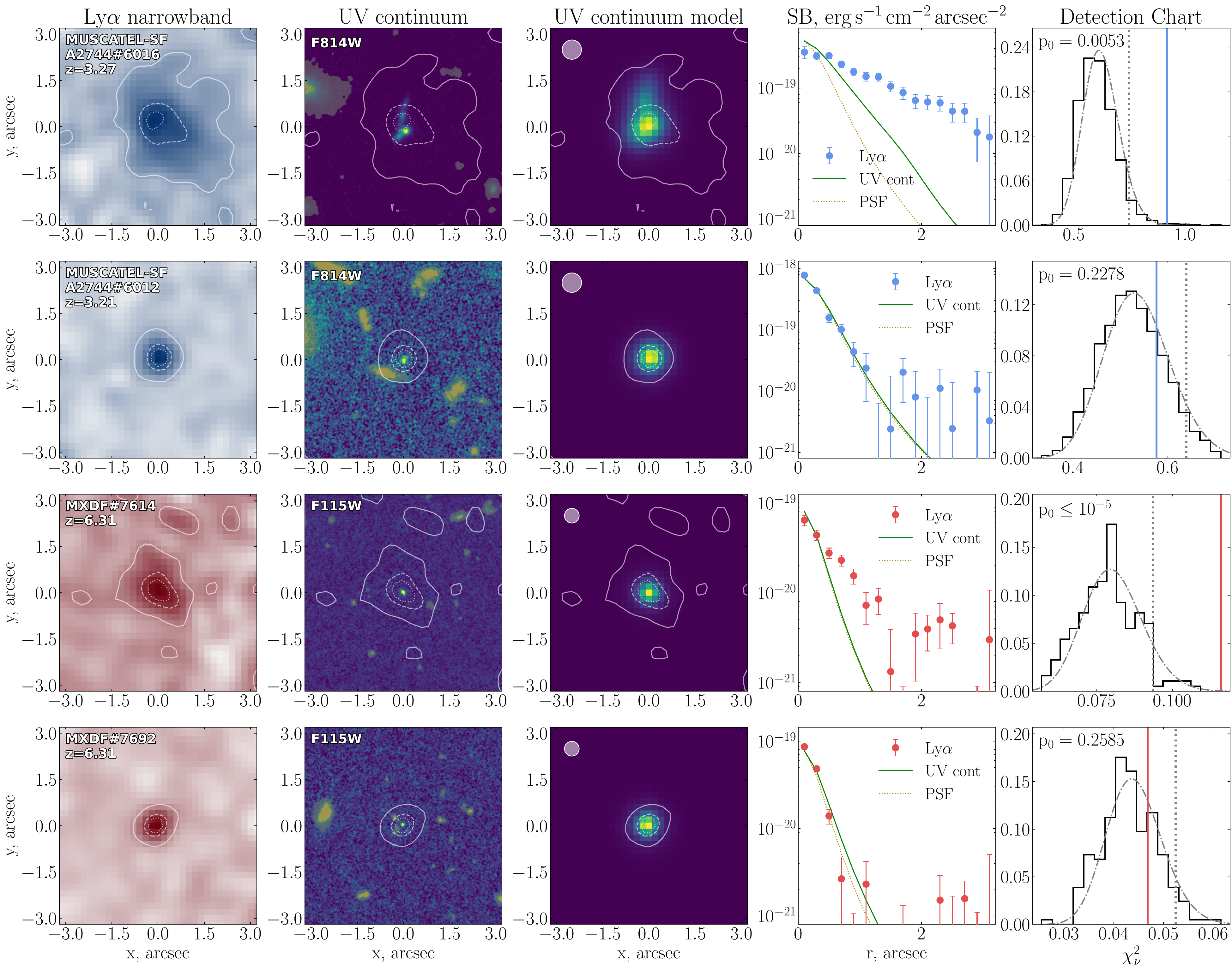}
\caption{Examples of LAEs in our sample. The upper two rows show objects at $z\simeq 3.2$ subset from MUSCATEL-SF, the lower rows show $z\geq6$ LAEs in the MXDF. Objects in the first and third rows are examples of clear \lya halo detections. \textit{First column}: \lya narrowband images (smoothed for display purposes), with overlayed SB$\times(1+z)^4$ contours of $10^{-15}$ (solid), $2\cdot10^{-15}$ (dashed) and $2.5\cdot10^{-15}$ (dotted) erg\,s$^{-1}$\,cm$^{-2}$\,arcsec$^{-2}$. \textit{Second column}: UV continuum images from either HST or JWST, with the filter specified in the top left corner. The overlay contours are the same as in the first column. \textit{Third column}: GALFIT model of the UV continuum image resampled and convolved to MUSE resolution. The circle in the top left corner of each panel indicates the FWHM of MUSE PSF at the \lya observed wavelength of each object. \textit{Fourth column}: Azimuthally averaged SB profile of the \lya emission (datapoints with error bars) compared to the UV continuum model fit (solid line) and to the PSF (dotted line). \textit{Fifth column}: Distribution of reduced $\chi^2_\nu$ values obtained in empty fields after inserting the ``true'' UV continuum model and then subtracting a flux-matched one. The thick vertical line shows $\chi^2_\nu$ obtained for the object itself after subtracting the flux-matched UV continuum model. The probability $p_0$ of the null hypothesis (\lya consistent with UV) is shown in the top left. The dotted line marks the $\chi^2_\nu$ threshold corresponding to $p_0=0.05$. The dash-dotted line shows the $\chi^2$ distribution fitted to the histogram.}
\label{ex_plot}
\end{figure*}

\section{Detection of \texorpdfstring{\lya}{Lyα }haloes}
\label{sec:big_det}

In order to study individual \lya haloes we need to distinguish between objects with and without significantly detected extended emission. In accordance with most previous work (e.g., \citetalias{2016A&A...587A..98W}, \citetalias{2017A&A...608A...8L}) we base this decision on a comparison with the spatial extent of the stellar light distribution: If the SB distribution of the \lya emission is consistent with the distribution of starlight (within the error bars and after accounting for PSF blurring) we adopt the null hypothesis that the observed \lya emission originates mainly in the ISM of the host galaxy, with no evidence for a circumgalactic component due to scattered light or added emission. Only if this null hypothesis is significantly rejected do we take the \lya halo as detected. As in our previous LAH studies using MUSE (e.g. \citetalias{2016A&A...587A..98W} and \citetalias{2017A&A...608A...8L}), a key ingredient to this test are the deep high-resolution space-based images available in our fields, allowing us to constrain the stellar light distribution much more accurately than possible from the MUSE data alone.

\subsection{Image construction}

\subsubsection{\texorpdfstring{\lya}{Lyα }pseudo-narrowband images}

For each object we extracted a $6\farcs6\times6\farcs6$ (33$\times$33 MUSE spaxels) pseudo-narrowband (NB) image from the corresponding MUSE datacube. The image size was chosen so that we could detect and model the extended \lya emission of all objects of interest while limiting the impact of nearby sources. We used the median filter subtracted cubes (see \citetalias{2016A&A...587A..98W} for details) to approximately remove the continuum of target galaxies and back-/foreground sources.

The size of each narrowband was chosen manually to incorporate most (but not strictly all) of the \lya flux in the band (see example in \autoref{FoV}). Since the blue peaks of \lya emission lines are increasingly lost to IGM absorption towards high redshifts (e.g., \citeads{2020MNRAS.499.1395M}, \citeads{2021ApJ...908...36H}) we treated our $z\sim 3.2$ objects similarly and deliberately excluded any blue peak emission also in those.  The resulting typical NB width for both samples was 9 MUSE spectral pixels (11.25\,\AA). In \autoref{sec:spec} we show that \lya lines are much narrower at $z\geq6$ than at $z\sim3$, this means that using a common rest frame spectral range for extracting NB images would either leave a considerable fraction of \lya flux out of the NB at $z\sim3$ or effectively add pure noise to the $z\geq6$ narrowbands. As the detection of extended emission is quite sensitive to the correct noise estimation of the extracted signal, we decide to define our NB limits in terms of encompassed flux rather than fixed velocity window. Examples of constructed \lya pseudo-narrowband images are shown in the left column of \autoref{ex_plot}.

\subsubsection{UV continuum images and model fits}
\label{sec:uvfits}

A typical \lya emitter detected in MUSE is a low-mass star-forming galaxy identified by its bright emission line, but with a very faint and often tiny continuum counterpart. Using the deepest available reference images (see \autoref{sec:uvdata}) we extracted matching $6\farcs6\times6\farcs6$ cutouts (220$\times$220 pixels at the 0\farcs03 sampling of the HST and JWST data). All sources except the assigned UV counterpart were masked. We used GALFIT (\citeads{2002AJ....124..266P}, \citeyearads{2010AJ....139.2097P}) to model each object as a linear combination of standard functions (point source, exponential, Sérsic profile) convolved with the PSF. For the MXDF sample we used a simulated JWST NIRCam PSF from the JWST PSF simulation library\footnote{\scriptsize\url{https://stsci.app.box.com/v/jwst-simulated-psf-library}}. For the HST data of the MUSCATEL sample we assumed a symmetric Gaussian PSF with FWHM adopted from \citetads{2018ApJS..235...14S}. Single component models were sufficient to fit the UV counterpart of all the $z\geq6$ objects. Of the lower redshift sample 61/90 objects could be modelled with just one component, 20 required two, 7 were modelled as three- and 2 as four-component systems. The 7  high-$z$ objects undetected in the continuum were treated as point sources in the remaining analysis. Second and third columns of \autoref{ex_plot} show examples of the UV continuum images and models after resampling to MUSE resolution and convolving with the MUSE PSF.

\subsection{Detection procedure}
\label{sec:det}

We now describe our LAH detection test procedure. In contrast to previous studies, our null hypothesis test for the presence of extended \lya emission is based on a spaxel-to-spaxel evaluation of the light distribution, combined with an empirical self-calibration of the noise properties using empty regions. For each object we performed the following steps:

\begin{enumerate}

\item We determined the centroid and flux scaling amplitude of the resampled and convolved UV continuum model to best match the \lya NB data. For the centroid we fitted the NB image with a 2D Gaussian using the photutils package \citepads{Photutils_17129028}. We then calculated the scaling factor for the continuum model placed at this location by $\chi^2$ minimisation and record the corresponding value of $\chi^2$ 

\item Next we defined a list of empty fields in the NB image of the full datacube. To do so we first smoothed the image with a Gaussian kernel with $\sigma$ of 1.5 pixels and constructed a binary mask to hide extended sources. Recall that the NB images were already continuum-subtracted using the prior median-filtered datacube, so this masking step mainly marked emission line sources (in particular by foreground [\ion{O}{ii}] emitters), but also cases of poor continuum subtraction. We projected a rectangular grid with a spacing of 16 pixels onto the image, and at each location we tested if a small ($3\farcs3\times3\farcs3$) NB image centred on this position intersected with the mask. Only if there was no such intersection did we record the location as an empty region. The resulting number of empty regions per object was around 300 in the MXDF and between 1500 and 2000 in the MUSCATEL-SF footprint.

\item To perform the null hypothesis test, we required a robust estimate of the noise inside each of the investigated apertures. While the ``effective noise'' approach \citepads{2019A&A...624A.141U} provides a good approximation for large-aperture measurements, it overestimates the noise level for the spaxel-by-spaxel comparison within small apertures around our LAEs. We therefore self-calibrated the goodness-of-fit test by the following iteration over the grid of empty regions, object by object: We inserted a copy of the UV continuum model scaled to the observed \lya flux at each grid point and then re-determined the best centroid and matching scale factor as described above. After that we subtracted the scaled UV model from this mock field. Combining the residuals in a single empty region provided a $\chi^2$ value (note that the noise level for faint objects in MUSE is entirely dominated by the background with a minor contribution of detector readout noise; shot noise from the object is negligible). For each object the histogram of $\chi^2$ values of all inserted sources in the empty regions then resulted in the expected $\chi^2$ distribution for the null hypothesis that the \lya light distribution follows the UV. This procedure fully includes the errors from centroiding uncertainties, continuum subtraction artefacts, and sky subtraction residuals leading to imperfect flatfielding (particularly relevant at $z\geq6$ where \lya is observed between the multiplets of the OH night sky emission forest). We also made sure to account for variations in the received exposure time with location, again mainly relevant for the MXDF with its steep exposure gradient outside the $1'\times 1'$ circular area \citepads{2023A&A...670A...4B}.
            
\item As the last step we assessed for each object whether the \lya spatial distribution is statistically distinguishable from the UV continuum model by determining the location of the measured $\chi^2$ value of the object relative to the self-calibrated cumulative $\chi^2$ distribution, expressed as a normalised frequency $p$. As in most previous studies we adopted a value of $p < p_0=0.05$ as the threshold for detecting extended emission; but now this threshold does not refer to a standard $\chi^2$ test but to the actually measured distribution of residuals. In \autoref{ex_plot} we show examples for these $\chi^2$ histograms together with the observed value of the object itself. From the fact that these histograms peak at $\chi^2$ values well below unity it can be seen that the effective noise indeed overestimates the pixel uncertainties, implying that our self-calibration procedure is both more realistic and more sensitive to detect faint \lya haloes than previous approaches.

\end{enumerate}

\subsection{Results}

\subsubsection{Radial profiles}

To visualise the spatial distribution of \lya emission, we extracted azimuthally averaged surface brightness profiles in concentric circular annuli for the objects in our sample. Our annuli have radial increments of 1 MUSE pixel (0\farcs2) to properly sample the inner LAH regions which are strongly affected by PSF blurring. Taking advantage of the procedure outlined in \autoref{sec:det}, we also estimated self-calibrated error bars for the SB profiles by constructing SB profiles of the empty fields using the same annuli and calculated the interquartile range (IQR) of all profiles in each radial bin. The values of IQR/1.34 were then adopted as 1$\sigma$ uncertainty at each radius. We used the IQR statistic rather than standard deviations as uncertainty estimator because of its greater robustness against outliers. The thus derived error bars are typically 1.2--1.7 $\times$ smaller than the ones formally propagated from the effective variances, but for objects located near strong sky emission lines (relevant for several of the $z\geq6$ objects) this ratio increases up to 5.

\subsubsection{LAH detection fractions}
\label{sec:detfrac}

Before the final assessment of which objects have significantly detected LAHs we performed a visual inspection of all \lya NB images and SB profiles to check for the integrity of the formal detections. We found and discarded four $z\sim 3.2$ objects with high relative $\chi^2$ values which however appeared to be driven by sky or continuum subtraction residuals. After this last cleaning step we count 40/90 ($(44\pm5)\%$) detected LAHs at $z\sim 3.2$, and 4/11 ($(36\pm15)\%$) among UV detected LAEs and 2/7 ($(29\pm17)\%$) among UV non-detected LAEs at $z\geq6$. As our study was designed to compensate for cosmological SB dimming, we a priori expect that the detection rate should depend mainly on the intrinsic prominence of LAHs in these two redshift domains. The fact that the LAH fractions are not statistically significantly different might be taken as indication that also the typical LAH properties are similar. We demonstrate in \autoref{sec:pars} below that this is however not the case.

At face value our LAH detection rates are much lower than the $\sim80\%$ reported in \citetalias{2016A&A...587A..98W} and \citetalias{2017A&A...608A...8L}. For the redshift $z\sim 3.2$ comparison sample this can be easily explained by the reduced sensitivity of our current data compared to the much deeper exposures used for those studies. \citetalias{2017A&A...608A...8L} demonstrated in their Appendix how a factor 3 in exposure time already makes a big difference for measuring LAH properties. Our MUSCATEL-SF data have a $6\times$ shorter exposure time than the UDF-Mosaic and $20\times$ shorter than the UDF-10 and HDFS used by \citetalias{2017A&A...608A...8L} and \citetalias{2016A&A...587A..98W}. Our stacking analysis (see \autoref{sec:stacks}) indeed demonstrates that the non-detections among our $z\sim 3.2$ objects typically have perfectly normal extended LAHs, which are just a little bit too faint to be detected individually at the sensitivity of the MUSCATEL shallow fields.

For our high redshift sample the comparison to \citetalias{2016A&A...587A..98W} and \citetalias{2017A&A...608A...8L} is not so easy to perform. On the one hand the effects of cosmological dimming make the intrinsic SB sensitivity (see \autoref{SB_sens}) of the MXDF at $z\geq6$ lower than that of the UDF-10 at $z<5.5$ and of the UDF-Mosaic at $z<4$, implying that a somewhat lower detection fraction should be expected. On the other hand, there could also be real differences in the properties of LAHs at these highest redshifts. In the following we show that a more detailed comparison of the LAH properties in our two samples, including a stacking analysis, strongly suggests that this is indeed so, and that even deeper data than the MXDF would probably not result in a much higher LAH detection rate at $z\geq6$.

\section{\texorpdfstring{\lya}{Lyα }halo properties}
\label{sec:pars}

\subsection{Surface brightness profiles and modelling}
\label{sec:sb}

\begin{figure*}
    \centering
    \sidecaption
    \includegraphics[width=12cm]{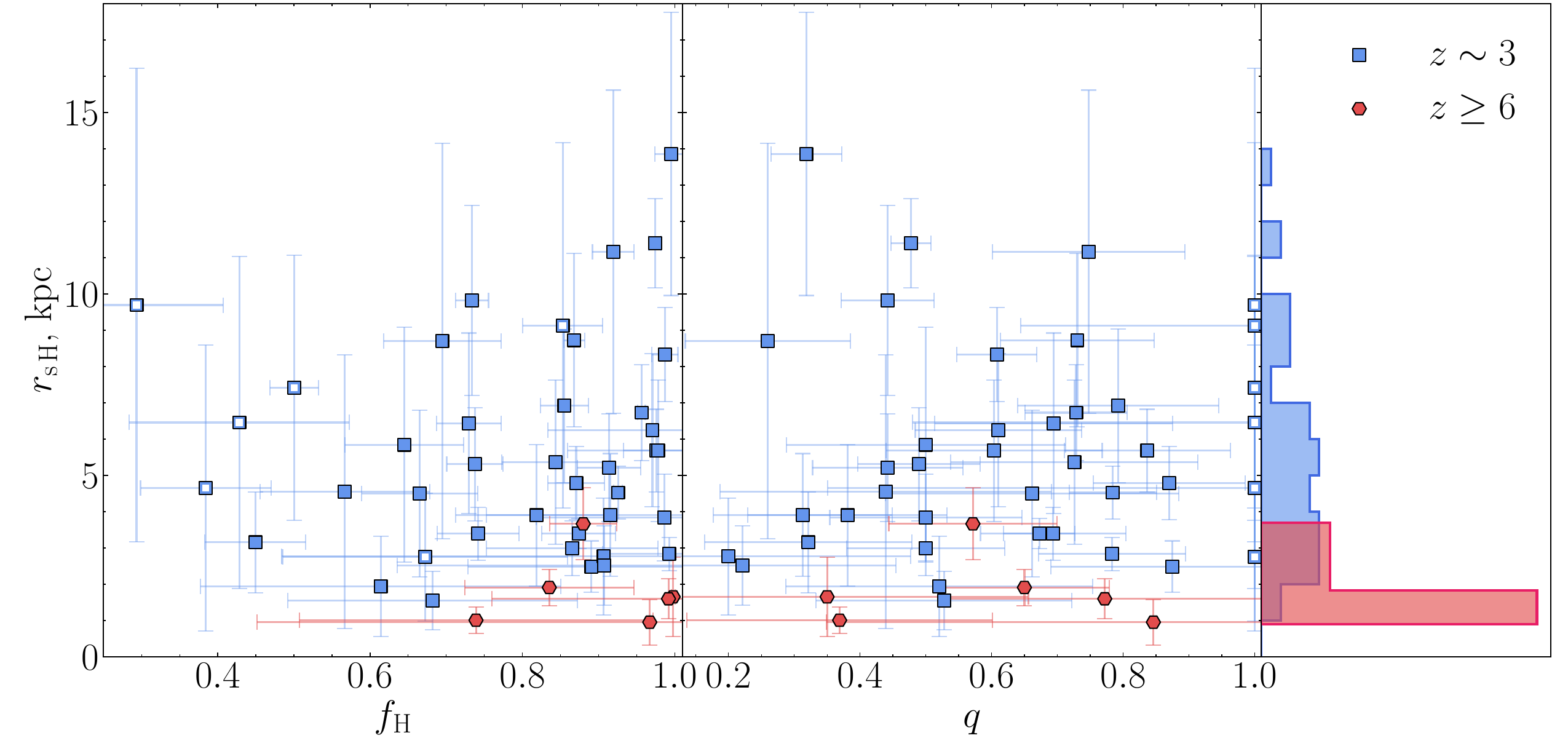}
    \caption{Comparison of main LAH parameters for the $z\sim3$ and $z\geq6$ samples. 
    The $z\sim3$ objects are shown as blue squares, the $z\geq6$ objects as red hexagons. The open markers indicate objects where the halo was forced to be circular in the fit. In the left panel we plot halo scale lengths against halo flux fractions, in the middle panel against halo axis ratios. The right panel shows the marginalised histograms of our fitted halo scale lengths, with the same colour coding as before.}
    \label{par_comp}
\end{figure*}

We now proceed to determine the main properties of our detected LAHs. In line with previous LAH studies (\citetalias{2016A&A...587A..98W}, \citetalias{2017A&A...608A...8L}) we approximate the \lya emission as the sum of a central component that follows the spatial distribution of the stellar UV continuum and an extended halo component that is usually well described by an exponential surface brightness profile.
        
For the central component models we adopted our HST- and JWST-based UV continuum fits from \autoref{sec:uvfits}. For each object we reconstructed the stellar light distribution in the UV, convolved with the MUSE PSF at the corresponding wavelength, keeping all parameters fixed except for the total \lya flux of the component ($F_\mathrm{c})$. The halo component was taken as a single exponential with four parameters: scale length ($\rsh$), axis ratio ($q_\mathrm{H}$), position angle (PA) and the total flux ($F_\mathrm{H}$). To fit the model to the \lya NB images, we again used GALFIT, with the additional constraint of forcing both components to have the same centroid. The fit was performed within a circular aperture of 10--14 pixels in radius, depending somewhat on the extent of the halo, plus a mask to avoid projected contaminating objects if required. For 6 LAHs in the $z\sim3$ sample we had to force circularity of the halo for the fit to converge. In the case of the $z\geq6$ objects, all halo fits converged with $q$ kept free (but recall that only 6/18 objects show a detectable LAH). From these fits we calculated the halo flux fractions, $f_\mathrm{H}=F_\mathrm{H}/(F_\mathrm{H}+F_\mathrm{c})$. 
        
\autoref{par_comp} compares the main halo properties of the $z\sim3$ and $z\geq6$ samples. Evidently, the two sets of objects have very similar distributions of $f_\mathrm{H}$ and $q_\mathrm{H}$; a Kolmogorov-Smirnov test gives $p$ values of 0.98 and 0.64 that both are drawn from the same parent distribution.
(See Appendix \autoref{up_lims} for a version of the left plot that includes upper limits for the individually undetected LAHs). There is a slight excess of relatively dim LAHs ($f_\mathrm{H} \la 0.5$) in the $z\sim3$ sample, which comes primarily from objects where halo circularity was forced, indicating a potentially unreliable fit. On the other hand, \autoref{par_comp} reveals a clear difference in the resulting halo scale lengths between the two samples: The LAHs at high redshifts are much more compact than their $z\sim3$ counterparts, with the latter having typical scale lengths $\sim$3 times larger than the former. Moreover, 39 out of 40 $z\sim 3$ LAHs have scale lengths larger than the mean (or median) of the $z\geq6$ objects in the MXDF halo, and 29/40 are more extended than any of the high-redshift haloes.

Comparing our lower redshift measurements with the results of \citetalias{2016A&A...587A..98W} and \citetalias{2017A&A...608A...8L} -- obtained with much deeper data -- we find that the recovered distributions of LAH properties at $z\sim3$ are in full agreement (see also \autoref{halo_evo_l} below). Our median halo scale length for the MUSCATEL sample is $5.3\pm1.8\,$kpc (with the error bar representing the standard uncertainty of the median), while that value is $4.7\pm0.6\,$kpc in the \citetalias{2016A&A...587A..98W} and $4.2\pm1.5\,$kpc in the \citetalias{2017A&A...608A...8L} samples, for the same redshift interval $3< z <3.4$. These values are statistically consistent with each other, although slightly higher in our data. That is, however, naturally explained by the lower sensitivity of MUSCATEL-Shallow, meaning that the detected LAHs have higher \lya luminosities, a property known to correlate with the haloes' scale lengths (\citetalias{2017A&A...608A...8L}, \citeads{2024A&A...688A..37G}, \citeads{2026ApJ..1001..150M}).
We obtain a distribution of halo to continuum scale length ratios for our $z\sim3$ sample which is also very similar to that seen by \citetalias{2016A&A...587A..98W} and \citetalias{2017A&A...608A...8L}, with a median value of $\sim$10 but a large scatter. 

For our $z\geq6$ sample there is almost no comparison data in the literature (see Introduction). The scale lengths of the only individually studied high-$z$ LAHs (CR7 \citepads{2020MNRAS.498.3043M}, VR7 \citepads{2020MNRAS.492.1778M} and z70-1 \citepads{2020ApJ...891..177Z}) are relatively close to our measured values, although it should be kept in mind that these objects are much more luminous (in \lya as well as in UV magnitude) than our mostly very faint LAEs. 4 of the 6 $z\geq6$ LAEs with detected haloes have identified UV counterparts in JWST, all of which are resolved beyond a point source. The ratios of their \lya to UV scale lengths range between 1.7 and 16 with a median of 6.8, which given the small sample, is fully compatible with the observed relation for lower redshift LAEs.

\subsection{Characteristics of \texorpdfstring{\lya}{Lyα }haloes at \texorpdfstring{$z\geq 6$}{z ≥ 6}}
        
\begin{figure*}
    \centering
    \includegraphics[width=\linewidth]{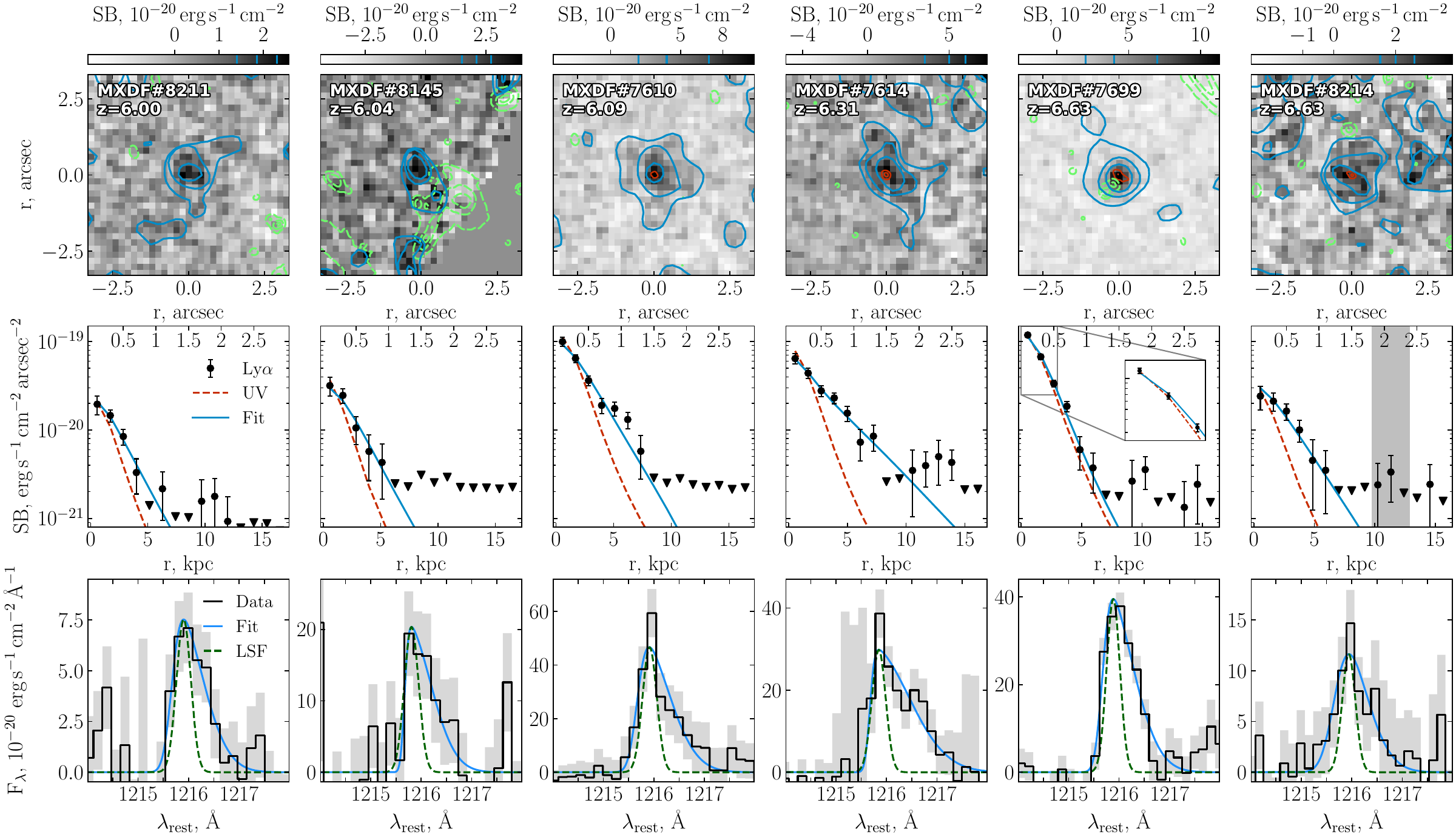}
    \caption{Gallery of the six individually detected $z\geq6$ \lya haloes.
    \textit{Top row:} \lya narrow band images overlayed with smoothed SB contours (solid blue line) outlining emission from the core and the extended halo with values indicated in the color bar on top. Red contours show the F115W images of the assigned UV counterparts at JWST resolution (their absence indicates no counterpart). Fore- and background objects from the same F115W image are shown in green dashed contours. AMUSED database ID and measured redshift are shown in the top left corner.
    \textit{Middle row:} SB profile of observed \lya emission (circles), scaled UV continuum model (dashed line) and the continuum + halo model 2D fit (solid line). 1$\sigma$ upper limits are shown with triangles. The three inner bins for MXDF\#7699 are shown in the inset on a large scale. The grey band at 12 kpc ($2^{\prime\prime}$) in the last column shows SB enhancement associated with a potential companion of MXDF\#8214.
    \textit{Bottom row:} observed \lya spectra (black step line with grey band), skewed gaussian fit to the data (blue solid) and MUSE LSF (green dashed).
    }
\label{halo_gallery}
\end{figure*}

The six spatially resolved LAHs in our high-redshift sample more than double the number of known individual LAH detections around star-forming galaxies at $z > 6$. We present them in \autoref{halo_gallery} and give their main structural parameters in \autoref{table:2}. In the following we briefly comment on each source separately, in order of increasing redshift:

\begin{itemize}

    \item MXDF\#8211 ($z=6.00$) has the lowest \lya luminosity among all known LAHs to date. It shows no continuum counterpart in any of the available deep JWST images, implying a high \lya equivalent width suggestive of a very metal-poor stellar population, a dust-free interstellar medium, and/or possibly contributions from other power sources than central photoionization (e.g., \citeads{2020MNRAS.495.1501C}, \citeads{2023A&A...678A..68S}).
    
    \item MXDF\#8145 ($z=6.04$) is located at the edge of the MXDF footprint and therefore has a significantly lower signal-to-noise ratio. In fact, a 1D SB profile test would have classified this object as not extended despite it showing clear signs of extended emission. Our self-calibrated 2D test reveals a \lya halo around this source ($p_0=0.03$). It also shows no  detectable counterpart in JWST. Notably, our image modelling with GALFIT prefers a single exponential with no separate central component.

    \item MXDF\#7610 ($z=6.09$) has a very similar \lya scale length as VR7 \citepads{2020MNRAS.492.1778M}, but at a $\sim40\times$ lower \lya luminosity. The UV counterpart of VR7 is even $\sim300\times$ brighter and $2.4\times$ more extended that that of MXDF\#7610. This highlights the diversity of the relation between galaxy and LAH properties. The inferred stellar metallicity of the UV counterpart is amongst the lowest in the $z\geq6$ sample.

    \item MXDF\#7614 ($z=6.31$) is located at the edge of the MXDF full exposure region, and its redshift places the \lya line very close in wavelength to a night sky emission line. Due to these restrictions the halo is only detected with our self-calibration method. The halo is notably elongated. This is the most extreme high-$z$ halo detected in the study. It is very similar to CR7 in terms of both line width and halo size (but not the \lya luminosity). Moreover, this object is the only $z\geq6$ LAH similar to the average $z\sim3$ halo in the MUSCATEL sample. This is the only object in our $z\geq6$ sample with available rest-frame spectroscopy from the JADES DR4 (\citeads{2026MNRAS.tmp..935C}, \citeads{2026MNRAS.tmp..886S}). This means we can measure its \lya velocity offset from systemic redshift $\Delta v_{\mathrm{Ly}\alpha} = c\,
    (z_{\mathrm{Ly}\alpha}-z_\mathrm{sys}) / (1 + z_\mathrm{sys})$, where $z_{\mathrm{Ly}\alpha}$ is the redshift estimated from the \lya peak position, and $z_\mathrm{sys}$ -- from rest-optical non resonant lines. We obtain the value of $220\pm55$\,km\,s$^{-1}$. Assuming Case-B recombination we calculate the \lya escape fraction ($f_\mathrm{esc}^{\mathrm{Ly}\alpha}=F_{\mathrm{Ly}\alpha}\, /\, 8.7\cdot F_{\mathrm{H}\alpha}$) to be $\approx0.3$. Both values are typical, given object's redshift and UV magnitude (e.g., \citeads{2024A&A...684A..84S}, \citeads{2024MNRAS.531.2701T}, \citeads{2024MNRAS.528.7052C}, \citeads{2026arXiv260213389G}).
    
    \item MXDF\#7699 ($z=6.63$) has the most compact halo detected in our sample despite being one of the most luminous objects. With a scale length of 1.2~kpc it is only marginally resolved, helped by the small error bars due to its relative brightness. The UV counterpart resembles an elongated U-warped edge-on galaxy, making the \lya to UV size ratio one of the lowest  recorded.
    
    \item MXDF\#8214 ($z=6.63$) is our highest redshift detected LAH. The NB image suggests a second emission source $\sim2''$ away, which was however masked and therefore did not influence the \lya halo detection (indicated by the grey band in the rightmost panel of \autoref{halo_gallery}). The secondary source is too faint for an independent classification, but if it is Ly$_\alpha$, then it would have a velocity offset of $\lesssim300\ \mathrm{km\,s^{-1}}$ from MXDF\#8214. It also has a weak JWST counterpart with $z_\mathrm{phot}=6.590_{-0.366}^{+1.459}$, making it quite likely that this is a close companion at a projected distance of $\sim12\ \mathrm{kpc}$. The extended LAH of MXDF\#8214 may thus be enhanced by gravitational interaction with its companion.

\end{itemize}

\begin{table*}
    \def\arraystretch{1.15}
    \caption{Summary of measured $z>$ \lya halo and continuum quantities.}
    \label{table:2}
    \centering
    \begin{tabular}{llllll}
        \hline\hline
          ID & $r_\mathrm{s,c}$ & $q_\mathrm{c}$ & $r_\mathrm{s,H}$ & $q_\mathrm{H}$ & $f_\mathrm{H}$ \\
             &     [kpc]        &                &     [kpc]        & & \\
          \hline
          8211\tablefootmark{a}  &        -        &         -       & $1.57 \pm 1.10$ & $0.78 \pm 0.48$ & $0.72 \pm 0.20$ \\
          8145\tablefootmark{a}  &        -        &         -       & $1.65 \pm 1.09$  & $0.35 \pm 0.31$ & $0.99 \pm 0.24$\\
          7610 & $0.33 \pm 0.04$ & $0.65 \pm 0.08$ & $1.91 \pm 0.50$ & $0.65 \pm 0.13$ & $0.84 \pm 0.11$ \\
          7614 & $0.23 \pm 0.01$ & $0.57 \pm 0.02$ & $3.66 \pm 0.99$ & $0.57 \pm 0.13$ & $0.88 \pm 0.04$ \\
          8214 & $0.20 \pm 0.02$ & $0.51 \pm 0.06$ & $1.60 \pm 0.55$ & $0.77 \pm 0.25$ & $0.99 \pm 0.23$ \\
          7699 & $0.59 \pm 0.03$ & $0.23 \pm 0.01$ & $1.18 \pm 0.36$ & $0.37 \pm 0.23$ & $0.74 \pm 0.23$ \\
          Halo Stack& $0.28 \pm 0.01$\tablefootmark{b} & $0.47 \pm 0.01$ & $1.52 \pm 0.21$ & $0.65 \pm 0.07$ & $0.99 \pm 0.02$ \\
          Full Stack& $0.23 \pm 0.01$\tablefootmark{b} & $0.26 \pm 0.02$ & $1.27 \pm 0.44$ & $0.54 \pm 0.14$ & $0.55 \pm 0.09$ \\
        \hline
    \end{tabular}
    \tablefoot{
    ID: source identifier in the AMUSED database. $r_\mathrm{s,c}$: UV continuum scale length (if available). $q_\mathrm{c}$: axis ratio of the UV continuum source. $r_\mathrm{s,H}$: \lya halo scale length. $q_\mathrm{H}$: halo axis ratio. $f_\mathrm{H}$: \lya halo flux fraction.
    \tablefoottext{a}{No counterpart detected, central component modelled as point source.} \tablefoottext{b}{Values converted from Sérsic half-light radii.}
    }

\end{table*}

\subsection{Stacking analysis}
\label{sec:stacks}

The vast difference in LAH scale lengths between $z\sim3$ and $z\geq6$ reported in \autoref{sec:sb} is necessarily based on only those objects where we could significantly detect an extended halo. Since LAH detectability depends strongly on the quality of the data, in particular depth and angular resolution, a non-detection by no means implies the absence of an extended halo. In \autoref{sec:up_lim} we address this point by estimating upper limits on the halo flux fractions and the halo scale lengths for all our undetected LAHs. 

A popular approach to increase the depth of imaging data, at least in a statistical sense, is stacking. We therefore stacked the \lya narrowband images of all LAEs where no individual halo could be detected, separately for each sample. From the centroid-aligned NB images with their associated variance maps and UV continuum model images we obtained pixel-by-pixel median and mean \lya and UV continuum images. We then extracted azimuthally averaged SB profiles from the stacked images and again performed a null hypothesis test for consistency between UV and \lya profile. Since our 2D self-calibration method could obviously not be applied here, we reverted to the simpler 1D profile test already adopted by \citetalias{2016A&A...587A..98W}. 

\begin{figure}
    \centering
    \includegraphics[width=\linewidth]{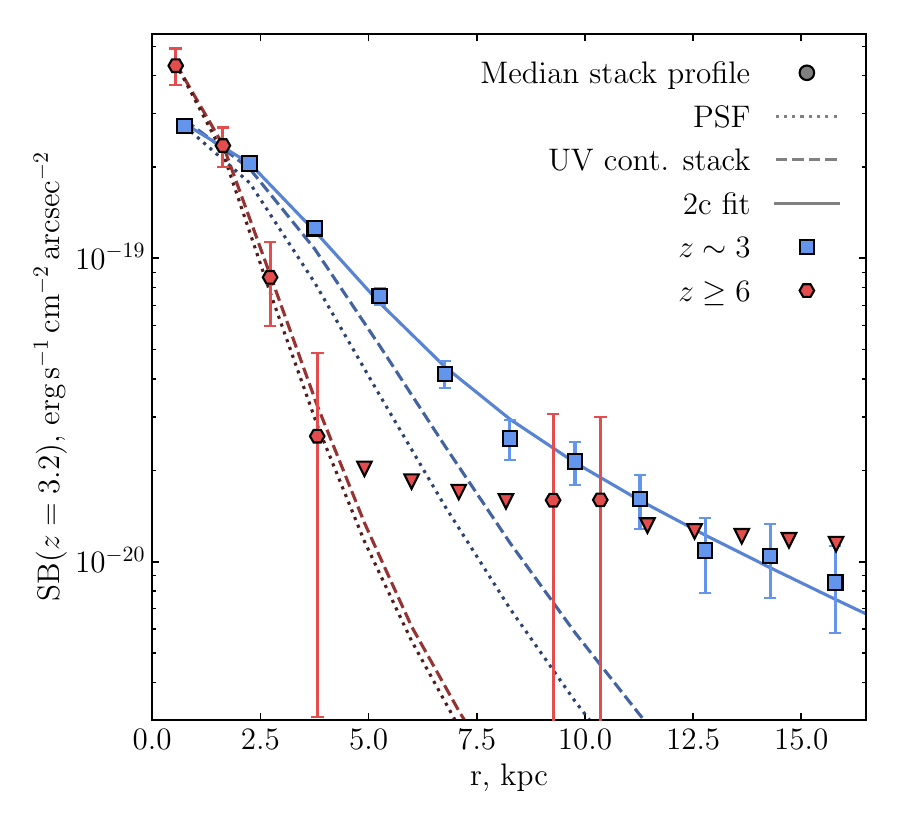}
    \caption{Median-stacked \lya surface brightness profiles of our LAEs with undetected \lya haloes. The blue squares represent the $z\sim3$ sample, red hexagons the $z\geq6$ sample. The latter values are corrected for cosmological dimming, such that the ordinate refers to surface brightnesses at $z=3.2$. 1$\sigma$ upper limits are shown with triangles. The profiles of the PSF and of the UV continuum stacks are shown in dotted and dashed lines, respectively. The blue solid line represents the profile of a 2-component fit to the $z\sim3$ stacked images.}
    \label{stack}
\end{figure}

\autoref{stack} contrasts the resulting median-stacked \lya SB profiles for the two samples. For this comparison we scaled the SB values of the $z\geq6$ stack by a factor of 9.12 to their equivalent values at redshift $z = 3.2$ (the median redshift of the MUSCATEL sample), thus approximately correcting for cosmological SB dimming. More details on the stacking results that include also the detected LAH objects are provided in \autoref{sec:stacks_ap}.

Remarkably, for the high-redshift undetected LAH sample, neither the mean nor the median stack shows any indication that the \lya SB profile differs significantly from the UV continuum profile, given the variance-propagated uncertainties. As discussed in \autoref{sec:up_lim}, this implies a halo flux fraction of less than 0.4 and/or a very compact stacked halo with $r_\mathrm{s, H}\lesssim 1.3\,$kpc.

In sharp contrast, the $z\sim3$ stack of the objects with individually undetected LAHs reveals highly significant \lya emission extending out to more than 10~kpc, with a SB profile very different from the stacked UV continuum. This strongly confirms the notion that most non-detections at $z\sim3$ are ``narrow misses'' that in only slightly deeper data would have been recognized as individual LAHs. To measure the scale length, we first fitted the UV continuum stack with a Sérsic profile. We then used this continuum model with fixed shape parameters plus an exponential profile to describe the extended halo. The resulting \lya scale length is $6.84 \pm 1.26$ kpc for the median and $7.84 \pm 1.23$ kpc for the mean stack, respectively.

The difference between the two samples is striking. We reiterate that the intrinsic (de-redshifted) SB sensitivities of MUSCATEL-Shallow at $z\sim3$ and of the MXDF at $z\geq6$ are essentially the same. The error bars of the $z\geq6$ stacked profiles are larger only because of the smaller number of objects in the high-redshift sample. The difference in the number of stacked objects is also inadequate to explain this result: stacking only 12 of the 50 $z\sim3$ LAEs would increase the individual error bars by a factor of $\approx2$, which is clearly insufficient to make the $z\sim3$ \lya SB profile consistent with the UV continuum profile. We thus conclude that \lya haloes of LAEs at $z\geq6$ are either generally much smaller and fainter than those at lower redshifts, or that a large fraction of LAEs at $z\geq6$ is even entirely devoid of any \lya halo.

\section{Spectral properties}
\label{sec:spec}

\begin{figure}
    \centering
    \includegraphics[width=\linewidth]{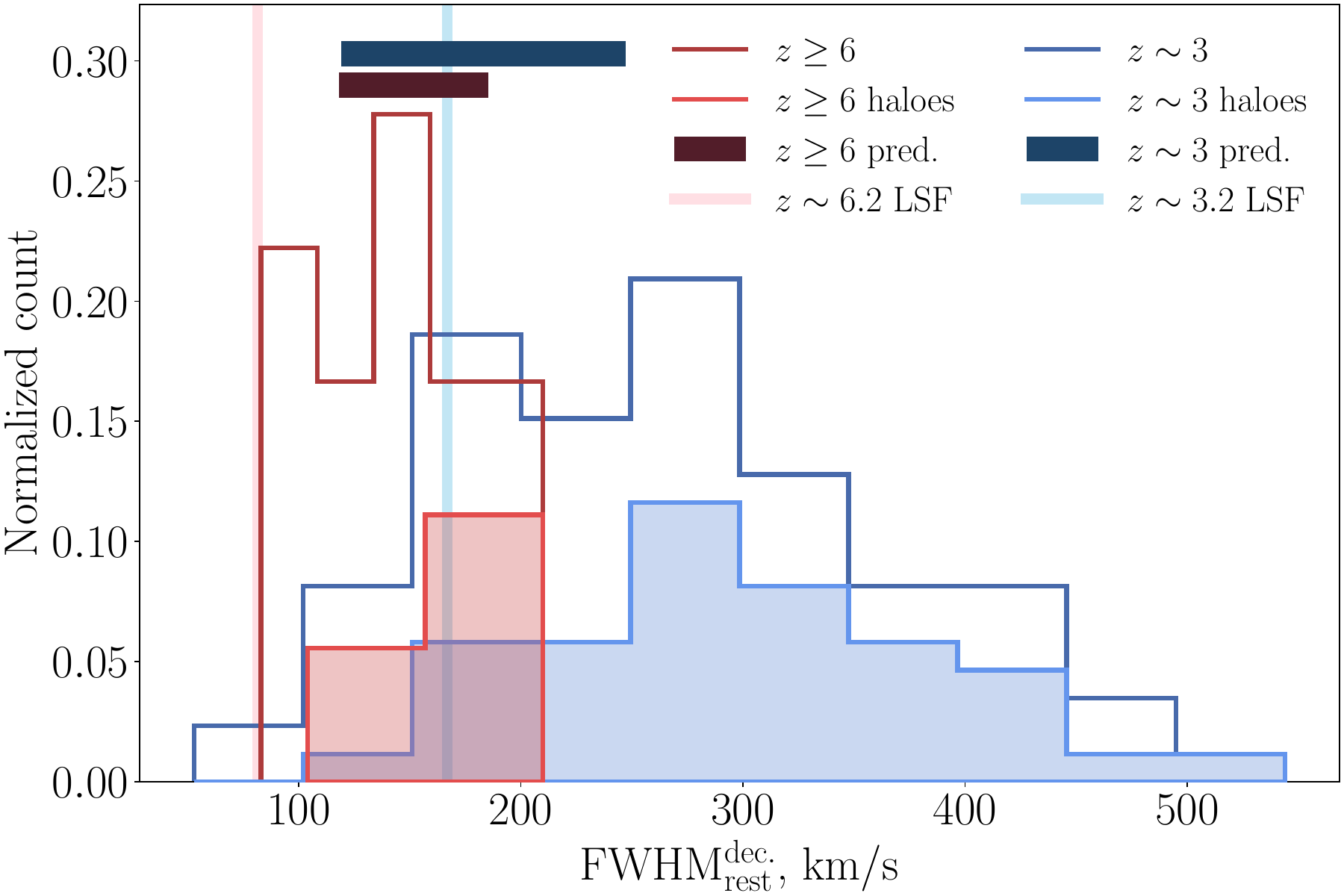}
    \caption{Distribution of \lya line widths (FWHM) for both samples, after
    deconvolving by the MUSE LSF. The open red and blue histograms give the results for the $z\geq6$ MXDF and the MUSCATEL $z\sim3$ samples, respectively. The filled red and blue histograms show this distribution for only LAEs with detected haloes at $z\geq6$ and $z\sim3$ respectively. The vertical lines indicate roughly the MUSE spectral resolution, for both samples. The horizontal bars at the top outline the range of expected intrinsic virial line widths from a simple stellar mass--halo mass conversion.}
\label{fwhm}
\end{figure}

To gain deeper insight into the physical origin of the observed halo differences, we investigated the \lya emission line spectra of the galaxies in each sample, since the \lya line profile is known to correlate with \lya halo properties, both from a theoretical perspective (\citeads{2006A&A...460..397V}, \citeads{2017A&A...608A.139G}) and empirically (\citetalias{2017A&A...608A...8L}, \citeads{2020A&A...635A..82L}) We fitted the red peak (the blue peak of $z\sim3$ LAEs was excluded to maintain consistency with the high-$z$ sample) of each \lya emission line with a ``skewed Gaussian'' profile (\citeads{2023A&A...670A...4B}), convolved with a Gaussian representing the MUSE line spread function (LSF) at line centre, using non-linear least-squares. The obtained \lya line widths (defined as the full width at half maximum, FWHM) are therefore approximate intrinsic widths with the instrumental broadening removed.

Figure~\ref{fwhm} compares the line width distributions of the LAEs in our two samples. Most of the lines at both high and low redshift are resolved, although some of them only marginally. Only $\sim20\%$ of the FWHM values fall below the LSF resolution, implying that their widths cannot be adequately constrained. The figure also shows the FWHM distributions for only the objects with individually detected haloes. We confirm the known trend \citepalias{2017A&A...608A...8L} that objects with significantly detected LAHs tend to exhibit larger line widths, for both samples. 

As is the case for the halo scale lengths, the distributions of line widths differ substantially between the two samples, with more than half of the lower-$z$ LAEs having lines broader than any of the $z\geq6$ objects. The median line widths and their interquartile ranges of two samples are $260\pm154$\,km\,s$^{-1}$ and $153\pm59$\,km\,s$^{-1}$ with unresolved lines treated as censored measurements for the Kaplan-Meier estimator. On the other hand, the skewness parameters of the line profile are very similar with a mean of $\gamma\simeq 9$ for both samples. This could however be a consequence of two effects conspiring. The skewed profile of \lya is usually believed to be a direct consequence of resonant scattering and photon escape through an optically thick, kinematically complex interstellar and circumgalactic medium of galaxies (e.g., \citeads{2006A&A...460..397V}, \citeads{2023MNRAS.523.3749B}). However, at high redshifts the line is further driven towards asymmetry by the evolving \lya forest opacity and the subsequently developing damping wing of intergalactic \ion{H}{i} absorption \citepads{2020MNRAS.499.1395M}. The impact of this second effect on the resulting profile shape depends strongly on the offset of the \lya peak from the systemic redshift, which in turn depends on the internal gas column density and kinematics. At the redshift range of our MXDF sample ($6<z<6.7$) we expect mainly a sharp cutoff at $z_{\mathrm{sys}}$, but not yet a substantial IGM damping wing. If the \lya peak offsets in our sample are around 200 km s$^{-1}$, similar to those observed for lower redshift LAEs (\citeads{2018MNRAS.478L..60V}, \citeads{2020MNRAS.496.1013M}, \citeads{2021A&A...654A..80S}), then the alteration of the profiles should be relatively modest, and it would be implausible to invoke IGM absorption to explain the observed lower FWHM values in our high-$z$ sample. This number could be directly tested using the empirical model by \citetads{2026arXiv260213389G} built to predict the emergent \lya spectra at $z\sim5-6$. Using their Eq. 15 we calculate the mean expected \lya velocity offset of the high-$z$ to be $202\pm18$\,km\,s$^{-1}$. We conclude that the differences in \lya line widths are most likely real and reflect intrinsic differences in the properties of the sources.

\citetads{2026PASA...43...21M} observed a similar trend of \lya line narrowing with redshift in the MAGPI survey. By studying the redshift evolution of FWHM of the red peak in LAE which feature double peaks, they were able to discriminate between the effects of the IGM attenuation and the intrinsic evolution of galaxies, favouring the second explanation.

We now explore whether these differences might be caused by systematically higher dynamical masses of LAH host galaxies at $z\sim 3.2$, leading to higher velocity dispersions and correspondingly  broader lines already before any scattering. To estimate the expected line broadening from the halo gas velocity dispersion, we first converted the stellar masses of all galaxies with detected HST or JWST counterparts into rough halo mass estimates using the stellar mass--halo mass relation from \citetads{2013ApJ...770...57B}. We estimated the one-dimensional velocity dispersion via the virial theorem using the calibration from \citetads{2008ApJ...672..122E} as
\begin{equation}
    \sigma_\mathrm{1D} = 1082.8\left[ \frac{h(z) \,M_{h}}{10^{15}\,M_\odot} \right]^{1/3}\ \mathrm{km\, s}^{-1},\label{eq:vd}
\end{equation}
where $M_{h}$ is the halo mass and $h(z)$ is the dimensionless Hubble parameter at redshift $z$. Finally, the velocity dispersion $\sigma_\mathrm{1D}$ was multiplied by 2.355 to obtain the predicted line width FWHM$_\mathrm{pred}$. Figure~\ref{fwhm} represents the 16th--84th percentile range of the FWHM$_\mathrm{pred}$ distribution as horizontal coloured bars for both samples. These values should only be taken as indicative as they depend heavily on the uncertain stellar masses and on the validity of the stellar mass to halo mass conversion. Moreover the emergent line widths also depend on the halo gas properties that can significantly alter the line profile, therefore \autoref{eq:vd} provides only a rough estimate on the expected line widths.

Nevertheless, \autoref{fwhm} reveals a different pattern for our two redshift ranges. Many of our $z\sim3.2$ LAEs have line widths substantially higher than what one would expect from their stellar masses, which would call for an additional broadening mechanism such as \lya\ radiative transfer. On the other hand, the observed line widths at $z\geq6$ lie within the predicted range, requiring no significant further broadening to explain them. As the two samples were treated in the same way we do not expect any strong systematic differences between the two. It is therefore this relative difference between the two samples, rather than the exact agreement with predicted line widths at either redshift, that we take forward into the discussion below.

\section{Discussion}
\label{sec:disc}

\begin{figure*}
    \centering
    \includegraphics[width=\linewidth]{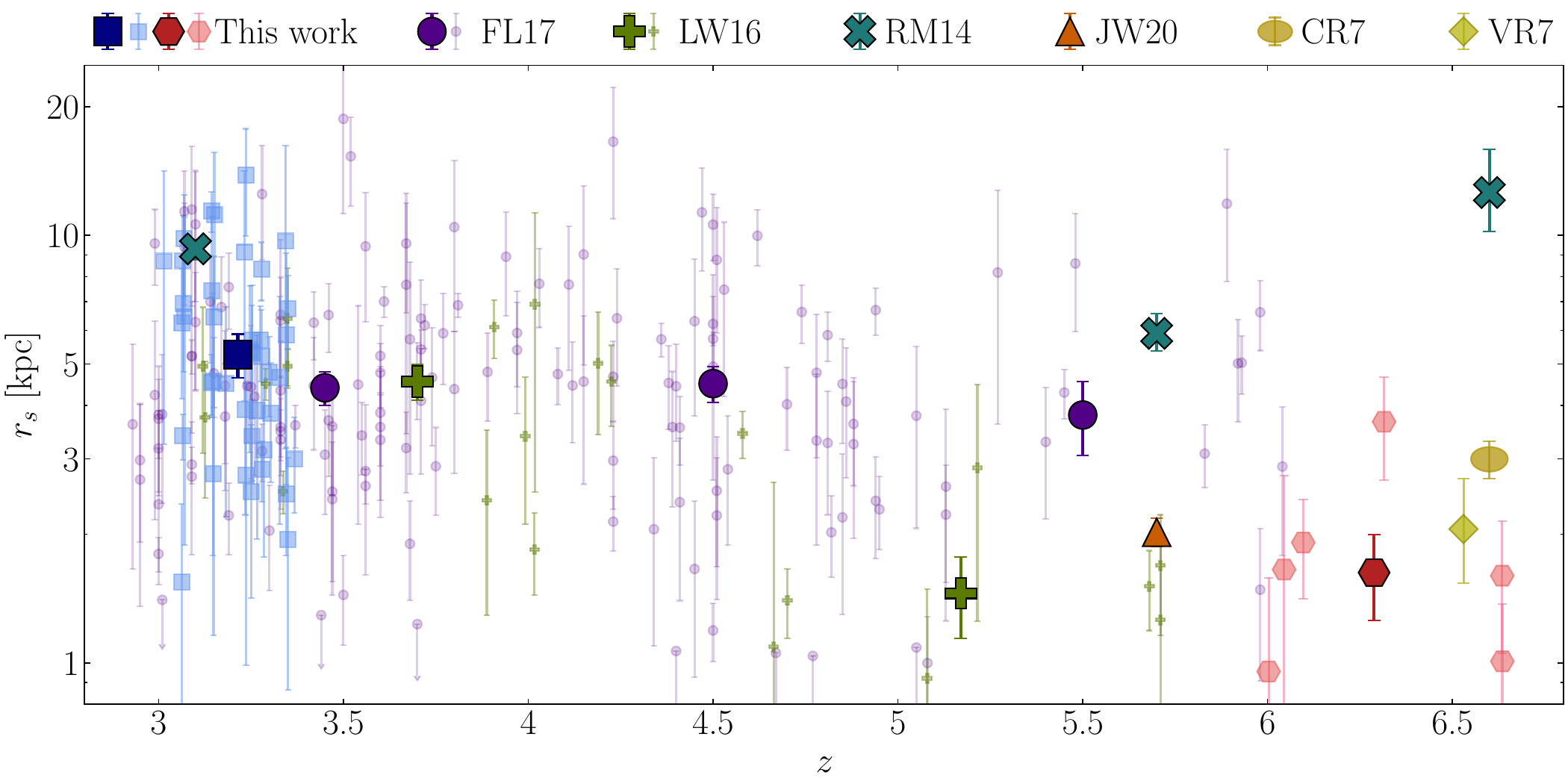}
    \caption{Measured \lya scale lengths as a function of redshift. The individually detected haloes from this study are shown as small squares ($z\sim 3.2$) and hexagons ($z\geq6$), while sample median values are given by the big dark markers. These are compared to results from previous MUSE observations (\citetalias{2016A&A...587A..98W}, \citetalias{2017A&A...608A...8L}) and from stacking (\citeads{2014MNRAS.442..110M} -- RM14, \citeads{2020ApJ...891..105W} -- JW20) as resolved in the legend, again with small symbols for individual measurements and big ones of the same marker type as median values. The MUSE measurements of CR7 \citepads{2020MNRAS.498.3043M} and VR7 \citepads{2020MNRAS.492.1778M} are also included.
    }
\label{halo_evo_l}
\end{figure*}

In the previous sections we presented evidence that the \lya halo and also the spectral properties of our two samples of LAEs are very different between $z\geq6$ and $z\sim 3$. We now discuss the implications of this finding in comparison with the literature and consider possible explanations.

\subsection{How different are \texorpdfstring{\lya}{Lyα }halo scale lengths at different redshifts?}
\label{sec:discussion-diff}

We start with a brief review of the available evidence for and against major changes in the observed properties of \lya haloes within the redshift range $3\la z \la 6$. We focus here on LAHs around normal star-forming galaxies, leaving out the much brighter, much larger, and often AGN-related \lya ``nebulae'' and ``blobs''. Figure~\ref{halo_evo_l} shows a compilation of measured \lya halo scale lengths from the literature plotted against redshift, including our new data. This diagram illustrates the conflicting trends seen in previous LAH observations: The stacking study by \citetads{2014MNRAS.442..110M} based on several thousands of LAEs in Subaru NB images found that the typical scale lengths stay essentially unchanged between $z=2.2$, 3.1, and 5.7, with values around $\sim$5--10~kpc, and rise to $\rsh \simeq 12.6$~kpc at $z=6.6$. In contrast, \citetads{2020ApJ...891..105W}, using similar data but a 3$\times$ larger sample, measured a much smaller mean scale length of only $\sim$2~kpc at $z=5.7$. \citetads{2021ApJ...916...22K} did not determine scale lengths, but reported no systematic differences in the stacked \lya radial profiles of LAEs between $z=5.7$ and 6.6 in their Subaru-HSC images.

The advent of individual LAH measurements with MUSE did initially not bring much more clarity. The first such study by \citetalias{2016A&A...587A..98W} of 21 LAHs in the Hubble Deep Field South suggested a tentative evolutionary trend that \lya haloes might evolve towards gradually smaller sizes with increasing redshifts up to $z\la 5.7$, but this trend was not deemed significant given the small sample. And indeed the nearly $10\times$ larger follow-up study by \citetalias{2017A&A...608A...8L} based on the MUSE-MOSAIC sample \citepads{2017A&A...608A...1B} did not confirm such a trend, but concluded that LAH scale lengths do \emph{not} change systematically within $3 \la z\la 6$. \citetads{2024A&A...688A..37G} found the same behaviour with stacking MUSE LAEs in bins of redshift from 3 to 6. That study also presented the first $z\geq6$ LAH measurement with $\rsh \simeq 3$~kpc. More recently, targeted MUSE observations of the outstandingly bright $z=6.6$ \lya emitter CR7 \citepads{2020MNRAS.498.3043M} and of the UV-luminous Lyman Break Galaxy VR7 at $z=6.5$ \citepads{2020MNRAS.492.1778M} gave similar \rsh values of around 3~kpc. 

A problem with all previous investigations is that results at different redshifts were obtained with roughly similar \emph{apparent} SB sensitivities. After accounting for cosmological dimming this implies vastly different intrinsic SB detection limits between different redshift domains, which could have led to $z$-dependent biases in the estimation of LAH properties. Other possible obstacles to a proper comparison of results between studies are differences in the angular resolution of the data, as well as incompatible analysis methods.

Here we aimed at accounting for all these effects in a consistent way. In particular, as explained in \autoref{sec:z6sample} our data at $z\sim 3$ and $z\geq6$ are fully matched in intrinsic surface brightness sensitivity (at the expense of giving up on sensitivity for the lower redshift range). The GLAO-improved PSF of the MUSE instrument provided the currently sharpest available view of LAEs, critical for our ability to constrain LAH scale lengths.

Figure~\ref{halo_evo_l} reveals that our results are strongly inconsistent with the null hypothesis of no evolution of LAH scale lengths from $z\geq6$ to $z\sim 3$. Instead, we find that $z\geq6$ LAHs are much smaller, by nearly a factor 3 in the mean, than their lower redshift siblings. In the right panel of \autoref{par_comp} we compare the distributions of scale lengths of the detected haloes between our two samples. The marginal histograms indicate that they are very different in shape and peak location. A Kolmogorov-Smirnov test rules out the null hypothesis with high significance ($p = 0.0008$) that both samples are drawn from the same underlying population. 

A possible caveat in this context could be differences in the physical characteristics -- in particular the stellar masses ($M_\star$) of the galaxies between the two samples, since these are known to correlate (weakly) with LAH properties (\citeads{2021MNRAS.506.5129B}, \citeads{2022A&A...662A..64R}). We use the Kaplan-Meier estimator to account for the UV non-detections in our statistics utilizing their stellar mass and SFR upper limits as censored measurements. The median\footnote{We use median and IQR to characterise the sample as they are more robust i n samples with censored measurements.} $\log_{10}(M_\star/\mathrm{M}_\odot)$ and its interquartile range for the lower and higher redshift samples are $9.13\pm0.98$ and $7.88\pm0.44$, respectively. The difference is mainly due to a larger fraction of relatively high mass galaxies ($M_\star \ga 10^9\,M_\odot$ in the $z\sim 3.2$ sample, while the lower mass bound is nearly the same for both datasets). 

We also checked that the \lya luminosities and star formation rates (SFR) of the two samples are broadly similar. Note that while both samples are selected by their \lya emission lines and thus approximately by SFR, the underlying surveys have very different depths so this similarity is not  trivially fulfilled. The mean values and rms scatter of $\log_{10}[L_{\mathrm{Ly}\alpha}/(\mathrm{erg}\:\mathrm{s}^{-1})]$ are $42.19\pm0.32$ and $41.58\pm0.35$, respectively. For the star formation rates $\log_{10}[\mathrm{SFR}/(M_\odot\,\mathrm{yr}^{-1})]$ resulting from our SED fitting we obtained median and IQR values of$0.06\pm0.92$ and $-0.56\pm0.73$. These shifts of a factor $\sim$2--3 in the mean are certainly insufficient to explain the differences between $z\geq6$ and $z\sim 3.2$, considering the known large scatter in the relations between LAH scale lengths and other properties (\citetalias{2017A&A...608A...8L}, \citeads{2022A&A...662A..64R}). In \autoref{sec:matched_samples} we investigate whether the samples matched in M$_*$, SFR, $M_{UV}$ or $\log_{10}$L$_\mathrm{Ly\alpha}$ produce different results. We find that contrast in both \lya line widths and halo scale lengths remain the same. This further proves that the difference in the galaxy properties between our samples is not responsible for the drastic change in LAE properties.

Not only sample properties, but also sample selection effects can play an important role in biasing our results. Although we built our samples based only on the properties of similar surveys, the LAE selection function may vary due to changing spatial and spectral resolution or adopted detection algorithms. We explore this possibility in \autoref{sec:fc}, using \lya selection function specifically tailored to MUSE deep fields. Our investigation shows, that the differences in completeness of our catalogues have minimal impact on recovered \lya line widths and halo sizes at $z\geq6$ and $z\sim3$.

While \autoref{par_comp} contains only the individually detected \lya haloes in our sample, we demonstrated in  \autoref{sec:stacks} and \autoref{stack} that the differences between $z\sim 3$ and $z\geq6$ LAHs become even more pronounced when comparing also the galaxies with undetected LAHs. We therefore conclude that the observed differences in \lya halo sizes between $z\sim3.2$ and $z\geq6$ are not an artefact of drawing our LAE samples from very different parts of the galaxy population, but physically real and highly significant.

\subsection{Are \texorpdfstring{\lya}{Lyα }haloes at \texorpdfstring{$z\geq 6$}{z ≥ 6} smaller because of reionisation?}
\label{sec:discussion-reion}

Since our high-redshift sample reaches into the late stages of the cosmic reionisation epoch (\citeads{2016A&A...596A.108P}, \citeads{2020MNRAS.494..600M}, \citeads{2026ApJ...997...86U})
which is known to affect the visibility of \lya \citepads{2014PASA...31...40D}, it is conceivable that the changing conditions in the intergalactic and circumgalactic media associated with this fundamental phase transition also lead to different emerging \lya haloes.

But how does a more neutral IGM and correspondingly higher \lya optical depth change the observed properties of LAHs? For \lya sources surrounded by a more or less homogeneous ambient medium following the Hubble flow the main effect will be to scatter the centrally emitted \lya photons until they are redshifted out of resonance \citepads{2014PASA...31...40D}. For the modest optical depths of the post-reionisation period one would therefore expect that LAH scale lengths and \lya line widths should get systematically larger with increasing neutral fraction and thus with redshift. When the reionisation epoch is reached this mechanism should result in very extended \lya haloes stretching over intergalactic scales, with a strongly broadened line profile (\citeads{1999ApJ...524..527L}, \citeads{2021ApJ...922..263P}). Such haloes are probably unobservable individually due to their extremely low surface brightnesses (but might be detectable in large-scale line intensity mapping experiments; \citeads{2024arXiv240816820P}), so that the scattered \lya photons would be essentially lost from the observational budget. 

This simple picture changes however when taking into account that \lya emission from galaxies in an incompletely reionized Universe is detected mainly from sources in ionized bubbles (e.g., \citeads{2023MNRAS.524.5891T}, \citeads{2024ApJ...977..250F}, \citeads{Neyer2026THESAN}). The appearance of extended \lya haloes will depend greatly on the properties of their host bubbles, in particular their sizes, which are still poorly constrained by observations. Numerical simulations have so far only provided a few indications of what to expect. In an early study, \citeads{2019MNRAS.484...39S} found that accounting for the evolving IGM transmission at $5<z<7$ actually steepens the emerging \lya SB profiles and reduces their scale lengths by about 20\% compared to the no IGM model. From their radiative transfer simulations, \citetads{2021ApJ...922..263P} concluded that large bubbles with radii of several comoving Mpc facilitate easier escape of \lya photons than much smaller ones where some of the radiation would still diffuse to IGM scales and get effectively lost. However, their simulations lacked the resolution required to properly model the circumgalactic \lya emission and make actual predictions for \lya halo properties.

In the light of these theoretical arguments it seems that our observational results could indeed be interpreted, at least qualitatively, as driven by reionisation. In this scenario, most galaxies would reside in ionized bubbles large enough to allow the central emission to at least partly escape, but too small (or not sufficiently ionized) to avoid the diffusion of a large fraction of \lya photons into the surrounding neutral regions, effectively suppressing the formation of detectable \lya haloes. The few cases where we actually observe a resolved halo could then correspond to a somewhat larger host bubble. 

Furthermore, recent deep observations of \lya haloes studying radial evolution of the spectra discovered that the peak of the red wing of the \lya line progressively moves towards the systemic velocity at larger radii (\citeads{2021MNRAS.504...16G}, \citeads{2024A&A...691A..66G}, \citeads{2026PASA...43...21M}, \citeads{2026arXiv260816665K}). \citetads{2024A&A...691A..66G} found tentative evidence that beyond 20\,kpc the red peak even shifts blueward of the systemic velocity. If this behaviour is real and ubiquitous at high-redshift, it would also affect the \lya halo sizes in the epoch of reionization. As the IGM transmission sharply drops to zero blueward of the systemic velocity, most of the blueshifted \lya photons would get absorbed and/or scattered away thus decreasing the observed SB at larger radii. This would naturally produce more compact haloes at redshifts where IGM is still significantly neutral. However, due to the lack of robustly measured systemic redshifts and statistical nature of the study such a scenario remains mostly speculative.

This scenario seems particularly plausible if for $z<6$ the properties of LAHs do \emph{not} evolve much with redshift, at least down to $z\sim 3$, as suggested by the majority of previous studies (\citeads{2014MNRAS.442..110M}, \citetalias{2017A&A...608A...8L}, \citeads{2021ApJ...916...22K} see \autoref{sec:discussion-diff}). Note that this would imply a substantial amount of \emph{intrinsic} evolution to counteract the strong population evolution of galaxies. \citetads{2021MNRAS.506.5129B} investigated this question in a cosmological context using the TNG50 simulation and found that self-regulated feedback balancing inflows and outflows produces nearly redshift-invariant LAH sizes. In such a framework, a sudden change in halo sizes for galaxies at $z\geq6$ would indeed have to be attributed mainly to the effects of reionisation.

Could the observed narrow \lya line widths of the LAEs in our $z\geq6$ sample also be an effect of the late stages of reionisation? A qualitatively similar phenomenon was already noted by \citetads{2022ApJ...935...52S} and \citetads{2024ApJ...971..136S} in a large spectroscopic follow-up study of bright LAEs selected from narrowband imaging. They showed that at fixed \lya luminosity, the \lya line widths are on average smaller at $z=6.6$ than at $z=5.7$, and that this difference is stronger for objects of lower \lya luminosities. An identical behaviour of LAEs in MUSE observations was reported by \citetads{2024PASA...41..105M} using MAGPI survey. Their data perfectly agrees findings of \citetads{2022ApJ...935...52S} both in the shape of $L_\mathrm{Ly\alpha}-$FWHM relation and its abrupt change between $z\sim5.7$ and $z\sim 6.5$. Since our LAE sample is much fainter than theirs, by a full order of magnitude, we cannot compare directly, but extrapolating the trend from \citetads{2024ApJ...971..136S} between $L_{\mathrm{Ly}\alpha}$ and FWHM$_{\mathrm{Ly}\alpha}$ to the luminosity range of our $z\geq6$ objects would yield a good agreement with our measured line widths. \citetads{2024PASA...41..105M} also estimated the typical ionised bubble sizes around their LAEs, assuming bubble age and Lyman continuum escape fraction, and considered its relation to the \lya line width. Their relation extrapolated to lower line widths observed in our sample predicts bubble sizes around our $z\geq6$ galaxies of $0.15-0.4$\,pMpc, assuming they exist in isolated environments with no ionising flux contributing from nearby sources.

To explain this fast change in FWHM from $z=6.6$ to 5.7, \citetads{2024ApJ...971..136S} suggested that the \lya line widths experienced different amounts of truncation of their blue wings due to the rapidly evolving intervening \lya Gunn-Peterson trough and damping wing. If this is correct then the same should apply to our sample. We find, however, much smaller line widths for our high-redshift objects, a decrease by a factor $\sim$1.7 compared to $z\sim 3.2$ (see \autoref{fwhm}). To achieve such a strong reduction in the emergent line width would require quite extreme conditions, with negligible \lya line offsets from systemic redshift as well as very small ionized bubble size, which seems somewhat unlikely (\citeads{2018MNRAS.478L..60V}, \citeads{2023ApJ...956..136P}, \citeads{2024A&A...684A..84S}, \citeads{2024MNRAS.528.4872L}). Assuming  typical \lya velocity offset of 200\,km\,s$^{-1}$ and bubble size of 0.2\,pMpc evaluated earlier as well as HI neutral fraction $x_\mathrm{HI}=0.2$, we estimate the effect of the IGM damping wing on the \lya line widths at $z\approx6.3$ using an analytic model by \citetads{1998ApJ...501...15M}. With the given parameters the \lya line narrows only by $\sim15\%$. Although this is only a crude estimate it shows that the IGM alone can not explain the observed evolution of the line widths.

\subsection{Or do \texorpdfstring{\lya}{Lyα }halo sizes change in lockstep with galaxy evolution?}
\label{sec:discussion-evol}

As an alternative scenario, we now consider the hypothesis that our finding of very different LAH sizes at $z\sim 3$ and $z\geq6$ is just a reflection of the overall evolution of galaxy properties between these redshifts and not specifically connected to reionisation. In that case the effects of LAH evolution should be observable not just at $z\geq6$ but at all redshifts, contradicting some -- but not all -- of the above mentioned studies. However, as already discussed in \autoref{sec:discussion-diff}, the many pitfalls and potential biases arising from data quality and analysis aspects could so far have obscured the true evolutionary behaviour.

As a starting point for such a halo co-evolution scenario, we consider one of the most robust measured scaling relations hitherto found for LAHs, namely the correlation between \lya and UV continuum scale lengths, $r_\mathrm{s,H}^{\mathrm{Ly}\alpha}\approx10\times r_\mathrm{s,c}^\mathrm{UV}$ (\citeads{2011ApJ...736..160S}, \citetalias{2016A&A...587A..98W}, \citetalias{2017A&A...608A...8L}, \citeads{2022A&A...666A..78C}). If this relation is equally valid at all redshifts of interest, with a non-evolving proportionality factor, then together with the well-constrained general size evolution of galaxies (e.g., \citeads{2015ApJS..219...15S}, \citeads{2017MNRAS.465.2717P}, \citeads{2024MNRAS.527.6110O}, \citeads{2025ApJ...982..200R}, \citeads{2025ApJS..281...68Y}), this necessarily implies a corresponding gradual growth of \lya halo scale lengths with decreasing redshift. In quantitative terms, these studies find that galaxies grow in size by a factor 1.7--2.3 over the redshift range of interest -- slightly smaller than the \lya halo size ratio that we find, but close enough to explain the bulk of the halo size growth.

In this way our observed decrease of LAH sizes for $z\geq6$ could thus simply be a consequence of the above correlation between halo and galaxy size holding also in the epoch of reionisation. Such a behaviour has already been observed for giant \lya nebulae around quasars, which show a gradual size growth (e.g., \citeads{2018MNRAS.476.2421G}, \citeads{2019MNRAS.488..120M}, \citeads{2019ApJ...887..196F}, \citeads{2020MNRAS.493.5336B}).

However, the growth of the galaxies' stellar bodies is directly driven by and strongly coupled to the assembly of their dark matter haloes \citepads{2019MNRAS.488.3143B}. In the same manner the CGM  properties (its extent most importantly) are tightly linked to the dark matter halo and its virial parameters (e.g., \citeads{2014ApJ...792....8W}, \citeads{2018ApJ...864..132B}, \citeads{2020MNRAS.493.1461L}, \citeads{2025MNRAS.543.1224C}). Therefore, the growth of the dark matter haloes from $z\geq6$ to $z\sim3$ would naturally explain the increase in the \lya halo sizes.

To test this hypothesis, we estimated the virial radii of galaxies in both samples based on the halo masses derived in \autoref{sec:spec}. We find that at $z\sim3$ the typical virial radius in our sample is 2.6 times larger than at $z\geq6$ (35\,kpc and 13\,kpc, respectively), a difference remarkably similar to the difference between the \lya scale lengths. We find the ratio between LAH scale length and virial radius to be $0.16$ across both samples. If our estimates of the halo masses (and hence virial radii) are correct, this similarity would naturally imply that the growth of the \lya haloes is naturally directly tied to the assembly of the dark matter halo reinforcing the gradual evolutionary scenario.

Since the spatial extent over which \lya gets scattered by circumgalactic gas depends strongly on the \ion{H}{i} fraction, we can try to link the expected evolution of \lya haloes to the general \ion{H}{i} content of the Universe. The latter is traced directly by the incidence rate of damped \lya absorbers (DLAs) that can be observed in quasar spectra up to $z\sim 5$ (but not beyond due to the ever increasing obscuration by the \lya forest). Earlier measurements suggested that the cosmic \ion{H}{i} density increases steadily towards higher redshifts (\citeads{2017MNRAS.466.2111B}, \citeads{2021MNRAS.507..704H}, \citeads{2023MNRAS.518.4646R}; see also the review by \citeads{2020ARA&A..58..363P}). In the absence of other effects, this would imply that \lya haloes would be larger at earlier cosmic times (see \autoref{sec:discussion-reion}), contrary to the size evolution trend of galaxies. Recently, however, \citetads{2025ApJ...983...10O} reported a maximum in the DLA incidence rate at $z\sim4-4.5$ and a decline beyond. The authors interpret this as evidence in support of a gradual build-up of the baryonic content of galaxies over cosmic time. If confirmed, then this should result in a likewise gradual growth of the \lya scattering medium, in lockstep with their central galaxies.

\section{Conclusions}
\label{sec:conc}

In this paper we took advantage of the unprecedented sensitivity of the MUSE Extremely Deep Field to perform the first systematic investigation of the properties of \lya haloes around \lya emitters at redshifts $z\geq6$. We find remarkably strong ensemble differences to a comparison sample at lower redshifts around $z\sim 3.2$, carefully selected to be observed at similar intrinsic surface brightness sensitivity after accounting for cosmological SB dimming. Our results can be summarized as follows:

\begin{enumerate}[(i)]
\item We identify significantly extended \lya emission around 6 individual high-redshift LAEs from our sample of 18 objects. This more than doubles the number of known \lya haloes beyond $z=6$ and provides the first statistical set of LAHs associated with low-luminosity, low-mass systems at these redshifts (\autoref{halo_gallery}).
\item The sizes (scale lengths) of these 6 high-$z$ \lya haloes are much smaller than those of typical lower redshift LAEs with similar \lya luminosities and stellar masses, by about a factor 3 in the mean. Their \lya halo flux fractions are however similarly high as at lower $z$ with values around 0.6 (\autoref{par_comp}).
\item Stacking the remaining 12 LAEs with individually undetected \lya haloes does not lead to the detection of a mean LAH signal, not even a tentative trace (\autoref{stack}). Thus, nearly 70\% of LAEs at these high redshifts have either no \lya halo at all, or only a very weak or small one.
\item In stark contrast, combining our $z\sim 3.2$ objects without individual LAH detections recovers a highly significant \lya halo in the stacked data (\autoref{stack}), indicating broadly similar LAH properties for the entire sample. Overall, our results for the lower redshift comparison sample are in excellent agreement with those of previous studies.
\item Comparing our $z\geq6$ measurements with the results of previous observations for this redshift range is not straightforward as the few available studies show a large dispersion. Our data rule out any previous suggestions that LAHs do not evolve at $z\geq6$ or are systematically larger (\autoref{halo_evo_l}). 
\item We also measure substantially lower \lya line widths of the $z\geq6$ LAEs, by a factor $\sim$1.7 compared to similar lower redshift objects (\autoref{fwhm}). The lines of the high-$z$ LAEs are often so narrow (but still resolved by MUSE) that the inferred amount of broadening by radiative transfer effects must be quite modest relative to the predicted intrinsic virial line widths based on stellar masses from SED fitting.
\end{enumerate}

Our results provide important constraints on the propagation of \lya photons during the late stages of reionisation, and on the evolution of the CGM in general. However, the question about the physical origin  for the observed differences in LAH properties must for the moment stay open. We considered two possible broad scenarios, one in which the transition from a partly neutral to a highly ionized IGM results in a rapid change of \lya scattering and visibility behaviour, and an alternative scenario of ``slow'' co-evolution between the general population of star forming galaxies and their \lya haloes. 

In the reionisation-dominated scenario (see \ \autoref{sec:discussion-reion}) it would be tempting to build a direct connection between the small sizes and frequent indetectability of \lya haloes and the measured narrow \lya line profiles, both of which are believed to be strongly shaped by resonant scattering. It could be that the \lya radiation that we receive from low-luminosity LAEs is composed mainly of those \lya photons that escaped with a relatively modest scattering history, both spatially (escaping closer to the sites of star formation) and spectrally (escaping closer to systemic redshift). The remainder of the total escaping \lya radiation would then be suffering much stronger scattering than at lower redshifts, producing a vast but probably unobservable halo due to its low surface brightness and strongly broadened line profile.

Such effects would also provide a natural explanation for the apparent non-evolution of the \lya luminosity function at $3 \leq z \leq 6$ and its rapid decrease beyond $z \approx 6$ \citepads{2025ApJS..277...37U}. A substantial loss of \lya photons to the still partly neutral IGM would result in systematically lower recovered \lya luminosities at $z \geq 6$. Another relevant finding of this study is that the two main drivers of LAE detectability -- halo size and spectral line width \citepads{2024A&A...690A.343P} -- show drastic redshift evolution, affecting the LAE selection function and hence strongly influencing the \lya luminosity function.
Thus, the prominence and sizes of \lya haloes has the potential to serve as an additional powerful diagnostic for the properties of the ionized regions where LAEs are found, especially in conjunction with high-resolution radiative transfer simulations. 

On the other hand, it is also possible that the previously held view, supported by a number of studies (e.g., \citeads{2014MNRAS.442..110M}, \citetalias{2017A&A...608A...8L}), of more or less unevolving \lya halo properties between $z\sim 3$ and $z\sim 6$ is incorrect, and that \lya halo sizes grow more or less in lockstep with their host galaxies (\autoref{sec:discussion-evol}). This is fortunately a testable hypothesis, as data of sufficient quality exist. A systematic investigation of \lya halo properties over the full redshift range accessible to MUSE, using improved analysis methods and taking advantage of the latest generation of GLAO-supported MUSE deep fields including the MXDF and the MUSCATEL survey, should firmly establish how the relation between galaxies and their surrounding LAHs evolve. We have started with such an analysis and aim to report on the outcome in a future paper.

Such a galaxy-LAH coevolution framework would certainly have the appeal of not requiring any special modelling treatment for LAHs at the highest redshifts. The question is whether this is physically plausible under the changing conditions during the late stages and aftermath of reionisation. Another possible problem with this scenario is that in general one should expect the \emph{minimum} \lya line widths to scale with the cosmologically growing host masses. The fact that our two LAE samples display very different FWHM$_{\mathrm{Ly}\alpha}$ values, at similar stellar mass distributions, is therefore not in obvious agreement with the simplest coevolution scenario. However, we emphasize that the two scenarios sketched above are by no means mutually exclusive. It is possible that both are required for a full explanation of our observations, a global evolutionary trend \emph{and} changes in \lya escape properties that are specific to the late stages of reionisation.

\begin{acknowledgements}
      The authors thank an anonymous referee for constructive comments and suggestions.
      This project has received funding from the European Research Council (ERC) under the European Union's Horizon 2020 research and innovation programme (grant agreement 101020943, SPECMAP-CGM). This work made use of the following software packages: \texttt{astropy} \citep{astropy:2013,astropy:2018,astropy:2022}, \texttt{matplotlib} \citep{Hunter:2007}, \texttt{numpy} \citep{numpy}, \texttt{pandas} \citep{mckinney-proc-scipy-2010,pandas_19340003}, \texttt{python} \citep{python}, \texttt{scipy} \citep{2020SciPy-NMeth,scipy_18736568}, \texttt{scikit-image} \citep{scikit-image}, and \texttt{TOPCAT} \citep{2005ASPC..347...29T}. This research made use of Photutils, an Astropy package for detection and photometry of astronomical sources \citep{Photutils_17129028}. Software citation information aggregated using \texttt{\href{https://www.tomwagg.com/software-citation-station/}{The Software Citation Station}} \citep{software-citation-station-paper,software-citation-station-zenodo}.
\end{acknowledgements}

\bibliographystyle{aa}
\bibliography{aa61652-26_bib}

@ARTICLE{2017A&A...608A...8L,
       author = {{Leclercq}, Floriane and {Bacon}, Roland and {Wisotzki}, Lutz and {Mitchell}, Peter and {Garel}, Thibault and {Verhamme}, Anne and {Blaizot}, J{\'e}r{\'e}my and {Hashimoto}, Takuya and {Herenz}, Edmund Christian and {Conseil}, Simon and {Cantalupo}, Sebastiano and {Inami}, Hanae and {Contini}, Thierry and {Richard}, Johan and {Maseda}, Michael and {Schaye}, Joop and {Marino}, Raffaella Anna and {Akhlaghi}, Mohammad and {Brinchmann}, Jarle and {Carollo}, Marcella},
        title = "{The MUSE Hubble Ultra Deep Field Survey. VIII. Extended Lyman-{\ensuremath{\alpha}} haloes around high-z star-forming galaxies}",
      journal = {\aap},
         year = 2017,
        month = dec,
       volume = {608},
          eid = {A8},
        pages = {A8},
          doi = {10.1051/0004-6361/201731480},
archivePrefix = {arXiv},
       eprint = {1710.10271},
 primaryClass = {astro-ph.GA},
       adsurl = {https://ui.adsabs.harvard.edu/abs/2017A&A...608A...8L}
}

@ARTICLE{2025ApJS..278...33K,
       author = {{Kageura}, Yuta and {Ouchi}, Masami and {Nakane}, Minami and {Umeda}, Hiroya and {Harikane}, Yuichi and {Yoshiura}, Shintaro and {Nakajima}, Kimihiko and {Yajima}, Hidenobu and {Thai}, Tran Thi},
        title = "{Census of Ly{\ensuremath{\alpha}} Emission from {\ensuremath{\sim}}600 Galaxies at z = 5{\textendash}14: Evolution of the Ly{\ensuremath{\alpha}} Luminosity Function and a Late Sharp Cosmic Reionization}",
      journal = {\apjs},
         year = 2025,
        month = jun,
       volume = {278},
       number = {2},
          eid = {33},
        pages = {33},
          doi = {10.3847/1538-4365/adc690},
archivePrefix = {arXiv},
       eprint = {2501.05834},
 primaryClass = {astro-ph.GA},
       adsurl = {https://ui.adsabs.harvard.edu/abs/2025ApJS..278...33K}
}

@ARTICLE{2019MNRAS.484...39S,
       author = {{Smith}, Aaron and {Ma}, Xiangcheng and {Bromm}, Volker and {Finkelstein}, Steven L. and {Hopkins}, Philip F. and {Faucher-Gigu{\`e}re}, Claude-Andr{\'e} and {Kere{\v{s}}}, Du{\v{s}}an},
        title = "{The physics of Lyman {\ensuremath{\alpha}} escape from high-redshift galaxies}",
      journal = {\mnras},
         year = 2019,
        month = mar,
       volume = {484},
       number = {1},
        pages = {39-59},
          doi = {10.1093/mnras/sty3483},
archivePrefix = {arXiv},
       eprint = {1810.08185},
 primaryClass = {astro-ph.GA},
       adsurl = {https://ui.adsabs.harvard.edu/abs/2019MNRAS.484...39S}
}

@ARTICLE{2020MNRAS.499.1395M,
       author = {{Mason}, Charlotte A. and {Gronke}, Max},
        title = "{Measuring the properties of reionized bubbles with resolved Ly{\ensuremath{\alpha}} spectra}",
      journal = {\mnras},
         year = 2020,
        month = nov,
       volume = {499},
       number = {1},
        pages = {1395-1405},
          doi = {10.1093/mnras/staa2910},
archivePrefix = {arXiv},
       eprint = {2004.13065},
 primaryClass = {astro-ph.GA},
       adsurl = {https://ui.adsabs.harvard.edu/abs/2020MNRAS.499.1395M}
}

@ARTICLE{2024A&A...690A.288B,
       author = {{Bunker}, Andrew J. and {Cameron}, Alex J. and {Curtis-Lake}, Emma and {Jakobsen}, Peter and {Carniani}, Stefano and {Curti}, Mirko and {Witstok}, Joris and {Maiolino}, Roberto and {D'Eugenio}, Francesco and {Looser}, Tobias J. and {Willott}, Chris and {Bonaventura}, Nina and {Hainline}, Kevin and {{\"U}bler}, Hannah and {Willmer}, Christopher N.~A. and {Saxena}, Aayush and {Smit}, Renske and {Alberts}, Stacey and {Arribas}, Santiago and {Baker}, William M. and {Baum}, Stefi and {Bhatawdekar}, Rachana and {Bowler}, Rebecca A.~A. and {Boyett}, Kristan and {Charlot}, Stephane and {Chen}, Zuyi and {Chevallard}, Jacopo and {Circosta}, Chiara and {DeCoursey}, Christa and {de Graaff}, Anna and {Egami}, Eiichi and {Eisenstein}, Daniel J. and {Endsley}, Ryan and {Ferruit}, Pierre and {Giardino}, Giovanna and {Hausen}, Ryan and {Helton}, Jakob M. and {Hviding}, Raphael E. and {Ji}, Zhiyuan and {Johnson}, Benjamin D. and {Jones}, Gareth C. and {Kumari}, Nimisha and {Laseter}, Isaac and {L{\"u}tzgendorf}, Nora and {Maseda}, Michael V. and {Nelson}, Erica and {Parlanti}, Eleonora and {Perna}, Michele and {Rauscher}, Bernard J. and {Rawle}, Tim and {Rix}, Hans-Walter and {Rieke}, Marcia and {Robertson}, Brant and {Rodr{\'\i}guez Del Pino}, Bruno and {Sandles}, Lester and {Scholtz}, Jan and {Sharpe}, Katherine and {Skarbinski}, Maya and {Stark}, Daniel P. and {Sun}, Fengwu and {Tacchella}, Sandro and {Topping}, Michael W. and {Villanueva}, Natalia C. and {Wallace}, Imaan E.~B. and {Williams}, Christina C. and {Woodrum}, Charity},
        title = "{JADES NIRSpec initial data release for the Hubble Ultra Deep Field: Redshifts and line fluxes of distant galaxies from the deepest JWST Cycle 1 NIRSpec multi-object spectroscopy}",
      journal = {\aap},
         year = 2024,
        month = oct,
       volume = {690},
          eid = {A288},
        pages = {A288},
          doi = {10.1051/0004-6361/202347094},
archivePrefix = {arXiv},
       eprint = {2306.02467},
 primaryClass = {astro-ph.GA},
       adsurl = {https://ui.adsabs.harvard.edu/abs/2024A&A...690A.288B}
}

@ARTICLE{2026ApJS..283....6E,
       author = {{Eisenstein}, Daniel J. and {Willott}, Chris and {Alberts}, Stacey and {Arribas}, Santiago and {Bonaventura}, Nina and {Bunker}, Andrew J. and {Cameron}, Alex J. and {Carniani}, Stefano and {Charlot}, Stephane and {Curtis-Lake}, Emma and {D'Eugenio}, Francesco and {Ferruit}, Pierre and {Giardino}, Giovanna and {Hainline}, Kevin and {Hausen}, Ryan and {Jakobsen}, Peter and {Johnson}, Benjamin D. and {Maiolino}, Roberto and {Rauscher}, Bernard J. and {Rieke}, Marcia and {Rieke}, George and {Rix}, Hans-Walter and {Robertson}, Brant and {Stark}, Daniel P. and {Tacchella}, Sandro and {Williams}, Christina C. and {Willmer}, Christopher N.~A. and {Baker}, William M. and {Baum}, Stefi and {Bhatawdekar}, Rachana and {Boyett}, Kristan and {Chen}, Zuyi and {Chevallard}, Jacopo and {Circosta}, Chiara and {Curti}, Mirko and {Danhaive}, A. Lola and {DeCoursey}, Christa and {Endsley}, Ryan and {de Graaff}, Anna and {Dressler}, Alan and {Egami}, Eiichi and {Helton}, Jakob M. and {Hviding}, Raphael E. and {Ji}, Zhiyuan and {Jones}, Gareth C. and {Kumari}, Nimisha and {L{\"u}tzgendorf}, Nora and {Laseter}, Isaac and {Looser}, Tobias J. and {Lyu}, Jianwei and {Maseda}, Michael V. and {Nelson}, Erica and {Parlanti}, Eleonora and {Perna}, Michele and {Pusk{\'a}s}, D{\'a}vid and {Rawle}, Tim and {Rodr{\'\i}guez Del Pino}, Bruno and {Rujopakarn}, Wiphu and {Sandles}, Lester and {Saxena}, Aayush and {Scholtz}, Jan and {Sharpe}, Katherine and {Shivaei}, Irene and {Silcock}, Maddie S. and {Simmonds}, Charlotte and {Skarbinski}, Maya and {Smit}, Renske and {Stone}, Meredith and {Suess}, Katherine A. and {Sun}, Fengwu and {Tang}, Mengtao and {Topping}, Michael W. and {{\"U}bler}, Hannah and {Villanueva}, Natalia C. and {Wallace}, Imaan E.~B. and {Whitler}, Lily and {Witstok}, Joris and {Woodrum}, Charity},
        title = "{Overview of the JWST Advanced Deep Extragalactic Survey (JADES)}",
      journal = {\apjs},
         year = 2026,
        month = mar,
       volume = {283},
       number = {1},
          eid = {6},
        pages = {6},
          doi = {10.3847/1538-4365/ae3163},
archivePrefix = {arXiv},
       eprint = {2306.02465},
 primaryClass = {astro-ph.GA},
       adsurl = {https://ui.adsabs.harvard.edu/abs/2026ApJS..283....6E}
}

@ARTICLE{2025ApJS..277....4D,
       author = {{D'Eugenio}, Francesco and {Cameron}, Alex J. and {Scholtz}, Jan and {Carniani}, Stefano and {Willott}, Chris J. and {Curtis-Lake}, Emma and {Bunker}, Andrew J. and {Parlanti}, Eleonora and {Maiolino}, Roberto and {Willmer}, Christopher N.~A. and {Jakobsen}, Peter and {Robertson}, Brant E. and {Johnson}, Benjamin D. and {Tacchella}, Sandro and {Cargile}, Phillip A. and {Rawle}, Tim and {Arribas}, Santiago and {Chevallard}, Jacopo and {Curti}, Mirko and {Egami}, Eiichi and {Eisenstein}, Daniel J. and {Kumari}, Nimisha and {Looser}, Tobias J. and {Rieke}, Marcia J. and {Rodr{\'\i}guez Del Pino}, Bruno and {Saxena}, Aayush and {{\"U}bler}, Hannah and {Venturi}, Giacomo and {Witstok}, Joris and {Baker}, William M. and {Bhatawdekar}, Rachana and {Bonaventura}, Nina and {Boyett}, Kristan and {Charlot}, Stephane and {Danhaive}, A. Lola and {Hainline}, Kevin N. and {Hausen}, Ryan and {Helton}, Jakob M. and {Ji}, Xihan and {Ji}, Zhiyuan and {Jones}, Gareth C. and {Juod{\v{z}}balis}, Ignas and {Maseda}, Michael V. and {P{\'e}rez-Gonz{\'a}lez}, Pablo G. and {Perna}, Michele and {Pusk{\'a}s}, D{\'a}vid and {Shivaei}, Irene and {Silcock}, Maddie S. and {Simmonds}, Charlotte and {Smit}, Renske and {Sun}, Fengwu and {Villanueva}, Natalia C. and {Williams}, Christina C. and {Zhu}, Yongda},
        title = "{JADES Data Release 3: NIRSpec/Microshutter Assembly Spectroscopy for 4000 Galaxies in the GOODS Fields}",
      journal = {\apjs},
         year = 2025,
        month = mar,
       volume = {277},
       number = {1},
          eid = {4},
        pages = {4},
          doi = {10.3847/1538-4365/ada148},
archivePrefix = {arXiv},
       eprint = {2404.06531},
 primaryClass = {astro-ph.GA},
       adsurl = {https://ui.adsabs.harvard.edu/abs/2025ApJS..277....4D}
}

@ARTICLE{2018ApJS..235...14S,
       author = {{Shipley}, Heath V. and {Lange-Vagle}, Daniel and {Marchesini}, Danilo and {Brammer}, Gabriel B. and {Ferrarese}, Laura and {Stefanon}, Mauro and {Kado-Fong}, Erin and {Whitaker}, Katherine E. and {Oesch}, Pascal A. and {Feinstein}, Adina D. and {Labb{\'e}}, Ivo and {Lundgren}, Britt and {Martis}, Nicholas and {Muzzin}, Adam and {Nedkova}, Kalina and {Skelton}, Rosalind and {van der Wel}, Arjen},
        title = "{HFF-DeepSpace Photometric Catalogs of the 12 Hubble Frontier Fields, Clusters, and Parallels: Photometry, Photometric Redshifts, and Stellar Masses}",
      journal = {\apjs},
         year = 2018,
        month = mar,
       volume = {235},
       number = {1},
          eid = {14},
        pages = {14},
          doi = {10.3847/1538-4365/aaacce},
archivePrefix = {arXiv},
       eprint = {1801.09734},
 primaryClass = {astro-ph.GA},
       adsurl = {https://ui.adsabs.harvard.edu/abs/2018ApJS..235...14S}
}

@ARTICLE{2010AJ....139.2097P,
       author = {{Peng}, Chien Y. and {Ho}, Luis C. and {Impey}, Chris D. and {Rix}, Hans-Walter},
        title = "{Detailed Decomposition of Galaxy Images. II. Beyond Axisymmetric Models}",
      journal = {\aj},
         year = 2010,
        month = jun,
       volume = {139},
       number = {6},
        pages = {2097-2129},
          doi = {10.1088/0004-6256/139/6/2097},
archivePrefix = {arXiv},
       eprint = {0912.0731},
 primaryClass = {astro-ph.CO},
       adsurl = {https://ui.adsabs.harvard.edu/abs/2010AJ....139.2097P}
}

@ARTICLE{2002AJ....124..266P,
       author = {{Peng}, Chien Y. and {Ho}, Luis C. and {Impey}, Chris D. and {Rix}, Hans-Walter},
        title = "{Detailed Structural Decomposition of Galaxy Images}",
      journal = {\aj},
         year = 2002,
        month = jul,
       volume = {124},
       number = {1},
        pages = {266-293},
          doi = {10.1086/340952},
archivePrefix = {arXiv},
       eprint = {astro-ph/0204182},
 primaryClass = {astro-ph},
       adsurl = {https://ui.adsabs.harvard.edu/abs/2002AJ....124..266P}
}

@ARTICLE{2020A&A...641A..28W,
       author = {{Weilbacher}, Peter M. and {Palsa}, Ralf and {Streicher}, Ole and {Bacon}, Roland and {Urrutia}, Tanya and {Wisotzki}, Lutz and {Conseil}, Simon and {Husemann}, Bernd and {Jarno}, Aur{\'e}lien and {Kelz}, Andreas and {P{\'e}contal-Rousset}, Arlette and {Richard}, Johan and {Roth}, Martin M. and {Selman}, Fernando and {Vernet}, Jo{\"e}l},
        title = "{The data processing pipeline for the MUSE instrument}",
      journal = {\aap},
         year = 2020,
        month = sep,
       volume = {641},
          eid = {A28},
        pages = {A28},
          doi = {10.1051/0004-6361/202037855},
archivePrefix = {arXiv},
       eprint = {2006.08638},
 primaryClass = {astro-ph.IM},
       adsurl = {https://ui.adsabs.harvard.edu/abs/2020A&A...641A..28W}
}

@ARTICLE{2024A&A...688A..37G,
       author = {{Guo}, Yucheng and {Bacon}, Roland and {Wisotzki}, Lutz and {Garel}, Thibault and {Blaizot}, J{\'e}r{\'e}my and {Schaye}, Joop and {Richard}, Johan and {Herrero Alonso}, Yohana and {Leclercq}, Floriane and {Boogaard}, Leindert and {Kusakabe}, Haruka and {Pharo}, John and {Vitte}, Elo{\"\i}se},
        title = "{Median surface-brightness profiles of Lyman-{\ensuremath{\alpha}} haloes in the MUSE Extremely Deep Field}",
      journal = {\aap},
         year = 2024,
        month = aug,
       volume = {688},
          eid = {A37},
        pages = {A37},
          doi = {10.1051/0004-6361/202347658},
archivePrefix = {arXiv},
       eprint = {2309.05513},
 primaryClass = {astro-ph.GA},
       adsurl = {https://ui.adsabs.harvard.edu/abs/2024A&A...688A..37G}
}

@ARTICLE{2024A&A...690A.343P,
       author = {{Pharo}, John and {Wisotzki}, Lutz and {Urrutia}, Tanya and {Bacon}, Roland and {Pessa}, Ismael and {Augustin}, Ramona and {Goovaerts}, Ilias and {Kozlova}, Daria and {Kusakabe}, Haruka and {Salas}, H{\'e}ctor and {Smirnov}, Daniil and {Thai}, Tran Thi and {Vitte}, Elo{\"\i}se},
        title = "{The intrinsic distribution of Lyman-{\ensuremath{\alpha}} halos}",
      journal = {\aap},
         year = 2024,
        month = oct,
       volume = {690},
          eid = {A343},
        pages = {A343},
          doi = {10.1051/0004-6361/202451318},
archivePrefix = {arXiv},
       eprint = {2409.04537},
 primaryClass = {astro-ph.GA},
       adsurl = {https://ui.adsabs.harvard.edu/abs/2024A&A...690A.343P}
}

@ARTICLE{2003PASP..115..763C,
       author = {{Chabrier}, Gilles},
        title = "{Galactic Stellar and Substellar Initial Mass Function}",
      journal = {\pasp},
         year = 2003,
        month = jul,
       volume = {115},
       number = {809},
        pages = {763-795},
          doi = {10.1086/376392},
archivePrefix = {arXiv},
       eprint = {astro-ph/0304382},
 primaryClass = {astro-ph},
       adsurl = {https://ui.adsabs.harvard.edu/abs/2003PASP..115..763C}
}

@ARTICLE{2009ApJ...696.1164O,
       author = {{Ouchi}, Masami and {Ono}, Yoshiaki and {Egami}, Eiichi and {Saito}, Tomoki and {Oguri}, Masamune and {McCarthy}, Patrick J. and {Farrah}, Duncan and {Kashikawa}, Nobunari and {Momcheva}, Ivelina and {Shimasaku}, Kazuhiro and {Nakanishi}, Kouichiro and {Furusawa}, Hisanori and {Akiyama}, Masayuki and {Dunlop}, James S. and {Mortier}, Angela M.~J. and {Okamura}, Sadanori and {Hayashi}, Masao and {Cirasuolo}, Michele and {Dressler}, Alan and {Iye}, Masanori and {Jarvis}, Matt J. and {Kodama}, Tadayuki and {Martin}, Crystal L. and {McLure}, Ross J. and {Ohta}, Kouji and {Yamada}, Toru and {Yoshida}, Michitoshi},
        title = "{Discovery of a Giant Ly{\ensuremath{\alpha}} Emitter Near the Reionization Epoch}",
      journal = {\apj},
         year = 2009,
        month = may,
       volume = {696},
       number = {2},
        pages = {1164-1175},
          doi = {10.1088/0004-637X/696/2/1164},
archivePrefix = {arXiv},
       eprint = {0807.4174},
 primaryClass = {astro-ph},
       adsurl = {https://ui.adsabs.harvard.edu/abs/2009ApJ...696.1164O}
}

@ARTICLE{2025ApJ...995..150K,
       author = {{Kiyota}, Tomokazu and {Ouchi}, Masami and {Xu}, Yi and {Nakazato}, Yurina and {Soga}, Kenta and {Yajima}, Hidenobu and {Fujimoto}, Seiji and {Harikane}, Yuichi and {Nakajima}, Kimihiko and {Ono}, Yoshiaki and {Sun}, Dongsheng and {Kusakabe}, Haruka and {Ceverino}, Daniel and {Hatsukade}, Bunyo and {Iono}, Daisuke and {Kohno}, Kotaro and {Nakanishi}, Kouichiro},
        title = "{Comprehensive JWST+ALMA Study on the Extended Ly{\ensuremath{\alpha}} Emitters, Himiko, and CR7 at z {\ensuremath{\sim}} 7: Blue Major Merger Systems in Stark Contrast to Submillimeter Galaxies}",
      journal = {\apj},
         year = 2025,
        month = dec,
       volume = {995},
       number = {2},
          eid = {150},
        pages = {150},
          doi = {10.3847/1538-4357/ae1cc3},
archivePrefix = {arXiv},
       eprint = {2504.03156},
 primaryClass = {astro-ph.GA},
       adsurl = {https://ui.adsabs.harvard.edu/abs/2025ApJ...995..150K}
}

@ARTICLE{2025A&A...699A.154M,
       author = {{Marconcini}, C. and {D'Eugenio}, F. and {Maiolino}, R. and {Arribas}, S. and {Bunker}, A. and {Carniani}, S. and {Charlot}, S. and {Perna}, M. and {Rodr{\'\i}guez Del Pino}, B. and {{\"U}bler}, H. and {P{\'e}rez-Gonz{\'a}lez}, P.~G. and {Willott}, C.~J. and {B{\"o}ker}, T. and {Cresci}, G. and {Curti}, M. and {Lamperti}, I. and {Scholtz}, J. and {Parlanti}, E. and {Venturi}, G.},
        title = "{GA-NIFS: Dissecting the multiple sub-structures and probing their complex interactions in the Ly{\ensuremath{\alpha}} emitter galaxy CR7 at z = 6.6 with JWST/NIRSpec}",
      journal = {\aap},
         year = 2025,
        month = jul,
       volume = {699},
          eid = {A154},
        pages = {A154},
          doi = {10.1051/0004-6361/202452994},
archivePrefix = {arXiv},
       eprint = {2411.08627},
 primaryClass = {astro-ph.GA},
       adsurl = {https://ui.adsabs.harvard.edu/abs/2025A&A...699A.154M}
}

@ARTICLE{2018PASJ...70S..15S,
       author = {{Shibuya}, Takatoshi and {Ouchi}, Masami and {Harikane}, Yuichi and {Rauch}, Michael and {Ono}, Yoshiaki and {Mukae}, Shiro and {Higuchi}, Ryo and {Kojima}, Takashi and {Yuma}, Suraphong and {Lee}, Chien-Hsiu and {Furusawa}, Hisanori and {Konno}, Akira and {Martin}, Crystal L. and {Shimasaku}, Kazuhiro and {Taniguchi}, Yoshiaki and {Kobayashi}, Masakazu A.~R. and {Kajisawa}, Masaru and {Nagao}, Tohru and {Goto}, Tomotsugu and {Kashikawa}, Nobunari and {Komiyama}, Yutaka and {Kusakabe}, Haruka and {Momose}, Rieko and {Nakajima}, Kimihiko and {Tanaka}, Masayuki and {Wang}, Shiang-Yu},
        title = "{SILVERRUSH. III. Deep optical and near-infrared spectroscopy for Ly{\ensuremath{\alpha}} and UV-nebular lines of bright Ly{\ensuremath{\alpha}} emitters at z = 6-7}",
      journal = {\pasj},
         year = 2018,
        month = jan,
       volume = {70},
          eid = {S15},
        pages = {S15},
          doi = {10.1093/pasj/psx107},
archivePrefix = {arXiv},
       eprint = {1705.00733},
 primaryClass = {astro-ph.GA},
       adsurl = {https://ui.adsabs.harvard.edu/abs/2018PASJ...70S..15S}
}

@ARTICLE{2022ApJ...931...97K,
       author = {{Kikuchihara}, Shotaro and {Harikane}, Yuichi and {Ouchi}, Masami and {Ono}, Yoshiaki and {Shibuya}, Takatoshi and {Itoh}, Ryohei and {Kakuma}, Ryota and {Inoue}, Akio K. and {Kusakabe}, Haruka and {Shimasaku}, Kazuhiro and {Momose}, Rieko and {Sugahara}, Yuma and {Kikuta}, Satoshi and {Saito}, Shun and {Kashikawa}, Nobunari and {Zhang}, Haibin and {Lee}, Chien-Hsiu},
        title = "{SILVERRUSH. XII. Intensity Mapping for Ly{\ensuremath{\alpha}} Emission Extending over 100-1000 Comoving Kpc around z   2-7 LAEs with Subaru HSC-SSP and CHORUS Data}",
      journal = {\apj},
         year = 2022,
        month = jun,
       volume = {931},
       number = {2},
          eid = {97},
        pages = {97},
          doi = {10.3847/1538-4357/ac69de},
archivePrefix = {arXiv},
       eprint = {2108.09288},
 primaryClass = {astro-ph.GA},
       adsurl = {https://ui.adsabs.harvard.edu/abs/2022ApJ...931...97K}
}

@ARTICLE{2021ApJ...908L..30P,
       author = {{Pelliccia}, Debora and {Strait}, Victoria and {Lemaux}, Brian C. and {Brada{\v{c}}}, Maru{\v{s}}a and {Coe}, Dan and {Bolan}, Patricia and {Bradley}, Larry D. and {Frye}, Brenda and {Gandhi}, Pratik J. and {Mainali}, Ramesh and {Mason}, Charlotte and {Ouchi}, Masami and {Sharon}, Keren and {Trenti}, Michele and {Zitrin}, Adi},
        title = "{RELICS-DP7: Spectroscopic Confirmation of a Dichromatic Primeval Galaxy at z {\ensuremath{\sim}} 7}",
      journal = {\apjl},
         year = 2021,
        month = feb,
       volume = {908},
       number = {2},
          eid = {L30},
        pages = {L30},
          doi = {10.3847/2041-8213/abdf56},
archivePrefix = {arXiv},
       eprint = {2011.08857},
 primaryClass = {astro-ph.GA},
       adsurl = {https://ui.adsabs.harvard.edu/abs/2021ApJ...908L..30P}
}

@ARTICLE{2015ApJS..219...15S,
       author = {{Shibuya}, Takatoshi and {Ouchi}, Masami and {Harikane}, Yuichi},
        title = "{Morphologies of {\ensuremath{\sim}}190,000 Galaxies at z = 0-10 Revealed with HST Legacy Data. I. Size Evolution}",
      journal = {\apjs},
         year = 2015,
        month = aug,
       volume = {219},
       number = {2},
          eid = {15},
        pages = {15},
          doi = {10.1088/0067-0049/219/2/15},
archivePrefix = {arXiv},
       eprint = {1503.07481},
 primaryClass = {astro-ph.GA},
       adsurl = {https://ui.adsabs.harvard.edu/abs/2015ApJS..219...15S}
}

@ARTICLE{2020ApJ...891..105W,
       author = {{Wu}, Jin and {Jiang}, Linhua and {Ning}, Yuanhang},
        title = "{Diffuse Ly{\ensuremath{\alpha}} Halos around 300 Spectroscopically Confirmed Ly{\ensuremath{\alpha}} Emitters at z {\ensuremath{\sim}} 5.7}",
      journal = {\apj},
         year = 2020,
        month = mar,
       volume = {891},
       number = {2},
          eid = {105},
        pages = {105},
          doi = {10.3847/1538-4357/ab7333},
archivePrefix = {arXiv},
       eprint = {2002.02029},
 primaryClass = {astro-ph.GA},
       adsurl = {https://ui.adsabs.harvard.edu/abs/2020ApJ...891..105W}
}

@ARTICLE{2021ApJ...916...22K,
       author = {{Kakuma}, Ryota and {Ouchi}, Masami and {Harikane}, Yuichi and {Ono}, Yoshiaki and {Inoue}, Akio K. and {Komiyama}, Yutaka and {Kusakabe}, Haruka and {Lee}, Chien-Hsiu and {Matsuda}, Yuichi and {Matsuoka}, Yoshiki and {Mawatari}, Ken and {Momose}, Rieko and {Shibuya}, Takatoshi and {Taniguchi}, Yoshiaki},
        title = "{SILVERRUSH. IX. Ly{\ensuremath{\alpha}} Intensity Mapping with Star-forming Galaxies at z = 5.7 and 6.6: A Possible Detection of Extended Ly{\ensuremath{\alpha}} Emission at {\ensuremath{\gtrsim}}100 Comoving Kiloparsecs around and beyond the Virial-radius Scale of Galaxy Dark Matter Halos}",
      journal = {\apj},
         year = 2021,
        month = jul,
       volume = {916},
       number = {1},
          eid = {22},
        pages = {22},
          doi = {10.3847/1538-4357/ac0725},
archivePrefix = {arXiv},
       eprint = {1906.00173},
 primaryClass = {astro-ph.GA},
       adsurl = {https://ui.adsabs.harvard.edu/abs/2021ApJ...916...22K}
}

@ARTICLE{2020MNRAS.498.3043M,
       author = {{Matthee}, Jorryt and {Pezzulli}, Gabriele and {Mackenzie}, Ruari and {Cantalupo}, Sebastiano and {Kusakabe}, Haruka and {Leclercq}, Floriane and {Sobral}, David and {Richard}, Johan and {Wisotzki}, Lutz and {Lilly}, Simon and {Boogaard}, Leindert and {Marino}, Raffaella and {Maseda}, Michael and {Nanayakkara}, Themiya},
        title = "{The nature of CR7 revealed with MUSE: a young starburst powering extended Ly {\ensuremath{\alpha}} emission at z = 6.6}",
      journal = {\mnras},
         year = 2020,
        month = oct,
       volume = {498},
       number = {2},
        pages = {3043-3059},
          doi = {10.1093/mnras/staa2550},
archivePrefix = {arXiv},
       eprint = {2008.01731},
 primaryClass = {astro-ph.GA},
       adsurl = {https://ui.adsabs.harvard.edu/abs/2020MNRAS.498.3043M}
}

@ARTICLE{2014MNRAS.442..110M,
       author = {{Momose}, Rieko and {Ouchi}, Masami and {Nakajima}, Kimihiko and {Ono}, Yoshiaki and {Shibuya}, Takatoshi and {Shimasaku}, Kazuhiro and {Yuma}, Suraphong and {Mori}, Masao and {Umemura}, Masayuki},
        title = "{Diffuse Ly{\ensuremath{\alpha}} haloes around galaxies at z = 2.2-6.6: implications for galaxy formation and cosmic reionization}",
      journal = {\mnras},
         year = 2014,
        month = jul,
       volume = {442},
       number = {1},
        pages = {110-120},
          doi = {10.1093/mnras/stu825},
archivePrefix = {arXiv},
       eprint = {1403.0732},
 primaryClass = {astro-ph.CO},
       adsurl = {https://ui.adsabs.harvard.edu/abs/2014MNRAS.442..110M}
}

@ARTICLE{2018MNRAS.476.2421G,
       author = {{Ginolfi}, M. and {Maiolino}, R. and {Carniani}, S. and {Arrigoni Battaia}, F. and {Cantalupo}, S. and {Schneider}, R.},
        title = "{Extended and broad Ly {\ensuremath{\alpha}} emission around a BAL quasar at z {\ensuremath{\sim}} 5}",
      journal = {\mnras},
         year = 2018,
        month = may,
       volume = {476},
       number = {2},
        pages = {2421-2431},
          doi = {10.1093/mnras/sty364},
archivePrefix = {arXiv},
       eprint = {1802.03400},
 primaryClass = {astro-ph.GA},
       adsurl = {https://ui.adsabs.harvard.edu/abs/2018MNRAS.476.2421G}
}

@ARTICLE{2019MNRAS.488..120M,
       author = {{Momose}, Rieko and {Goto}, Tomotsugu and {Utsumi}, Yousuke and {Hashimoto}, Tetsuya and {Chiang}, Chia-Ying and {Kim}, Seong-Jin and {Kashikawa}, Nobunari and {Shimasaku}, Kazuhiro and {Miyazaki}, Satoshi},
        title = "{Possible evolution of the circum-galactic medium around QSOs with QSO age and cosmic time revealed by Ly {\ensuremath{\alpha}} haloes}",
      journal = {\mnras},
         year = 2019,
        month = sep,
       volume = {488},
       number = {1},
        pages = {120-134},
          doi = {10.1093/mnras/stz1707},
archivePrefix = {arXiv},
       eprint = {1809.10916},
 primaryClass = {astro-ph.GA},
       adsurl = {https://ui.adsabs.harvard.edu/abs/2019MNRAS.488..120M}
}

@ARTICLE{2019ApJ...887..196F,
       author = {{Farina}, Emanuele Paolo and {Arrigoni-Battaia}, Fabrizio and {Costa}, Tiago and {Walter}, Fabian and {Hennawi}, Joseph F. and {Drake}, Alyssa B. and {Decarli}, Roberto and {Gutcke}, Thales A. and {Mazzucchelli}, Chiara and {Neeleman}, Marcel and {Georgiev}, Iskren and {Eilers}, Anna-Christina and {Davies}, Frederick B. and {Ba{\~n}ados}, Eduardo and {Fan}, Xiaohui and {Onoue}, Masafusa and {Schindler}, Jan-Torge and {Venemans}, Bram P. and {Wang}, Feige and {Yang}, Jinyi and {Rabien}, Sebastian and {Busoni}, Lorenzo},
        title = "{The REQUIEM Survey. I. A Search for Extended Ly{\ensuremath{\alpha}} Nebular Emission Around 31 z > 5.7 Quasars}",
      journal = {\apj},
         year = 2019,
        month = dec,
       volume = {887},
       number = {2},
          eid = {196},
        pages = {196},
          doi = {10.3847/1538-4357/ab5847},
archivePrefix = {arXiv},
       eprint = {1911.08498},
 primaryClass = {astro-ph.GA},
       adsurl = {https://ui.adsabs.harvard.edu/abs/2019ApJ...887..196F}
}

@ARTICLE{2020MNRAS.493.5336B,
       author = {{Bielby}, Richard M. and {Fumagalli}, Michele and {Fossati}, Matteo and {Rafelski}, Marc and {Oppenheimer}, Benjamin and {Cantalupo}, Sebastiano and {Christensen}, Lise and {Fynbo}, J.~P.~U. and {Lopez}, Sebastian and {Morris}, Simon L. and {D'Odorico}, Valentina and {Peroux}, Celine},
        title = "{Into the Ly {\ensuremath{\alpha}} jungle: exploring the circumgalactic medium of galaxies at z {\ensuremath{\sim}} 4-5 with MUSE}",
      journal = {\mnras},
         year = 2020,
        month = apr,
       volume = {493},
       number = {4},
        pages = {5336-5356},
          doi = {10.1093/mnras/staa546},
archivePrefix = {arXiv},
       eprint = {2001.09058},
 primaryClass = {astro-ph.GA},
       adsurl = {https://ui.adsabs.harvard.edu/abs/2020MNRAS.493.5336B}
}

@ARTICLE{2021MNRAS.506.5129B,
       author = {{Byrohl}, Chris and {Nelson}, Dylan and {Behrens}, Christoph and {Kostyuk}, Ivan and {Glatzle}, Martin and {Pillepich}, Annalisa and {Hernquist}, Lars and {Marinacci}, Federico and {Vogelsberger}, Mark},
        title = "{The physical origins and dominant emission mechanisms of Lyman alpha haloes: results from the TNG50 simulation in comparison to MUSE observations}",
      journal = {\mnras},
         year = 2021,
        month = oct,
       volume = {506},
       number = {4},
        pages = {5129-5152},
          doi = {10.1093/mnras/stab1958},
archivePrefix = {arXiv},
       eprint = {2009.07283},
 primaryClass = {astro-ph.GA},
       adsurl = {https://ui.adsabs.harvard.edu/abs/2021MNRAS.506.5129B}
}

@ARTICLE{2025ApJS..277...37U,
       author = {{Umeda}, Hiroya and {Ouchi}, Masami and {Kikuta}, Satoshi and {Harikane}, Yuichi and {Ono}, Yoshiaki and {Shibuya}, Takatoshi and {Inoue}, Akio K. and {Shimasaku}, Kazuhiro and {Liang}, Yongming and {Matsumoto}, Akinori and {Saito}, Shun and {Kusakabe}, Haruka and {Kageura}, Yuta and {Nakane}, Minami},
        title = "{SILVERRUSH. XIV. Ly{\ensuremath{\alpha}} Luminosity Functions and Angular Correlation Functions from 20,000 Ly{\ensuremath{\alpha}} Emitters at z {\ensuremath{\sim}} 2.2{\textendash}7.3 from up to 24 deg$^{2}$ HSC-SSP and CHORUS Surveys: Linking the Postreionization Epoch to the Heart of Reionization}",
      journal = {\apjs},
         year = 2025,
        month = apr,
       volume = {277},
       number = {2},
          eid = {37},
        pages = {37},
          doi = {10.3847/1538-4365/adb1c0},
archivePrefix = {arXiv},
       eprint = {2411.15495},
 primaryClass = {astro-ph.GA},
       adsurl = {https://ui.adsabs.harvard.edu/abs/2025ApJS..277...37U}
}

@ARTICLE{2016A&A...587A..98W,
       author = {{Wisotzki}, L. and {Bacon}, R. and {Blaizot}, J. and {Brinchmann}, J. and {Herenz}, E.~C. and {Schaye}, J. and {Bouch{\'e}}, N. and {Cantalupo}, S. and {Contini}, T. and {Carollo}, C.~M. and {Caruana}, J. and {Courbot}, J. -B. and {Emsellem}, E. and {Kamann}, S. and {Kerutt}, J. and {Leclercq}, F. and {Lilly}, S.~J. and {Patr{\'\i}cio}, V. and {Sandin}, C. and {Steinmetz}, M. and {Straka}, L.~A. and {Urrutia}, T. and {Verhamme}, A. and {Weilbacher}, P.~M. and {Wendt}, M.},
        title = "{Extended Lyman {\ensuremath{\alpha}} haloes around individual high-redshift galaxies revealed by MUSE}",
      journal = {\aap},
         year = 2016,
        month = mar,
       volume = {587},
          eid = {A98},
        pages = {A98},
          doi = {10.1051/0004-6361/201527384},
archivePrefix = {arXiv},
       eprint = {1509.05143},
 primaryClass = {astro-ph.GA},
       adsurl = {https://ui.adsabs.harvard.edu/abs/2016A&A...587A..98W}
}

@ARTICLE{2023A&A...670A...4B,
       author = {{Bacon}, Roland and {Brinchmann}, Jarle and {Conseil}, Simon and {Maseda}, Michael and {Nanayakkara}, Themiya and {Wendt}, Martin and {Bacher}, Raphael and {Mary}, David and {Weilbacher}, Peter M. and {Krajnovi{\'c}}, Davor and {Boogaard}, Leindert and {Bouch{\'e}}, Nicolas and {Contini}, Thierry and {Epinat}, Beno{\^\i}t and {Feltre}, Anna and {Guo}, Yucheng and {Herenz}, Christian and {Kollatschny}, Wolfram and {Kusakabe}, Haruka and {Leclercq}, Floriane and {Michel-Dansac}, L{\'e}o and {Pello}, Roser and {Richard}, Johan and {Roth}, Martin and {Salvignol}, Gregory and {Schaye}, Joop and {Steinmetz}, Matthias and {Tresse}, Laurence and {Urrutia}, Tanya and {Verhamme}, Anne and {Vitte}, Eloise and {Wisotzki}, Lutz and {Zoutendijk}, Sebastiaan L.},
        title = "{The MUSE Hubble Ultra Deep Field surveys: Data release II}",
      journal = {\aap},
         year = 2023,
        month = feb,
       volume = {670},
          eid = {A4},
        pages = {A4},
          doi = {10.1051/0004-6361/202244187},
archivePrefix = {arXiv},
       eprint = {2211.08493},
 primaryClass = {astro-ph.GA},
       adsurl = {https://ui.adsabs.harvard.edu/abs/2023A&A...670A...4B}
}

@ARTICLE{2017ApJ...837...97L,
       author = {{Lotz}, J.~M. and {Koekemoer}, A. and {Coe}, D. and {Grogin}, N. and {Capak}, P. and {Mack}, J. and {Anderson}, J. and {Avila}, R. and {Barker}, E.~A. and {Borncamp}, D. and {Brammer}, G. and {Durbin}, M. and {Gunning}, H. and {Hilbert}, B. and {Jenkner}, H. and {Khandrika}, H. and {Levay}, Z. and {Lucas}, R.~A. and {MacKenty}, J. and {Ogaz}, S. and {Porterfield}, B. and {Reid}, N. and {Robberto}, M. and {Royle}, P. and {Smith}, L.~J. and {Storrie-Lombardi}, L.~J. and {Sunnquist}, B. and {Surace}, J. and {Taylor}, D.~C. and {Williams}, R. and {Bullock}, J. and {Dickinson}, M. and {Finkelstein}, S. and {Natarajan}, P. and {Richard}, J. and {Robertson}, B. and {Tumlinson}, J. and {Zitrin}, A. and {Flanagan}, K. and {Sembach}, K. and {Soifer}, B.~T. and {Mountain}, M.},
        title = "{The Frontier Fields: Survey Design and Initial Results}",
      journal = {\apj},
         year = 2017,
        month = mar,
       volume = {837},
       number = {1},
          eid = {97},
        pages = {97},
          doi = {10.3847/1538-4357/837/1/97},
archivePrefix = {arXiv},
       eprint = {1605.06567},
 primaryClass = {astro-ph.GA},
       adsurl = {https://ui.adsabs.harvard.edu/abs/2017ApJ...837...97L}
}

@ARTICLE{2014PASA...31...40D,
       author = {{Dijkstra}, Mark},
        title = "{Ly{\ensuremath{\alpha}} Emitting Galaxies as a Probe of Reionisation}",
      journal = {\pasa},
         year = 2014,
        month = oct,
       volume = {31},
          eid = {e040},
        pages = {e040},
          doi = {10.1017/pasa.2014.33},
archivePrefix = {arXiv},
       eprint = {1406.7292},
 primaryClass = {astro-ph.CO},
       adsurl = {https://ui.adsabs.harvard.edu/abs/2014PASA...31...40D}
}

@ARTICLE{2011ApJ...736..160S,
       author = {{Steidel}, Charles C. and {Bogosavljevi{\'c}}, Milan and {Shapley}, Alice E. and {Kollmeier}, Juna A. and {Reddy}, Naveen A. and {Erb}, Dawn K. and {Pettini}, Max},
        title = "{Diffuse Ly{\ensuremath{\alpha}} Emitting Halos: A Generic Property of High-redshift Star-forming Galaxies}",
      journal = {\apj},
         year = 2011,
        month = aug,
       volume = {736},
       number = {2},
          eid = {160},
        pages = {160},
          doi = {10.1088/0004-637X/736/2/160},
archivePrefix = {arXiv},
       eprint = {1101.2204},
 primaryClass = {astro-ph.CO},
       adsurl = {https://ui.adsabs.harvard.edu/abs/2011ApJ...736..160S}
}

@ARTICLE{2024A&A...689A..10M,
       author = {{Melia}, F.},
        title = "{The cosmic timeline implied by the JWST reionization crisis}",
      journal = {\aap},
         year = 2024,
        month = sep,
       volume = {689},
          eid = {A10},
        pages = {A10},
          doi = {10.1051/0004-6361/202450835},
archivePrefix = {arXiv},
       eprint = {2407.01581},
 primaryClass = {astro-ph.CO},
       adsurl = {https://ui.adsabs.harvard.edu/abs/2024A&A...689A..10M}
}

@ARTICLE{2021ApJ...914...44A,
       author = {{Ahn}, Kyungjin and {Shapiro}, Paul R.},
        title = "{Cosmic Reionization May Still Have Started Early and Ended Late: Confronting Early Onset with Cosmic Microwave Background Anisotropy and 21 cm Global Signals}",
      journal = {\apj},
         year = 2021,
        month = jun,
       volume = {914},
       number = {1},
          eid = {44},
        pages = {44},
          doi = {10.3847/1538-4357/abf3bf},
archivePrefix = {arXiv},
       eprint = {2011.03582},
 primaryClass = {astro-ph.CO},
       adsurl = {https://ui.adsabs.harvard.edu/abs/2021ApJ...914...44A}
}

@ARTICLE{2020ARA&A..58..617O,
       author = {{Ouchi}, Masami and {Ono}, Yoshiaki and {Shibuya}, Takatoshi},
        title = "{Observations of the Lyman-{\ensuremath{\alpha}} Universe}",
      journal = {\araa},
         year = 2020,
        month = aug,
       volume = {58},
        pages = {617-659},
          doi = {10.1146/annurev-astro-032620-021859},
archivePrefix = {arXiv},
       eprint = {2012.07960},
 primaryClass = {astro-ph.GA},
       adsurl = {https://ui.adsabs.harvard.edu/abs/2020ARA&A..58..617O}
}

@ARTICLE{2021MNRAS.501.5757M,
       author = {{Mitchell}, Peter D. and {Blaizot}, J{\'e}r{\'e}my and {Cadiou}, Corentin and {Dubois}, Yohan and {Garel}, Thibault and {Rosdahl}, Joakim},
        title = "{Tracing the simulated high-redshift circumgalactic medium with Lyman {\ensuremath{\alpha}} emission}",
      journal = {\mnras},
         year = 2021,
        month = mar,
       volume = {501},
       number = {4},
        pages = {5757-5775},
          doi = {10.1093/mnras/stab035},
archivePrefix = {arXiv},
       eprint = {2008.12790},
 primaryClass = {astro-ph.GA},
       adsurl = {https://ui.adsabs.harvard.edu/abs/2021MNRAS.501.5757M}
}

@ARTICLE{2017A&A...608A...1B,
       author = {{Bacon}, Roland and {Conseil}, Simon and {Mary}, David and {Brinchmann}, Jarle and {Shepherd}, Martin and {Akhlaghi}, Mohammad and {Weilbacher}, Peter M. and {Piqueras}, Laure and {Wisotzki}, Lutz and {Lagattuta}, David and {Epinat}, Benoit and {Guerou}, Adrien and {Inami}, Hanae and {Cantalupo}, Sebastiano and {Courbot}, Jean Baptiste and {Contini}, Thierry and {Richard}, Johan and {Maseda}, Michael and {Bouwens}, Rychard and {Bouch{\'e}}, Nicolas and {Kollatschny}, Wolfram and {Schaye}, Joop and {Marino}, Raffaella Anna and {Pello}, Roser and {Herenz}, Christian and {Guiderdoni}, Bruno and {Carollo}, Marcella},
        title = "{The MUSE Hubble Ultra Deep Field Survey. I. Survey description, data reduction, and source detection}",
      journal = {\aap},
         year = 2017,
        month = dec,
       volume = {608},
          eid = {A1},
        pages = {A1},
          doi = {10.1051/0004-6361/201730833},
archivePrefix = {arXiv},
       eprint = {1710.03002},
 primaryClass = {astro-ph.GA},
       adsurl = {https://ui.adsabs.harvard.edu/abs/2017A&A...608A...1B}
}

@ARTICLE{2019A&A...624A.141U,
       author = {{Urrutia}, T. and {Wisotzki}, L. and {Kerutt}, J. and {Schmidt}, K.~B. and {Herenz}, E.~C. and {Klar}, J. and {Saust}, R. and {Werhahn}, M. and {Diener}, C. and {Caruana}, J. and {Krajnovi{\'c}}, D. and {Bacon}, R. and {Boogaard}, L. and {Brinchmann}, J. and {Enke}, H. and {Maseda}, M. and {Nanayakkara}, T. and {Richard}, J. and {Steinmetz}, M. and {Weilbacher}, P.~M.},
        title = "{The MUSE-Wide Survey: survey description and first data release}",
      journal = {\aap},
         year = 2019,
        month = apr,
       volume = {624},
          eid = {A141},
        pages = {A141},
          doi = {10.1051/0004-6361/201834656},
archivePrefix = {arXiv},
       eprint = {1811.06549},
 primaryClass = {astro-ph.GA},
       adsurl = {https://ui.adsabs.harvard.edu/abs/2019A&A...624A.141U}
}

@INPROCEEDINGS{2010SPIE.7735E..08B,
       author = {{Bacon}, R. and {Accardo}, M. and {Adjali}, L. and {Anwand}, H. and {Bauer}, S. and {Biswas}, I. and {Blaizot}, J. and {Boudon}, D. and {Brau-Nogue}, S. and {Brinchmann}, J. and {Caillier}, P. and {Capoani}, L. and {Carollo}, C.~M. and {Contini}, T. and {Couderc}, P. and {Daguis{\'e}}, E. and {Deiries}, S. and {Delabre}, B. and {Dreizler}, S. and {Dubois}, J. and {Dupieux}, M. and {Dupuy}, C. and {Emsellem}, E. and {Fechner}, T. and {Fleischmann}, A. and {Fran{\c{c}}ois}, M. and {Gallou}, G. and {Gharsa}, T. and {Glindemann}, A. and {Gojak}, D. and {Guiderdoni}, B. and {Hansali}, G. and {Hahn}, T. and {Jarno}, A. and {Kelz}, A. and {Koehler}, C. and {Kosmalski}, J. and {Laurent}, F. and {Le Floch}, M. and {Lilly}, S.~J. and {Lizon}, J. -L. and {Loupias}, M. and {Manescau}, A. and {Monstein}, C. and {Nicklas}, H. and {Olaya}, J. -C. and {Pares}, L. and {Pasquini}, L. and {P{\'e}contal-Rousset}, A. and {Pell{\'o}}, R. and {Petit}, C. and {Popow}, E. and {Reiss}, R. and {Remillieux}, A. and {Renault}, E. and {Roth}, M. and {Rupprecht}, G. and {Serre}, D. and {Schaye}, J. and {Soucail}, G. and {Steinmetz}, M. and {Streicher}, O. and {Stuik}, R. and {Valentin}, H. and {Vernet}, J. and {Weilbacher}, P. and {Wisotzki}, L. and {Yerle}, N.},
        title = "{The MUSE second-generation VLT instrument}",
    booktitle = {Ground-based and Airborne Instrumentation for Astronomy III},
         year = 2010,
       editor = {{McLean}, Ian S. and {Ramsay}, Suzanne K. and {Takami}, Hideki},
       series = {Society of Photo-Optical Instrumentation Engineers (SPIE) Conference Series},
       volume = {7735},
        month = jul,
          eid = {773508},
        pages = {773508},
          doi = {10.1117/12.856027},
archivePrefix = {arXiv},
       eprint = {2211.16795},
 primaryClass = {astro-ph.IM},
       adsurl = {https://ui.adsabs.harvard.edu/abs/2010SPIE.7735E..08B}
}

@ARTICLE{2020MNRAS.492.1778M,
       author = {{Matthee}, Jorryt and {Sobral}, David and {Gronke}, Max and {Pezzulli}, Gabriele and {Cantalupo}, Sebastiano and {R{\"o}ttgering}, Huub and {Darvish}, Behnam and {Santos}, S{\'e}rgio},
        title = "{Resolved Lyman-{\ensuremath{\alpha}} properties of a luminous Lyman-break galaxy in a large ionized bubble at z = 6.53}",
      journal = {\mnras},
         year = 2020,
        month = feb,
       volume = {492},
       number = {2},
        pages = {1778-1790},
          doi = {10.1093/mnras/stz3554},
archivePrefix = {arXiv},
       eprint = {1909.06376},
 primaryClass = {astro-ph.GA},
       adsurl = {https://ui.adsabs.harvard.edu/abs/2020MNRAS.492.1778M}
}

@ARTICLE{2021ApJS..256...27P,
       author = {{Pagul}, A. and {S{\'a}nchez}, F.~J. and {Davidzon}, I. and {Mobasher}, Bahram},
        title = "{Hubble Frontier Field Clusters and Their Parallel Fields: Photometric and Photometric Redshift Catalogs}",
      journal = {\apjs},
         year = 2021,
        month = oct,
       volume = {256},
       number = {2},
          eid = {27},
        pages = {27},
          doi = {10.3847/1538-4365/abea9d},
archivePrefix = {arXiv},
       eprint = {2103.01952},
 primaryClass = {astro-ph.GA},
       adsurl = {https://ui.adsabs.harvard.edu/abs/2021ApJS..256...27P}
}

@ARTICLE{2025ApJ...983...10O,
       author = {{Oyarz{\'u}n}, Grecco A. and {Rafelski}, Marc and {Christensen}, Lise and {Ozyurt}, Fiona and {Jorgenson}, Regina A. and {Neeleman}, M. and {Fumagalli}, Michele and {Prochaska}, J. Xavier and {Worseck}, G. and {Wisz}, M.~E. and {Becker}, George D. and {L{\'o}pez}, Sebasti{\'a}n},
        title = "{The Qz5 Survey. I. How the H I Mass Density of the Universe Evolves with Cosmic Time}",
      journal = {\apj},
         year = 2025,
        month = apr,
       volume = {983},
       number = {1},
          eid = {10},
        pages = {10},
          doi = {10.3847/1538-4357/adb430},
archivePrefix = {arXiv},
       eprint = {2502.05261},
 primaryClass = {astro-ph.GA},
       adsurl = {https://ui.adsabs.harvard.edu/abs/2025ApJ...983...10O}
}

@ARTICLE{2023MNRAS.518.4646R,
       author = {{Rhee}, Jonghwan and {Meyer}, Martin and {Popping}, Attila and {Bellstedt}, Sabine and {Driver}, Simon P. and {Robotham}, Aaron S.~G. and {Whiting}, Matthew and {Baldry}, Ivan K. and {Brough}, Sarah and {Brown}, Michael J.~I. and {Bunton}, John D. and {Dodson}, Richard and {Holwerda}, Benne W. and {Hopkins}, Andrew M. and {Koribalski}, B{\"a}rbel S. and {Lee-Waddell}, Karen and {L{\'o}pez-S{\'a}nchez}, {\'A}ngel R. and {Loveday}, Jon and {Mahony}, Elizabeth and {Roychowdhury}, Sambit and {Rozgonyi}, Krist{\'o}f and {Staveley-Smith}, Lister},
        title = "{Deep investigation of neutral gas origins (DINGO): H I stacking experiments with early science data}",
      journal = {\mnras},
         year = 2023,
        month = jan,
       volume = {518},
       number = {3},
        pages = {4646-4671},
          doi = {10.1093/mnras/stac3065},
archivePrefix = {arXiv},
       eprint = {2210.09697},
 primaryClass = {astro-ph.GA},
       adsurl = {https://ui.adsabs.harvard.edu/abs/2023MNRAS.518.4646R}
}

@ARTICLE{2020ARA&A..58..363P,
       author = {{P{\'e}roux}, C{\'e}line and {Howk}, J. Christopher},
        title = "{The Cosmic Baryon and Metal Cycles}",
      journal = {\araa},
         year = 2020,
        month = aug,
       volume = {58},
        pages = {363-406},
          doi = {10.1146/annurev-astro-021820-120014},
archivePrefix = {arXiv},
       eprint = {2011.01935},
 primaryClass = {astro-ph.GA},
       adsurl = {https://ui.adsabs.harvard.edu/abs/2020ARA&A..58..363P}
}

@ARTICLE{2017MNRAS.466.2111B,
       author = {{Bird}, Simeon and {Garnett}, Roman and {Ho}, Shirley},
        title = "{Statistical properties of damped Lyman-alpha systems from Sloan Digital Sky Survey DR12}",
      journal = {\mnras},
         year = 2017,
        month = apr,
       volume = {466},
       number = {2},
        pages = {2111-2122},
          doi = {10.1093/mnras/stw3246},
archivePrefix = {arXiv},
       eprint = {1610.01165},
 primaryClass = {astro-ph.GA},
       adsurl = {https://ui.adsabs.harvard.edu/abs/2017MNRAS.466.2111B}
}

@ARTICLE{2021MNRAS.507..704H,
       author = {{Ho}, Ming-Feng and {Bird}, Simeon and {Garnett}, Roman},
        title = "{Damped Lyman-{\ensuremath{\alpha}} absorbers from Sloan digital sky survey DR16Q with Gaussian processes}",
      journal = {\mnras},
         year = 2021,
        month = oct,
       volume = {507},
       number = {1},
        pages = {704-719},
          doi = {10.1093/mnras/stab2169},
archivePrefix = {arXiv},
       eprint = {2103.10964},
 primaryClass = {astro-ph.GA},
       adsurl = {https://ui.adsabs.harvard.edu/abs/2021MNRAS.507..704H}
}

@ARTICLE{2006A&A...460..397V,
       author = {{Verhamme}, A. and {Schaerer}, D. and {Maselli}, A.},
        title = "{3D Ly{\ensuremath{\alpha}} radiation transfer. I. Understanding Ly{\ensuremath{\alpha}} line profile morphologies}",
      journal = {\aap},
         year = 2006,
        month = dec,
       volume = {460},
       number = {2},
        pages = {397-413},
          doi = {10.1051/0004-6361:20065554},
archivePrefix = {arXiv},
       eprint = {astro-ph/0608075},
 primaryClass = {astro-ph},
       adsurl = {https://ui.adsabs.harvard.edu/abs/2006A&A...460..397V}
}

@ARTICLE{2017A&A...608A.139G,
       author = {{Gronke}, Max},
        title = "{Modeling 237 Lyman-{\ensuremath{\alpha}} spectra of the MUSE-Wide survey}",
      journal = {\aap},
         year = 2017,
        month = dec,
       volume = {608},
          eid = {A139},
        pages = {A139},
          doi = {10.1051/0004-6361/201731791},
archivePrefix = {arXiv},
       eprint = {1709.07008},
 primaryClass = {astro-ph.GA},
       adsurl = {https://ui.adsabs.harvard.edu/abs/2017A&A...608A.139G}
}

@ARTICLE{2017A&A...602A.111H,
       author = {{Herenz}, Edmund Christian and {Wisotzki}, Lutz},
        title = "{LSDCat: Detection and cataloguing of emission-line sources in integral-field spectroscopy datacubes}",
      journal = {\aap},
         year = 2017,
        month = jun,
       volume = {602},
          eid = {A111},
        pages = {A111},
          doi = {10.1051/0004-6361/201629507},
archivePrefix = {arXiv},
       eprint = {1703.05166},
 primaryClass = {astro-ph.IM},
       adsurl = {https://ui.adsabs.harvard.edu/abs/2017A&A...602A.111H}
}

@ARTICLE{2022A&A...666A..78C,
       author = {{Claeyssens}, A. and {Richard}, J. and {Blaizot}, J. and {Garel}, T. and {Kusakabe}, H. and {Bacon}, R. and {Bauer}, F.~E. and {Guaita}, L. and {Jeanneau}, A. and {Lagattuta}, D. and {Leclercq}, F. and {Maseda}, M. and {Matthee}, J. and {Nanayakkara}, T. and {Pello}, R. and {Thai}, T.~T. and {Tuan-Anh}, P. and {Verhamme}, A. and {Vitte}, E. and {Wisotzki}, L.},
        title = "{The Lensed Lyman-Alpha MUSE Arcs Sample (LLAMAS). I. Characterisation of extended Lyman-alpha halos and spatial offsets}",
      journal = {\aap},
         year = 2022,
        month = oct,
       volume = {666},
          eid = {A78},
        pages = {A78},
          doi = {10.1051/0004-6361/202142320},
archivePrefix = {arXiv},
       eprint = {2201.04674},
 primaryClass = {astro-ph.GA},
       adsurl = {https://ui.adsabs.harvard.edu/abs/2022A&A...666A..78C}
}

@ARTICLE{2024MNRAS.527.6110O,
       author = {{Ormerod}, K. and {Conselice}, C.~J. and {Adams}, N.~J. and {Harvey}, T. and {Austin}, D. and {Trussler}, J. and {Ferreira}, L. and {Caruana}, J. and {Lucatelli}, G. and {Li}, Q. and {Roper}, W.~J.},
        title = "{EPOCHS VI: the size and shape evolution of galaxies since z   8 with JWST Observations}",
      journal = {\mnras},
         year = 2024,
        month = jan,
       volume = {527},
       number = {3},
        pages = {6110-6125},
          doi = {10.1093/mnras/stad3597},
archivePrefix = {arXiv},
       eprint = {2309.04377},
 primaryClass = {astro-ph.GA},
       adsurl = {https://ui.adsabs.harvard.edu/abs/2024MNRAS.527.6110O}
}

@ARTICLE{2017MNRAS.465.2717P,
       author = {{Paulino-Afonso}, Ana and {Sobral}, David and {Buitrago}, Fernando and {Afonso}, Jos{\'e}},
        title = "{The structural and size evolution of star-forming galaxies over the last 11 Gyr}",
      journal = {\mnras},
         year = 2017,
        month = mar,
       volume = {465},
       number = {3},
        pages = {2717-2733},
          doi = {10.1093/mnras/stw2933},
archivePrefix = {arXiv},
       eprint = {1611.05039},
 primaryClass = {astro-ph.GA},
       adsurl = {https://ui.adsabs.harvard.edu/abs/2017MNRAS.465.2717P}
}

@ARTICLE{2025ApJ...982..200R,
       author = {{Ren}, Jian and {Liu}, F.~S. and {Li}, Nan and {Zhao}, Pinsong and {Cui}, Qifan and {Song}, Qi and {Li}, Yubin and {Mo}, Hao and {Yesuf}, Hassen M. and {Wang}, Weichen and {An}, Fangxia and {Zheng}, Xian Zhong},
        title = "{The Evolution of the Size and Merger Fraction of Submillimeter Galaxies across 1 < z {\ensuremath{\lesssim}} 6 as Observed by JWST}",
      journal = {\apj},
         year = 2025,
        month = apr,
       volume = {982},
       number = {2},
          eid = {200},
        pages = {200},
          doi = {10.3847/1538-4357/adb961},
archivePrefix = {arXiv},
       eprint = {2502.15569},
 primaryClass = {astro-ph.GA},
       adsurl = {https://ui.adsabs.harvard.edu/abs/2025ApJ...982..200R}
}

@ARTICLE{2025ApJS..281...68Y,
       author = {{Yang}, Lilan and {Kartaltepe}, Jeyhan S. and {Franco}, Maximilien and {Ding}, Xuheng and {Achenbach}, Mark J. and {Arango-Toro}, Rafael C. and {Casey}, Caitlin M. and {Drakos}, Nicole E. and {Faisst}, Andreas L. and {Gillman}, Steven and {Gozaliasl}, Ghassem and {Huertas-Company}, Marc and {Jin}, Shuowen and {Liu}, Daizhong and {Magdis}, Georgios and {Massey}, Richard and {Silverman}, John D. and {Tanaka}, Takumi S. and {Yu}, Si-Yue and {Akins}, Hollis B. and {Allen}, Natalie and {Ilbert}, Olivier and {Koekemoer}, Anton M. and {McCracken}, Henry Joy and {Paquereau}, Louise and {Rhodes}, Jason and {Robertson}, Brant E. and {Shuntov}, Marko and {Toft}, Sune},
        title = "{COSMOS-Web: Unraveling the Evolution of Galaxy Size and Related Properties at 2 < z < 10}",
      journal = {\apjs},
         year = 2025,
        month = dec,
       volume = {281},
       number = {2},
          eid = {68},
        pages = {68},
          doi = {10.3847/1538-4365/ae0e1b},
archivePrefix = {arXiv},
       eprint = {2504.07185},
 primaryClass = {astro-ph.GA},
       adsurl = {https://ui.adsabs.harvard.edu/abs/2025ApJS..281...68Y}
}

@ARTICLE{2013ApJ...770...57B,
       author = {{Behroozi}, Peter S. and {Wechsler}, Risa H. and {Conroy}, Charlie},
        title = "{The Average Star Formation Histories of Galaxies in Dark Matter Halos from z = 0-8}",
      journal = {\apj},
         year = 2013,
        month = jun,
       volume = {770},
       number = {1},
          eid = {57},
        pages = {57},
          doi = {10.1088/0004-637X/770/1/57},
archivePrefix = {arXiv},
       eprint = {1207.6105},
 primaryClass = {astro-ph.CO},
       adsurl = {https://ui.adsabs.harvard.edu/abs/2013ApJ...770...57B}
}

@ARTICLE{2024ApJ...971..136S,
       author = {{Songaila}, A. and {Cowie}, L.~L. and {Barger}, A.~J. and {Hu}, E.~M. and {Taylor}, A.~J.},
        title = "{A Spectral Atlas of Ly{\ensuremath{\alpha}} Emitters at z = 5.7 and z = 6.6}",
      journal = {\apj},
         year = 2024,
        month = aug,
       volume = {971},
       number = {2},
          eid = {136},
        pages = {136},
          doi = {10.3847/1538-4357/ad5674},
archivePrefix = {arXiv},
       eprint = {2407.08772},
 primaryClass = {astro-ph.GA},
       adsurl = {https://ui.adsabs.harvard.edu/abs/2024ApJ...971..136S}
}

@ARTICLE{2022ApJ...935...52S,
       author = {{Songaila}, A. and {Barger}, A.~J. and {Cowie}, L.~L. and {Hu}, E.~M. and {Taylor}, A.~J.},
        title = "{The Evolution of Ly{\ensuremath{\alpha}} Emitter Line Widths from z = 5.7 to z = 6.6}",
      journal = {\apj},
         year = 2022,
        month = aug,
       volume = {935},
       number = {1},
          eid = {52},
        pages = {52},
          doi = {10.3847/1538-4357/ac8051},
archivePrefix = {arXiv},
       eprint = {2207.05758},
 primaryClass = {astro-ph.GA},
       adsurl = {https://ui.adsabs.harvard.edu/abs/2022ApJ...935...52S}
}

@ARTICLE{2023MNRAS.523.3749B,
       author = {{Blaizot}, J{\'e}r{\'e}my and {Garel}, Thibault and {Verhamme}, Anne and {Katz}, Harley and {Kimm}, Taysun and {Michel-Dansac}, L{\'e}o and {Mitchell}, Peter D. and {Rosdahl}, Joakim and {Trebitsch}, Maxime},
        title = "{Simulating the diversity of shapes of the Lyman-{\ensuremath{\alpha}} line}",
      journal = {\mnras},
         year = 2023,
        month = aug,
       volume = {523},
       number = {3},
        pages = {3749-3772},
          doi = {10.1093/mnras/stad1523},
archivePrefix = {arXiv},
       eprint = {2305.10047},
 primaryClass = {astro-ph.GA},
       adsurl = {https://ui.adsabs.harvard.edu/abs/2023MNRAS.523.3749B}
}

@ARTICLE{2008ApJ...672..122E,
       author = {{Evrard}, A.~E. and {Bialek}, J. and {Busha}, M. and {White}, M. and {Habib}, S. and {Heitmann}, K. and {Warren}, M. and {Rasia}, E. and {Tormen}, G. and {Moscardini}, L. and {Power}, C. and {Jenkins}, A.~R. and {Gao}, L. and {Frenk}, C.~S. and {Springel}, V. and {White}, S.~D.~M. and {Diemand}, J.},
        title = "{Virial Scaling of Massive Dark Matter Halos: Why Clusters Prefer a High Normalization Cosmology}",
      journal = {\apj},
         year = 2008,
        month = jan,
       volume = {672},
       number = {1},
        pages = {122-137},
          doi = {10.1086/521616},
archivePrefix = {arXiv},
       eprint = {astro-ph/0702241},
 primaryClass = {astro-ph},
       adsurl = {https://ui.adsabs.harvard.edu/abs/2008ApJ...672..122E}
}

@ARTICLE{2024MNRAS.528.4872L,
       author = {{Lu}, Ting-Yi and {Mason}, Charlotte A. and {Hutter}, Anne and {Mesinger}, Andrei and {Qin}, Yuxiang and {Stark}, Daniel P. and {Endsley}, Ryan},
        title = "{The reionizing bubble size distribution around galaxies}",
      journal = {\mnras},
         year = 2024,
        month = mar,
       volume = {528},
       number = {3},
        pages = {4872-4890},
          doi = {10.1093/mnras/stae266},
archivePrefix = {arXiv},
       eprint = {2304.11192},
 primaryClass = {astro-ph.GA},
       adsurl = {https://ui.adsabs.harvard.edu/abs/2024MNRAS.528.4872L}
}

@ARTICLE{2023ApJ...956..136P,
       author = {{Prieto-Lyon}, Gonzalo and {Mason}, Charlotte and {Mascia}, Sara and {Merlin}, Emiliano and {Roy}, Namrata and {Henry}, Alaina and {Roberts-Borsani}, Guido and {Morishita}, Takahiro and {Wang}, Xin and {Boyett}, Kit and {Bolan}, Patricia and {Bradac}, Marusa and {Castellano}, Marco and {Mercurio}, Amata and {Nanayakkara}, Themiya and {Paris}, Diego and {Pentericci}, Laura and {Scarlata}, Claudia and {Trenti}, Michele and {Treu}, Tommaso and {Vanzella}, Eros},
        title = "{Early Results from GLASS-JWST. XXIII. The Transmission of Ly{\ensuremath{\alpha}} from UV-faint z   3-6 Galaxies}",
      journal = {\apj},
         year = 2023,
        month = oct,
       volume = {956},
       number = {2},
          eid = {136},
        pages = {136},
          doi = {10.3847/1538-4357/acf715},
archivePrefix = {arXiv},
       eprint = {2304.02666},
 primaryClass = {astro-ph.GA},
       adsurl = {https://ui.adsabs.harvard.edu/abs/2023ApJ...956..136P}
}

@ARTICLE{2024A&A...684A..84S,
       author = {{Saxena}, Aayush and {Bunker}, Andrew J. and {Jones}, Gareth C. and {Stark}, Daniel P. and {Cameron}, Alex J. and {Witstok}, Joris and {Arribas}, Santiago and {Baker}, William M. and {Baum}, Stefi and {Bhatawdekar}, Rachana and {Bowler}, Rebecca and {Boyett}, Kristan and {Carniani}, Stefano and {Charlot}, Stephane and {Chevallard}, Jacopo and {Curti}, Mirko and {Curtis-Lake}, Emma and {Eisenstein}, Daniel J. and {Endsley}, Ryan and {Hainline}, Kevin and {Helton}, Jakob M. and {Johnson}, Benjamin D. and {Kumari}, Nimisha and {Looser}, Tobias J. and {Maiolino}, Roberto and {Rieke}, Marcia and {Rix}, Hans-Walter and {Robertson}, Brant E. and {Sandles}, Lester and {Simmonds}, Charlotte and {Smit}, Renske and {Tacchella}, Sandro and {Williams}, Christina C. and {Willmer}, Christopher N.~A. and {Willott}, Chris},
        title = "{JADES: The production and escape of ionizing photons from faint Lyman-alpha emitters in the epoch of reionization}",
      journal = {\aap},
         year = 2024,
        month = apr,
       volume = {684},
          eid = {A84},
        pages = {A84},
          doi = {10.1051/0004-6361/202347132},
archivePrefix = {arXiv},
       eprint = {2306.04536},
 primaryClass = {astro-ph.GA},
       adsurl = {https://ui.adsabs.harvard.edu/abs/2024A&A...684A..84S}
}

@ARTICLE{2021ApJ...922..263P,
       author = {{Park}, Hyunbae and {Jung}, Intae and {Song}, Hyunmi and {Ocvirk}, Pierre and {Shapiro}, Paul R. and {Dawoodbhoy}, Taha and {Iliev}, Ilian T. and {Ahn}, Kyungjin and {Bianco}, Michele and {Kim}, Hyo Jeong},
        title = "{Crucial Factors for Ly{\ensuremath{\alpha}} Transmission in the Reionizing Intergalactic Medium: Infall Motion, H II Bubble Size, and Self-shielded Systems}",
      journal = {\apj},
         year = 2021,
        month = dec,
       volume = {922},
       number = {2},
          eid = {263},
        pages = {263},
          doi = {10.3847/1538-4357/ac2f4b},
archivePrefix = {arXiv},
       eprint = {2105.10770},
 primaryClass = {astro-ph.GA},
       adsurl = {https://ui.adsabs.harvard.edu/abs/2021ApJ...922..263P}
}

@ARTICLE{2016A&A...596A.108P,
       author = {{Planck Collaboration} and {Adam}, R. and {Aghanim}, N. and {Ashdown}, M. and {Aumont}, J. and {Baccigalupi}, C. and {Ballardini}, M. and {Banday}, A.~J. and {Barreiro}, R.~B. and {Bartolo}, N. and {Basak}, S. and {Battye}, R. and {Benabed}, K. and {Bernard}, J.-P. and {Bersanelli}, M. and {Bielewicz}, P. and {Bock}, J.~J. and {Bonaldi}, A. and {Bonavera}, L. and {Bond}, J.~R. and {Borrill}, J. and {Bouchet}, F.~R. and {Boulanger}, F. and {Bucher}, M. and {Burigana}, C. and {Calabrese}, E. and {Cardoso}, J.-F. and {Carron}, J. and {Chiang}, H.~C. and {Colombo}, L.~P.~L. and {Combet}, C. and {Comis}, B. and {Couchot}, F. and {Coulais}, A. and {Crill}, B.~P. and {Curto}, A. and {Cuttaia}, F. and {Davis}, R.~J. and {de Bernardis}, P. and {de Rosa}, A. and {de Zotti}, G. and {Delabrouille}, J. and {Di Valentino}, E. and {Dickinson}, C. and {Diego}, J.~M. and {Dor{\'e}}, O. and {Douspis}, M. and {Ducout}, A. and {Dupac}, X. and {Elsner}, F. and {En{\ss}lin}, T.~A. and {Eriksen}, H.~K. and {Falgarone}, E. and {Fantaye}, Y. and {Finelli}, F. and {Forastieri}, F. and {Frailis}, M. and {Fraisse}, A.~A. and {Franceschi}, E. and {Frolov}, A. and {Galeotta}, S. and {Galli}, S. and {Ganga}, K. and {G{\'e}nova-Santos}, R.~T. and {Gerbino}, M. and {Ghosh}, T. and {Gonz{\'a}lez-Nuevo}, J. and {G{\'o}rski}, K.~M. and {Gruppuso}, A. and {Gudmundsson}, J.~E. and {Hansen}, F.~K. and {Helou}, G. and {Henrot-Versill{\'e}}, S. and {Herranz}, D. and {Hivon}, E. and {Huang}, Z. and {Ili{\'c}}, S. and {Jaffe}, A.~H. and {Jones}, W.~C. and {Keih{\"a}nen}, E. and {Keskitalo}, R. and {Kisner}, T.~S. and {Knox}, L. and {Krachmalnicoff}, N. and {Kunz}, M. and {Kurki-Suonio}, H. and {Lagache}, G. and {L{\"a}hteenm{\"a}ki}, A. and {Lamarre}, J.-M. and {Langer}, M. and {Lasenby}, A. and {Lattanzi}, M. and {Lawrence}, C.~R. and {Le Jeune}, M. and {Levrier}, F. and {Lewis}, A. and {Liguori}, M. and {Lilje}, P.~B. and {L{\'o}pez-Caniego}, M. and {Ma}, Y.-Z. and {Mac{\'\i}as-P{\'e}rez}, J.~F. and {Maggio}, G. and {Mangilli}, A. and {Maris}, M. and {Martin}, P.~G. and {Mart{\'\i}nez-Gonz{\'a}lez}, E. and {Matarrese}, S. and {Mauri}, N. and {McEwen}, J.~D. and {Meinhold}, P.~R. and {Melchiorri}, A. and {Mennella}, A. and {Migliaccio}, M. and {Miville-Desch{\^e}nes}, M.-A. and {Molinari}, D. and {Moneti}, A. and {Montier}, L. and {Morgante}, G. and {Moss}, A. and {Naselsky}, P. and {Natoli}, P. and {Oxborrow}, C.~A. and {Pagano}, L. and {Paoletti}, D. and {Partridge}, B. and {Patanchon}, G. and {Patrizii}, L. and {Perdereau}, O. and {Perotto}, L. and {Pettorino}, V. and {Piacentini}, F. and {Plaszczynski}, S. and {Polastri}, L. and {Polenta}, G. and {Puget}, J.-L. and {Rachen}, J.~P. and {Racine}, B. and {Reinecke}, M. and {Remazeilles}, M. and {Renzi}, A. and {Rocha}, G. and {Rossetti}, M. and {Roudier}, G. and {Rubi{\~n}o-Mart{\'\i}n}, J.~A. and {Ruiz-Granados}, B. and {Salvati}, L. and {Sandri}, M. and {Savelainen}, M. and {Scott}, D. and {Sirri}, G. and {Sunyaev}, R. and {Suur-Uski}, A.-S. and {Tauber}, J.~A. and {Tenti}, M. and {Toffolatti}, L. and {Tomasi}, M. and {Tristram}, M. and {Trombetti}, T. and {Valiviita}, J. and {Van Tent}, F. and {Vielva}, P. and {Villa}, F. and {Vittorio}, N. and {Wandelt}, B.~D. and {Wehus}, I.~K. and {White}, M. and {Zacchei}, A. and {Zonca}, A.},
        title = "{Planck intermediate results. XLVII. Planck constraints on reionization history}",
      journal = {\aap},
         year = 2016,
        month = dec,
       volume = {596},
          eid = {A108},
        pages = {A108},
          doi = {10.1051/0004-6361/201628897},
archivePrefix = {arXiv},
       eprint = {1605.03507},
 primaryClass = {astro-ph.CO},
       adsurl = {https://ui.adsabs.harvard.edu/abs/2016A&A...596A.108P}
}

@ARTICLE{2026ApJ...997...86U,
       author = {{Umeda}, Hiroya and {Ouchi}, Masami and {Kageura}, Yuta and {Harikane}, Yuichi and {Nakane}, Minami and {Thai}, Tran Thi and {Nakajima}, Kimihiko},
        title = "{Probing the Cosmic Reionization History with JWST: Gunn─Peterson and Ly{\ensuremath{\alpha}} Damping Wing Absorption at 4.5 < z < 13}",
      journal = {\apj},
         year = 2026,
        month = jan,
       volume = {997},
       number = {1},
          eid = {86},
        pages = {86},
          doi = {10.3847/1538-4357/ae232b},
archivePrefix = {arXiv},
       eprint = {2504.04683},
 primaryClass = {astro-ph.GA},
       adsurl = {https://ui.adsabs.harvard.edu/abs/2026ApJ...997...86U}
}

@ARTICLE{2020MNRAS.494..600M,
       author = {{Mangena}, Tumelo and {Hassan}, Sultan and {Santos}, Mario G.},
        title = "{Constraining the reionization history using deep learning from 21-cm tomography with the Square Kilometre Array}",
      journal = {\mnras},
         year = 2020,
        month = may,
       volume = {494},
       number = {1},
        pages = {600-606},
          doi = {10.1093/mnras/staa750},
archivePrefix = {arXiv},
       eprint = {2003.04905},
 primaryClass = {astro-ph.CO},
       adsurl = {https://ui.adsabs.harvard.edu/abs/2020MNRAS.494..600M}
}

@ARTICLE{1999ApJ...524..527L,
       author = {{Loeb}, Abraham and {Rybicki}, George B.},
        title = "{Scattered Ly{\ensuremath{\alpha}} Radiation around Sources before Cosmological Reionization}",
      journal = {\apj},
         year = 1999,
        month = oct,
       volume = {524},
       number = {2},
        pages = {527-535},
          doi = {10.1086/307844},
archivePrefix = {arXiv},
       eprint = {astro-ph/9902180},
 primaryClass = {astro-ph},
       adsurl = {https://ui.adsabs.harvard.edu/abs/1999ApJ...524..527L}
}

@ARTICLE{2024arXiv240816820P,
       author = {{Padmanabhan}, Hamsa and {Loeb}, Abraham},
        title = "{Intensity mapping of Loeb-Rybicki haloes from scattering of galactic Lyman-$\alpha$ emission by the diffuse intergalactic medium before reionization}",
      journal = {arXiv e-prints},
         year = 2024,
        month = aug,
          eid = {arXiv:2408.16820},
        pages = {arXiv:2408.16820},
          doi = {10.48550/arXiv.2408.16820},
archivePrefix = {arXiv},
       eprint = {2408.16820},
 primaryClass = {astro-ph.CO},
         note = {accepted for publication in \aap},
       adsurl = {https://ui.adsabs.harvard.edu/abs/2024arXiv240816820P}
}

@ARTICLE{2024ApJ...977..250F,
       author = {{Fujimoto}, Seiji and {Wang}, Bingjie and {Weaver}, John R. and {Kokorev}, Vasily and {Atek}, Hakim and {Bezanson}, Rachel and {Labbe}, Ivo and {Brammer}, Gabriel and {Greene}, Jenny E. and {Chemerynska}, Iryna and {Dayal}, Pratika and {de Graaff}, Anna and {Furtak}, Lukas J. and {Oesch}, Pascal A. and {Setton}, David J. and {Price}, Sedona H. and {Miller}, Tim B. and {Williams}, Christina C. and {Whitaker}, Katherine E. and {Zitrin}, Adi and {Cutler}, Sam E. and {Leja}, Joel and {Pan}, Richard and {Coe}, Dan and {van Dokkum}, Pieter and {Feldmann}, Robert and {Fudamoto}, Yoshinobu and {Goulding}, Andy D. and {Khullar}, Gourav and {Marchesini}, Danilo and {Maseda}, Michael and {Nanayakkara}, Themiya and {Nelson}, Erica J. and {Smit}, Renske and {Stefanon}, Mauro and {Weibel}, Andrea},
        title = "{UNCOVER: A NIRSpec Census of Lensed Galaxies at z = 8.50{\textendash}13.08 Probing a High-AGN Fraction and Ionized Bubbles in the Shadow}",
      journal = {\apj},
         year = 2024,
        month = dec,
       volume = {977},
       number = {2},
          eid = {250},
        pages = {250},
          doi = {10.3847/1538-4357/ad9027},
archivePrefix = {arXiv},
       eprint = {2308.11609},
 primaryClass = {astro-ph.GA},
       adsurl = {https://ui.adsabs.harvard.edu/abs/2024ApJ...977..250F}
}

@ARTICLE{2023MNRAS.524.5891T,
       author = {{Trapp}, A.~C. and {Furlanetto}, Steven R. and {Davies}, Frederick B.},
        title = "{Lyman {\ensuremath{\alpha}} emitters in ionized bubbles: constraining the environment and ionized fraction}",
      journal = {\mnras},
         year = 2023,
        month = oct,
       volume = {524},
       number = {4},
        pages = {5891-5903},
          doi = {10.1093/mnras/stad2228},
archivePrefix = {arXiv},
       eprint = {2210.06504},
 primaryClass = {astro-ph.CO},
       adsurl = {https://ui.adsabs.harvard.edu/abs/2023MNRAS.524.5891T}
}

@article{Neyer2026THESAN,
	author = {Neyer, Meredith and Smith, Aaron and Vogelsberger, Mark and Garc{\' i}a, Luz {\' A}ngela and Kannan, Rahul and Garaldi, Enrico and Keating, Laura},
	journal = {The Open Journal of Astrophysics},
	doi = {10.33232/001c.160756},
	year = {2026},
	month = {apr 17},
	publisher = {Maynooth Academic Publishing},
	title = {The {THESAN} project: Lyman-alpha emitters as probes of ionized bubble sizes},
	volume = {9},
}

@ARTICLE{2024A&A...688A.106N,
       author = {{Napolitano}, L. and {Pentericci}, L. and {Santini}, P. and {Calabr{\`o}}, A. and {Mascia}, S. and {Llerena}, M. and {Castellano}, M. and {Dickinson}, M. and {Finkelstein}, S.~L. and {Amor{\'\i}n}, R. and {Arrabal Haro}, P. and {Bagley}, M. and {Bhatawdekar}, R. and {Cleri}, N.~J. and {Davis}, K. and {Gardner}, J.~P. and {Gawiser}, E. and {Giavalisco}, M. and {Hathi}, N. and {Holwerda}, B.~W. and {Hu}, W. and {Jung}, I. and {Kartaltepe}, J.~S. and {Koekemoer}, A.~M. and {Larson}, R.~L. and {Merlin}, E. and {Mobasher}, B. and {Papovich}, C. and {Park}, H. and {Pirzkal}, N. and {Trump}, J.~R. and {Wilkins}, S.~M. and {Yung}, L.~Y.~A.},
        title = "{Peering into cosmic reionization: Ly{\ensuremath{\alpha}} visibility evolution from galaxies at z = 4.5-8.5 with JWST}",
      journal = {\aap},
         year = 2024,
        month = aug,
       volume = {688},
          eid = {A106},
        pages = {A106},
          doi = {10.1051/0004-6361/202449644},
archivePrefix = {arXiv},
       eprint = {2402.11220},
 primaryClass = {astro-ph.GA},
       adsurl = {https://ui.adsabs.harvard.edu/abs/2024A&A...688A.106N}
}

@ARTICLE{2025ApJ...984...95R,
       author = {{Runnholm}, Axel and {Hayes}, Matthew J. and {Mehta}, Vihang and {Malkan}, Matthew A. and {Scarlata}, Claudia and {Nedkova}, Kalina V. and {Rafelski}, Marc and {Vulcani}, Benedetta and {Huberty}, Mason and {Herenz}, E. Christian and {Hutter}, Anne and {Bruton}, Sean and {Acharyya}, Ayan and {Atek}, Hakim and {Baronchelli}, Ivano and {Battisti}, Andrew J. and {Brada{\v{c}}}, Maru{\v{s}}a and {Bunker}, Andrew J. and {Dai}, Y. Sophia and {Hannahs}, Clea and {Hasan}, Farhanul and {Kim}, Keunho J. and {Leethochawalit}, Nicha and {Lin}, Yu-Heng and {Rutkowski}, Michael J. and {Saldana-Lopez}, Alberto and {Sattari}, Zahra and {Wang}, Xin},
        title = "{The JWST/PASSAGE Survey: Testing Reionization Histories with JWST's First Unbiased Survey for Ly{\ensuremath{\alpha}} Emitters at Redshifts 7.5─9.5}",
      journal = {\apj},
         year = 2025,
        month = may,
       volume = {984},
       number = {1},
          eid = {95},
        pages = {95},
          doi = {10.3847/1538-4357/adc008},
archivePrefix = {arXiv},
       eprint = {2502.19174},
 primaryClass = {astro-ph.GA},
       adsurl = {https://ui.adsabs.harvard.edu/abs/2025ApJ...984...95R}
}

@ARTICLE{2025Natur.639..897W,
       author = {{Witstok}, Joris and {Jakobsen}, Peter and {Maiolino}, Roberto and {Helton}, Jakob M. and {Johnson}, Benjamin D. and {Robertson}, Brant E. and {Tacchella}, Sandro and {Cameron}, Alex J. and {Smit}, Renske and {Bunker}, Andrew J. and {Saxena}, Aayush and {Sun}, Fengwu and {Alberts}, Stacey and {Arribas}, Santiago and {Baker}, William M. and {Bhatawdekar}, Rachana and {Boyett}, Kristan and {Cargile}, Phillip A. and {Carniani}, Stefano and {Charlot}, St{\'e}phane and {Chevallard}, Jacopo and {Curti}, Mirko and {Curtis-Lake}, Emma and {D'Eugenio}, Francesco and {Eisenstein}, Daniel J. and {Hainline}, Kevin N. and {Jones}, Gareth C. and {Kumari}, Nimisha and {Maseda}, Michael V. and {P{\'e}rez-Gonz{\'a}lez}, Pablo G. and {Rinaldi}, Pierluigi and {Scholtz}, Jan and {{\"U}bler}, Hannah and {Williams}, Christina C. and {Willmer}, Christopher N.~A. and {Willott}, Chris and {Zhu}, Yongda},
        title = "{Witnessing the onset of reionization through Lyman-{\ensuremath{\alpha}} emission at redshift 13}",
      journal = {\nat},
         year = 2025,
        month = mar,
       volume = {639},
       number = {8056},
        pages = {897-901},
          doi = {10.1038/s41586-025-08779-5},
archivePrefix = {arXiv},
       eprint = {2408.16608},
 primaryClass = {astro-ph.GA},
       adsurl = {https://ui.adsabs.harvard.edu/abs/2025Natur.639..897W}
}

@ARTICLE{2019MNRAS.482.3162A,
       author = {{Arrigoni Battaia}, Fabrizio and {Hennawi}, Joseph F. and {Prochaska}, J. Xavier and {O{\~n}orbe}, Jose and {Farina}, Emanuele P. and {Cantalupo}, Sebastiano and {Lusso}, Elisabeta},
        title = "{QSO MUSEUM I: a sample of 61 extended Ly {\ensuremath{\alpha}}-emission nebulae surrounding z {\ensuremath{\sim}} 3 quasars}",
      journal = {\mnras},
         year = 2019,
        month = jan,
       volume = {482},
       number = {3},
        pages = {3162-3205},
          doi = {10.1093/mnras/sty2827},
archivePrefix = {arXiv},
       eprint = {1808.10857},
 primaryClass = {astro-ph.GA},
       adsurl = {https://ui.adsabs.harvard.edu/abs/2019MNRAS.482.3162A}
}

@ARTICLE{2017ApJ...848...78F,
       author = {{Farina}, Emanuele P. and {Venemans}, Bram P. and {Decarli}, Roberto and {Hennawi}, Joseph F. and {Walter}, Fabian and {Ba{\~n}ados}, Eduardo and {Mazzucchelli}, Chiara and {Cantalupo}, Sebastiano and {Arrigoni-Battaia}, Fabrizio and {McGreer}, Ian D.},
        title = "{Mapping the Ly{\ensuremath{\alpha}} Emission around a z {\ensuremath{\sim}} 6.6 QSO with MUSE: Extended Emission and a Companion at a Close Separation}",
      journal = {\apj},
         year = 2017,
        month = oct,
       volume = {848},
       number = {2},
          eid = {78},
        pages = {78},
          doi = {10.3847/1538-4357/aa8df4},
archivePrefix = {arXiv},
       eprint = {1709.06096},
 primaryClass = {astro-ph.GA},
       adsurl = {https://ui.adsabs.harvard.edu/abs/2017ApJ...848...78F}
}

@ARTICLE{2020A&A...635A..82L,
       author = {{Leclercq}, Floriane and {Bacon}, Roland and {Verhamme}, Anne and {Garel}, Thibault and {Blaizot}, J{\'e}r{\'e}my and {Brinchmann}, Jarle and {Cantalupo}, Sebastiano and {Claeyssens}, Ad{\'e}la{\"\i}de and {Conseil}, Simon and {Contini}, Thierry and {Hashimoto}, Takuya and {Herenz}, Edmund Christian and {Kusakabe}, Haruka and {Marino}, Raffaella Anna and {Maseda}, Michael and {Matthee}, Jorryt and {Mitchell}, Peter and {Pezzulli}, Gabriele and {Richard}, Johan and {Schmidt}, Kasper Borello and {Wisotzki}, Lutz},
        title = "{The MUSE Hubble Ultra Deep Field Survey. XIII. Spatially resolved spectral properties of Lyman {\ensuremath{\alpha}} haloes around star-forming galaxies at z > 3}",
      journal = {\aap},
         year = 2020,
        month = mar,
       volume = {635},
          eid = {A82},
        pages = {A82},
          doi = {10.1051/0004-6361/201937339},
archivePrefix = {arXiv},
       eprint = {2002.05731},
 primaryClass = {astro-ph.GA},
       adsurl = {https://ui.adsabs.harvard.edu/abs/2020A&A...635A..82L}
}

@ARTICLE{2018MNRAS.478L..60V,
       author = {{Verhamme}, A. and {Garel}, T. and {Ventou}, E. and {Contini}, T. and {Bouch{\'e}}, N. and {Herenz}, EC and {Richard}, J. and {Bacon}, R. and {Schmidt}, KB and {Maseda}, M. and {Marino}, RA and {Brinchmann}, J. and {Cantalupo}, S. and {Caruana}, J. and {Cl{\'e}ment}, B. and {Diener}, C. and {Drake}, AB and {Hashimoto}, T. and {Inami}, H. and {Kerutt}, J. and {Kollatschny}, W. and {Leclercq}, F. and {Patr{\'\i}cio}, V. and {Schaye}, J. and {Wisotzki}, L. and {Zabl}, J.},
        title = "{Recovering the systemic redshift of galaxies from their Lyman alpha line profile}",
      journal = {\mnras},
         year = 2018,
        month = jul,
       volume = {478},
       number = {1},
        pages = {L60-L65},
          doi = {10.1093/mnrasl/sly058},
archivePrefix = {arXiv},
       eprint = {1804.01883},
 primaryClass = {astro-ph.GA},
       adsurl = {https://ui.adsabs.harvard.edu/abs/2018MNRAS.478L..60V}
}

@ARTICLE{2020MNRAS.496.1013M,
       author = {{Muzahid}, Sowgat and {Schaye}, Joop and {Marino}, Raffaella Anna and {Cantalupo}, Sebastiano and {Brinchmann}, Jarle and {Contini}, Thierry and {Wendt}, Martin and {Wisotzki}, Lutz and {Zabl}, Johannes and {Bouch{\'e}}, Nicolas and {Akhlaghi}, Mohammad and {Chen}, Hsiao-Wen and {Claeyssens}, Ad{\'e}la{\^\i}de and {Johnson}, Sean and {Leclercq}, Floriane and {Maseda}, Michael and {Matthee}, Jorryt and {Richard}, Johan and {Urrutia}, Tanya and {Verhamme}, Anne},
        title = "{MUSEQuBES: calibrating the redshifts of Ly {\ensuremath{\alpha}} emitters using stacked circumgalactic medium absorption profiles}",
      journal = {\mnras},
         year = 2020,
        month = aug,
       volume = {496},
       number = {2},
        pages = {1013-1022},
          doi = {10.1093/mnras/staa1347},
archivePrefix = {arXiv},
       eprint = {1910.03593},
 primaryClass = {astro-ph.GA},
       adsurl = {https://ui.adsabs.harvard.edu/abs/2020MNRAS.496.1013M}
}

@ARTICLE{2021A&A...654A..80S,
       author = {{Schmidt}, K.~B. and {Kerutt}, J. and {Wisotzki}, L. and {Urrutia}, T. and {Feltre}, A. and {Maseda}, M.~V. and {Nanayakkara}, T. and {Bacon}, R. and {Boogaard}, L.~A. and {Conseil}, S. and {Contini}, T. and {Herenz}, E.~C. and {Kollatschny}, W. and {Krumpe}, M. and {Leclercq}, F. and {Mahler}, G. and {Matthee}, J. and {Mauerhofer}, V. and {Richard}, J. and {Schaye}, J.},
        title = "{Recovery and analysis of rest-frame UV emission lines in 2052 galaxies observed with MUSE at 1.5 < z < 6.4}",
      journal = {\aap},
         year = 2021,
        month = oct,
       volume = {654},
          eid = {A80},
        pages = {A80},
          doi = {10.1051/0004-6361/202140876},
archivePrefix = {arXiv},
       eprint = {2108.01713},
 primaryClass = {astro-ph.GA},
       adsurl = {https://ui.adsabs.harvard.edu/abs/2021A&A...654A..80S}
}

@ARTICLE{2022A&A...662A..64R,
       author = {{Rasekh}, A. and {Melinder}, J. and {{\"O}stlin}, G. and {Hayes}, M. and {Herenz}, E.~C. and {Runnholm}, A. and {Kunth}, D. and {Mas Hesse}, J.~M. and {Verhamme}, A. and {Cannon}, J.~M.},
        title = "{The Lyman Alpha Reference Sample. XII. Morphology of extended Lyman alpha emission in star-forming galaxies}",
      journal = {\aap},
         year = 2022,
        month = jun,
       volume = {662},
          eid = {A64},
        pages = {A64},
          doi = {10.1051/0004-6361/202140734},
archivePrefix = {arXiv},
       eprint = {2110.01626},
 primaryClass = {astro-ph.GA},
       adsurl = {https://ui.adsabs.harvard.edu/abs/2022A&A...662A..64R}
}

@article{astropy:2013,
Adsurl = {http://adsabs.harvard.edu/abs/2013A%26A...558A..33A},
Archiveprefix = {arXiv},
Author = {{Astropy Collaboration} and {Robitaille}, T.~P. and {Tollerud}, E.~J. and {Greenfield}, P. and {Droettboom}, M. and {Bray}, E. and {Aldcroft}, T. and {Davis}, M. and {Ginsburg}, A. and {Price-Whelan}, A.~M. and {Kerzendorf}, W.~E. and {Conley}, A. and {Crighton}, N. and {Barbary}, K. and {Muna}, D. and {Ferguson}, H. and {Grollier}, F. and {Parikh}, M.~M. and {Nair}, P.~H. and {Unther}, H.~M. and {Deil}, C. and {Woillez}, J. and {Conseil}, S. and {Kramer}, R. and {Turner}, J.~E.~H. and {Singer}, L. and {Fox}, R. and {Weaver}, B.~A. and {Zabalza}, V. and {Edwards}, Z.~I. and {Azalee Bostroem}, K. and {Burke}, D.~J. and {Casey}, A.~R. and {Crawford}, S.~M. and {Dencheva}, N. and {Ely}, J. and {Jenness}, T. and {Labrie}, K. and {Lim}, P.~L. and {Pierfederici}, F. and {Pontzen}, A. and {Ptak}, A. and {Refsdal}, B. and {Servillat}, M. and {Streicher}, O.},
Doi = {10.1051/0004-6361/201322068},
Eid = {A33},
Eprint = {1307.6212},
Journal = {\aap},
Month = oct,
Pages = {A33},
Primaryclass = {astro-ph.IM},
Title = {{Astropy: A community Python package for astronomy}},
Volume = 558,
Year = 2013}

@ARTICLE{astropy:2018,
       author = {{Astropy Collaboration} and {Price-Whelan}, A.~M. and
         {Sip{\H{o}}cz}, B.~M. and {G{\"u}nther}, H.~M. and {Lim}, P.~L. and
         {Crawford}, S.~M. and {Conseil}, S. and {Shupe}, D.~L. and
         {Craig}, M.~W. and {Dencheva}, N. and {Ginsburg}, A. and {Vand
        erPlas}, J.~T. and {Bradley}, L.~D. and {P{\'e}rez-Su{\'a}rez}, D. and
         {de Val-Borro}, M. and {Aldcroft}, T.~L. and {Cruz}, K.~L. and
         {Robitaille}, T.~P. and {Tollerud}, E.~J. and {Ardelean}, C. and
         {Babej}, T. and {Bach}, Y.~P. and {Bachetti}, M. and {Bakanov}, A.~V. and
         {Bamford}, S.~P. and {Barentsen}, G. and {Barmby}, P. and
         {Baumbach}, A. and {Berry}, K.~L. and {Biscani}, F. and {Boquien}, M. and
         {Bostroem}, K.~A. and {Bouma}, L.~G. and {Brammer}, G.~B. and
         {Bray}, E.~M. and {Breytenbach}, H. and {Buddelmeijer}, H. and
         {Burke}, D.~J. and {Calderone}, G. and {Cano Rodr{\'\i}guez}, J.~L. and
         {Cara}, M. and {Cardoso}, J.~V.~M. and {Cheedella}, S. and {Copin}, Y. and
         {Corrales}, L. and {Crichton}, D. and {D'Avella}, D. and {Deil}, C. and
         {Depagne}, {\'E}. and {Dietrich}, J.~P. and {Donath}, A. and
         {Droettboom}, M. and {Earl}, N. and {Erben}, T. and {Fabbro}, S. and
         {Ferreira}, L.~A. and {Finethy}, T. and {Fox}, R.~T. and
         {Garrison}, L.~H. and {Gibbons}, S.~L.~J. and {Goldstein}, D.~A. and
         {Gommers}, R. and {Greco}, J.~P. and {Greenfield}, P. and
         {Groener}, A.~M. and {Grollier}, F. and {Hagen}, A. and {Hirst}, P. and
         {Homeier}, D. and {Horton}, A.~J. and {Hosseinzadeh}, G. and {Hu}, L. and
         {Hunkeler}, J.~S. and {Ivezi{\'c}}, {\v{Z}}. and {Jain}, A. and
         {Jenness}, T. and {Kanarek}, G. and {Kendrew}, S. and {Kern}, N.~S. and
         {Kerzendorf}, W.~E. and {Khvalko}, A. and {King}, J. and {Kirkby}, D. and
         {Kulkarni}, A.~M. and {Kumar}, A. and {Lee}, A. and {Lenz}, D. and
         {Littlefair}, S.~P. and {Ma}, Z. and {Macleod}, D.~M. and
         {Mastropietro}, M. and {McCully}, C. and {Montagnac}, S. and
         {Morris}, B.~M. and {Mueller}, M. and {Mumford}, S.~J. and {Muna}, D. and
         {Murphy}, N.~A. and {Nelson}, S. and {Nguyen}, G.~H. and
         {Ninan}, J.~P. and {N{\"o}the}, M. and {Ogaz}, S. and {Oh}, S. and
         {Parejko}, J.~K. and {Parley}, N. and {Pascual}, S. and {Patil}, R. and
         {Patil}, A.~A. and {Plunkett}, A.~L. and {Prochaska}, J.~X. and
         {Rastogi}, T. and {Reddy Janga}, V. and {Sabater}, J. and
         {Sakurikar}, P. and {Seifert}, M. and {Sherbert}, L.~E. and
         {Sherwood-Taylor}, H. and {Shih}, A.~Y. and {Sick}, J. and
         {Silbiger}, M.~T. and {Singanamalla}, S. and {Singer}, L.~P. and
         {Sladen}, P.~H. and {Sooley}, K.~A. and {Sornarajah}, S. and
         {Streicher}, O. and {Teuben}, P. and {Thomas}, S.~W. and
         {Tremblay}, G.~R. and {Turner}, J.~E.~H. and {Terr{\'o}n}, V. and
         {van Kerkwijk}, M.~H. and {de la Vega}, A. and {Watkins}, L.~L. and
         {Weaver}, B.~A. and {Whitmore}, J.~B. and {Woillez}, J. and
         {Zabalza}, V. and {Astropy Contributors}},
        title = "{The Astropy Project: Building an Open-science Project and Status of the v2.0 Core Package}",
      journal = {\aj},
         year = 2018,
        month = sep,
       volume = {156},
       number = {3},
          eid = {123},
        pages = {123},
          doi = {10.3847/1538-3881/aabc4f},
archivePrefix = {arXiv},
       eprint = {1801.02634},
 primaryClass = {astro-ph.IM},
       adsurl = {https://ui.adsabs.harvard.edu/abs/2018AJ....156..123A}
}

@ARTICLE{astropy:2022,
       author = {{Astropy Collaboration} and {Price-Whelan}, Adrian M. and {Lim}, Pey Lian and {Earl}, Nicholas and {Starkman}, Nathaniel and {Bradley}, Larry and {Shupe}, David L. and {Patil}, Aarya A. and {Corrales}, Lia and {Brasseur}, C.~E. and {N{"o}the}, Maximilian and {Donath}, Axel and {Tollerud}, Erik and {Morris}, Brett M. and {Ginsburg}, Adam and {Vaher}, Eero and {Weaver}, Benjamin A. and {Tocknell}, James and {Jamieson}, William and {van Kerkwijk}, Marten H. and {Robitaille}, Thomas P. and {Merry}, Bruce and {Bachetti}, Matteo and {G{"u}nther}, H. Moritz and {Aldcroft}, Thomas L. and {Alvarado-Montes}, Jaime A. and {Archibald}, Anne M. and {B{'o}di}, Attila and {Bapat}, Shreyas and {Barentsen}, Geert and {Baz{'a}n}, Juanjo and {Biswas}, Manish and {Boquien}, M{'e}d{'e}ric and {Burke}, D.~J. and {Cara}, Daria and {Cara}, Mihai and {Conroy}, Kyle E. and {Conseil}, Simon and {Craig}, Matthew W. and {Cross}, Robert M. and {Cruz}, Kelle L. and {D'Eugenio}, Francesco and {Dencheva}, Nadia and {Devillepoix}, Hadrien A.~R. and {Dietrich}, J{"o}rg P. and {Eigenbrot}, Arthur Davis and {Erben}, Thomas and {Ferreira}, Leonardo and {Foreman-Mackey}, Daniel and {Fox}, Ryan and {Freij}, Nabil and {Garg}, Suyog and {Geda}, Robel and {Glattly}, Lauren and {Gondhalekar}, Yash and {Gordon}, Karl D. and {Grant}, David and {Greenfield}, Perry and {Groener}, Austen M. and {Guest}, Steve and {Gurovich}, Sebastian and {Handberg}, Rasmus and {Hart}, Akeem and {Hatfield-Dodds}, Zac and {Homeier}, Derek and {Hosseinzadeh}, Griffin and {Jenness}, Tim and {Jones}, Craig K. and {Joseph}, Prajwel and {Kalmbach}, J. Bryce and {Karamehmetoglu}, Emir and {Ka{l}uszy{'n}ski}, Miko{l}aj and {Kelley}, Michael S.~P. and {Kern}, Nicholas and {Kerzendorf}, Wolfgang E. and {Koch}, Eric W. and {Kulumani}, Shankar and {Lee}, Antony and {Ly}, Chun and {Ma}, Zhiyuan and {MacBride}, Conor and {Maljaars}, Jakob M. and {Muna}, Demitri and {Murphy}, N.~A. and {Norman}, Henrik and {O'Steen}, Richard and {Oman}, Kyle A. and {Pacifici}, Camilla and {Pascual}, Sergio and {Pascual-Granado}, J. and {Patil}, Rohit R. and {Perren}, Gabriel I. and {Pickering}, Timothy E. and {Rastogi}, Tanuj and {Roulston}, Benjamin R. and {Ryan}, Daniel F. and {Rykoff}, Eli S. and {Sabater}, Jose and {Sakurikar}, Parikshit and {Salgado}, Jes{'u}s and {Sanghi}, Aniket and {Saunders}, Nicholas and {Savchenko}, Volodymyr and {Schwardt}, Ludwig and {Seifert-Eckert}, Michael and {Shih}, Albert Y. and {Jain}, Anany Shrey and {Shukla}, Gyanendra and {Sick}, Jonathan and {Simpson}, Chris and {Singanamalla}, Sudheesh and {Singer}, Leo P. and {Singhal}, Jaladh and {Sinha}, Manodeep and {Sip{H{o}}cz}, Brigitta M. and {Spitler}, Lee R. and {Stansby}, David and {Streicher}, Ole and {{{S}}umak}, Jani and {Swinbank}, John D. and {Taranu}, Dan S. and {Tewary}, Nikita and {Tremblay}, Grant R. and {Val-Borro}, Miguel de and {Van Kooten}, Samuel J. and {Vasovi{'c}}, Zlatan and {Verma}, Shresth and {de Miranda Cardoso}, Jos{'e} Vin{'i}cius and {Williams}, Peter K.~G. and {Wilson}, Tom J. and {Winkel}, Benjamin and {Wood-Vasey}, W.~M. and {Xue}, Rui and {Yoachim}, Peter and {Zhang}, Chen and {Zonca}, Andrea and {Astropy Project Contributors}},
        title = "{The Astropy Project: Sustaining and Growing a Community-oriented Open-source Project and the Latest Major Release (v5.0) of the Core Package}",
      journal = {\apj},
         year = 2022,
        month = aug,
       volume = {935},
       number = {2},
          eid = {167},
        pages = {167},
          doi = {10.3847/1538-4357/ac7c74},
archivePrefix = {arXiv},
       eprint = {2206.14220},
 primaryClass = {astro-ph.IM},
       adsurl = {https://ui.adsabs.harvard.edu/abs/2022ApJ...935..167A}
}

@Article{Hunter:2007,
  Author    = {Hunter, J. D.},
  Title     = {Matplotlib: A 2D graphics environment},
  Journal   = {Computing in Science \& Engineering},
  Volume    = {9},
  Number    = {3},
  Pages     = {90--95},
  publisher = {IEEE COMPUTER SOC},
  doi       = {10.1109/MCSE.2007.55},
  year      = 2007
}

@article{numpy,
 title         = {Array programming with {NumPy}},
 author        = {Charles R. Harris and K. Jarrod Millman and St{\'{e}}fan J.
                 van der Walt and Ralf Gommers and Pauli Virtanen and David
                 Cournapeau and Eric Wieser and Julian Taylor and Sebastian
                 Berg and Nathaniel J. Smith and Robert Kern and Matti Picus
                 and Stephan Hoyer and Marten H. van Kerkwijk and Matthew
                 Brett and Allan Haldane and Jaime Fern{\'{a}}ndez del
                 R{\'{i}}o and Mark Wiebe and Pearu Peterson and Pierre
                 G{\'{e}}rard-Marchant and Kevin Sheppard and Tyler Reddy and
                 Warren Weckesser and Hameer Abbasi and Christoph Gohlke and
                 Travis E. Oliphant},
 year          = {2020},
 month         = sep,
 journal       = {Nature},
 volume        = {585},
 number        = {7825},
 pages         = {357--362},
 doi           = {10.1038/s41586-020-2649-2},
 publisher     = {Springer Science and Business Media {LLC}},
 url           = {https://doi.org/10.1038/s41586-020-2649-2}
}

@misc{pandas_19340003,
  author       = { The pandas development team},
  title        = {pandas-dev/pandas: Pandas},
  month        = mar,
  year         = 2026,
  publisher    = {Zenodo},
  version      = {v3.0.2},
  doi          = {10.5281/zenodo.19340003},
  url          = {https://doi.org/10.5281/zenodo.19340003},
}

@InProceedings{mckinney-proc-scipy-2010,
  author    = { {W}es {M}c{K}inney },
  title     = { {D}ata {S}tructures for {S}tatistical {C}omputing in {P}ython },
  booktitle = { {P}roceedings of the 9th {P}ython in {S}cience {C}onference },
  pages     = { 56 - 61 },
  year      = { 2010 },
  editor    = { {S}t\'efan van der {W}alt and {J}arrod {M}illman },
  doi       = { 10.25080/Majora-92bf1922-00a }
}

@book{python,
  author    = {Van Rossum, Guido and Drake, Fred L.},
  title     = {Python 3 Reference Manual},
  year      = {2009},
  isbn      = {1441412697},
  publisher = {CreateSpace},
  address   = {Scotts Valley, CA}
}

@misc{scipy_18736568,
  author       = {Ralf Gommers and
                  Pauli Virtanen and
                  Matt Haberland and
                  Evgeni Burovski and
                  Tyler Reddy and
                  Warren Weckesser and
                  Travis E. Oliphant and
                  Andrew Nelson and
                  David Cournapeau and
                  Ilhan Polat and
                  alexbrc and
                  Pamphile Roy and
                  Pearu Peterson and
                  Lucas Colley and
                  Josh Wilson and
                  endolith and
                  Nikolay Mayorov and
                  Jake Bowhay and
                  Stefan van der Walt and
                  Albert Steppi and
                  Matthew Brett and
                  Denis Laxalde and
                  Eric Larson and
                  Atsushi Sakai and
                  Jarrod Millman and
                  Lars and
                  peterbell10 and
                  CJ Carey and
                  Paul van Mulbregt and
                  eric-jones},
  title        = {scipy/scipy: SciPy 1.17.1},
  month        = feb,
  year         = 2026,
  publisher    = {Zenodo},
  version      = {v1.17.1},
  doi          = {10.5281/zenodo.18736568},
  url          = {https://doi.org/10.5281/zenodo.18736568},
  swhid        = {swh:1:dir:934360229a7c597c39d811831bb6c020ef7f8151
                   ;origin=https://doi.org/10.5281/zenodo.595738;visi
                   t=swh:1:snp:2125cfcb4a989632c0e87df14bff9dc83bc816
                   e7;anchor=swh:1:rel:66e87815fdd0136976a21a8a0284e5
                   5c4e5b156e;path=scipy-scipy-c59ed93
                  },
}

@ARTICLE{2020SciPy-NMeth,
  author  = {Virtanen, Pauli and Gommers, Ralf and Oliphant, Travis E. and
            Haberland, Matt and Reddy, Tyler and Cournapeau, David and
            Burovski, Evgeni and Peterson, Pearu and Weckesser, Warren and
            Bright, Jonathan and {van der Walt}, St{\'e}fan J. and
            Brett, Matthew and Wilson, Joshua and Millman, K. Jarrod and
            Mayorov, Nikolay and Nelson, Andrew R. J. and Jones, Eric and
            Kern, Robert and Larson, Eric and Carey, C J and
            Polat, {\.I}lhan and Feng, Yu and Moore, Eric W. and
            {VanderPlas}, Jake and Laxalde, Denis and Perktold, Josef and
            Cimrman, Robert and Henriksen, Ian and Quintero, E. A. and
            Harris, Charles R. and Archibald, Anne M. and
            Ribeiro, Ant{\^o}nio H. and Pedregosa, Fabian and
            {van Mulbregt}, Paul and {SciPy 1.0 Contributors}},
  title   = {{{SciPy} 1.0: Fundamental Algorithms for Scientific
            Computing in Python}},
  journal = {Nature Methods},
  year    = {2020},
  volume  = {17},
  pages   = {261--272},
  adsurl  = {https://rdcu.be/b08Wh},
  doi     = {10.1038/s41592-019-0686-2},
}

@misc{Photutils_17129028,
  author       = {Larry Bradley and
                  Brigitta Sipőcz and
                  Thomas Robitaille and
                  Erik Tollerud and
                  Zé Vinícius and
                  Christoph Deil and
                  Kyle Barbary and
                  Tom J Wilson and
                  Ivo Busko and
                  Axel Donath and
                  Hans Moritz Günther and
                  Mihai Cara and
                  P. L. Lim and
                  Sebastian Meßlinger and
                  Simon Conseil and
                  Michael Droettboom and
                  Azalee Bostroem and
                  E. M. Bray and
                  Lars Andersen Bratholm and
                  Zach Burnett and
                  William Jamieson and
                  Adam Ginsburg and
                  Dan Taranu and
                  Geert Barentsen and
                  Matt Craig and
                  Brett M. Morris and
                  Marshall Perrin and
                  Shivangee Rathi},
  title        = {astropy/photutils: 2.3.0},
  month        = sep,
  year         = 2025,
  publisher    = {Zenodo},
  version      = {2.3.0},
  doi          = {10.5281/zenodo.17129028},
  url          = {https://doi.org/10.5281/zenodo.17129028},
  swhid        = {swh:1:dir:dd51869167d76d722ba87e3f80f9f4199ec08c3f
                   ;origin=https://doi.org/10.5281/zenodo.596036;visi
                   t=swh:1:snp:30a5f50b0586911dc674668853d9abc352a2bc
                   22;anchor=swh:1:rel:e97861da904cf010c499a4211cd8a6
                   12373e912a;path=astropy-photutils-2294e35
                  },
}

@article{scikit-image,
 title = {scikit-image: image processing in {P}ython},
 author = {van der Walt, {S}t\'efan and {S}ch\"onberger, {J}ohannes {L}. and
           {Nunez-Iglesias}, {J}uan and {B}oulogne, {F}ran\c{c}ois and {W}arner,
           {J}oshua {D}. and {Y}ager, {N}eil and {G}ouillart, {E}mmanuelle and
           {Y}u, {T}ony and the scikit-image contributors},
 year = {2014},
 month = {6},
 volume = {2},
 pages = {e453},
 journal = {PeerJ},
 issn = {2167-8359},
 url = {https://doi.org/10.7717/peerj.453},
 doi = {10.7717/peerj.453}
}

@INPROCEEDINGS{2005ASPC..347...29T,
       author = {{Taylor}, M.~B.},
        title = "{TOPCAT \& STIL: Starlink Table/VOTable Processing Software}",
    booktitle = {Astronomical Data Analysis Software and Systems XIV},
         year = 2005,
       editor = {{Shopbell}, P. and {Britton}, M. and {Ebert}, R.},
       series = {Astronomical Society of the Pacific Conference Series},
       volume = {347},
        month = dec,
        pages = {29},
       adsurl = {https://ui.adsabs.harvard.edu/abs/2005ASPC..347...29T}
}

@ARTICLE{software-citation-station-paper,
       author = {{Wagg}, Tom and {Broekgaarden}, Floor S.},
        title = "{Streamlining and standardizing software citations with The Software Citation Station}",
      journal = {arXiv e-prints},
         year = 2024,
        month = jun,
          eid = {arXiv:2406.04405},
        pages = {arXiv:2406.04405},
archivePrefix = {arXiv},
       eprint = {2406.04405},
 primaryClass = {astro-ph.IM},
       adsurl = {https://ui.adsabs.harvard.edu/abs/2024arXiv240604405W}
}

@misc{software-citation-station-zenodo,
  author       = {Tom Wagg and
                  Floor Broekgaarden and
                  Phil Van-Lane and
                  Kai Wu and
                  Kayhan Gültekin},
  title        = {TomWagg/software-citation-station: v1.4},
  month        = nov,
  year         = 2025,
  publisher    = {Zenodo},
  version      = {v1.4},
  doi          = {10.5281/zenodo.17654855},
  url          = {https://doi.org/10.5281/zenodo.17654855},
  swhid        = {swh:1:dir:9a009430037c791424a572f542e9a5d5c1fb44ff
                   ;origin=https://doi.org/10.5281/zenodo.13225526;vi
                   sit=swh:1:snp:ef11f058d718d691f0661c9445c2251328cb
                   ac95;anchor=swh:1:rel:84cde4e532032537b7f014c4467c
                   102430a53fa2;path=TomWagg-software-citation-
                   station-61a588a
                  },
}

@ARTICLE{2000ApJ...533..682C,
       author = {{Calzetti}, Daniela and {Armus}, Lee and {Bohlin}, Ralph C. and {Kinney}, Anne L. and {Koornneef}, Jan and {Storchi-Bergmann}, Thaisa},
        title = "{The Dust Content and Opacity of Actively Star-forming Galaxies}",
      journal = {\apj},
         year = 2000,
        month = apr,
       volume = {533},
       number = {2},
        pages = {682-695},
          doi = {10.1086/308692},
archivePrefix = {arXiv},
       eprint = {astro-ph/9911459},
 primaryClass = {astro-ph},
       adsurl = {https://ui.adsabs.harvard.edu/abs/2000ApJ...533..682C}
}

@ARTICLE{2020ApJ...904...33L,
       author = {{Lower}, Sidney and {Narayanan}, Desika and {Leja}, Joel and {Johnson}, Benjamin D. and {Conroy}, Charlie and {Dav{\'e}}, Romeel},
        title = "{How Well Can We Measure the Stellar Mass of a Galaxy: The Impact of the Assumed Star Formation History Model in SED Fitting}",
      journal = {\apj},
         year = 2020,
        month = nov,
       volume = {904},
       number = {1},
          eid = {33},
        pages = {33},
          doi = {10.3847/1538-4357/abbfa7},
archivePrefix = {arXiv},
       eprint = {2006.03599},
 primaryClass = {astro-ph.GA},
       adsurl = {https://ui.adsabs.harvard.edu/abs/2020ApJ...904...33L}
}

@ARTICLE{2021ApJS..254...22J,
       author = {{Johnson}, Benjamin D. and {Leja}, Joel and {Conroy}, Charlie and {Speagle}, Joshua S.},
        title = "{Stellar Population Inference with Prospector}",
      journal = {\apjs},
         year = 2021,
        month = jun,
       volume = {254},
       number = {2},
          eid = {22},
        pages = {22},
          doi = {10.3847/1538-4365/abef67},
archivePrefix = {arXiv},
       eprint = {2012.01426},
 primaryClass = {astro-ph.GA},
       adsurl = {https://ui.adsabs.harvard.edu/abs/2021ApJS..254...22J}
}

@ARTICLE{2017ApJ...837..170L,
       author = {{Leja}, Joel and {Johnson}, Benjamin D. and {Conroy}, Charlie and {van Dokkum}, Pieter G. and {Byler}, Nell},
        title = "{Deriving Physical Properties from Broadband Photometry with Prospector: Description of the Model and a Demonstration of its Accuracy Using 129 Galaxies in the Local Universe}",
      journal = {\apj},
         year = 2017,
        month = mar,
       volume = {837},
       number = {2},
          eid = {170},
        pages = {170},
          doi = {10.3847/1538-4357/aa5ffe},
archivePrefix = {arXiv},
       eprint = {1609.09073},
 primaryClass = {astro-ph.GA},
       adsurl = {https://ui.adsabs.harvard.edu/abs/2017ApJ...837..170L}
}

@ARTICLE{2021ApJ...908...36H,
       author = {{Hayes}, Matthew J. and {Runnholm}, Axel and {Gronke}, Max and {Scarlata}, Claudia},
        title = "{Spectral Shapes of the Ly{\ensuremath{\alpha}} Emission from Galaxies. I. Blueshifted Emission and Intrinsic Invariance with Redshift}",
      journal = {\apj},
         year = 2021,
        month = feb,
       volume = {908},
       number = {1},
          eid = {36},
        pages = {36},
          doi = {10.3847/1538-4357/abd246},
archivePrefix = {arXiv},
       eprint = {2006.03232},
 primaryClass = {astro-ph.GA},
       adsurl = {https://ui.adsabs.harvard.edu/abs/2021ApJ...908...36H}
}

@ARTICLE{2009ApJ...699..486C,
       author = {{Conroy}, Charlie and {Gunn}, James E. and {White}, Martin},
        title = "{The Propagation of Uncertainties in Stellar Population Synthesis Modeling. I. The Relevance of Uncertain Aspects of Stellar Evolution and the Initial Mass Function to the Derived Physical Properties of Galaxies}",
      journal = {\apj},
         year = 2009,
        month = jul,
       volume = {699},
       number = {1},
        pages = {486-506},
          doi = {10.1088/0004-637X/699/1/486},
archivePrefix = {arXiv},
       eprint = {0809.4261},
 primaryClass = {astro-ph},
       adsurl = {https://ui.adsabs.harvard.edu/abs/2009ApJ...699..486C}
}

@ARTICLE{2010ApJ...712..833C,
       author = {{Conroy}, Charlie and {Gunn}, James E.},
        title = "{The Propagation of Uncertainties in Stellar Population Synthesis Modeling. III. Model Calibration, Comparison, and Evaluation}",
      journal = {\apj},
         year = 2010,
        month = apr,
       volume = {712},
       number = {2},
        pages = {833-857},
          doi = {10.1088/0004-637X/712/2/833},
archivePrefix = {arXiv},
       eprint = {0911.3151},
 primaryClass = {astro-ph.CO},
       adsurl = {https://ui.adsabs.harvard.edu/abs/2010ApJ...712..833C}
}

@ARTICLE{2017ApJ...840...44B,
       author = {{Byler}, Nell and {Dalcanton}, Julianne J. and {Conroy}, Charlie and {Johnson}, Benjamin D.},
        title = "{Nebular Continuum and Line Emission in Stellar Population Synthesis Models}",
      journal = {\apj},
         year = 2017,
        month = may,
       volume = {840},
       number = {1},
          eid = {44},
        pages = {44},
          doi = {10.3847/1538-4357/aa6c66},
archivePrefix = {arXiv},
       eprint = {1611.08305},
 primaryClass = {astro-ph.GA},
       adsurl = {https://ui.adsabs.harvard.edu/abs/2017ApJ...840...44B}
}

@ARTICLE{1967ApJ...147..868P,
       author = {{Partridge}, R.~B. and {Peebles}, P.~J.~E.},
        title = "{Are Young Galaxies Visible?}",
      journal = {\apj},
         year = 1967,
        month = mar,
       volume = {147},
        pages = {868},
          doi = {10.1086/149079},
       adsurl = {https://ui.adsabs.harvard.edu/abs/1967ApJ...147..868P}
}

@ARTICLE{2019PASJ...71...55K,
       author = {{Kusakabe}, Haruka and {Shimasaku}, Kazuhiro and {Momose}, Rieko and {Ouchi}, Masami and {Nakajima}, Kimihiko and {Hashimoto}, Takuya and {Harikane}, Yuichi and {Silverman}, John D. and {Capak}, Peter L.},
        title = "{The dominant origin of diffuse Ly{\ensuremath{\alpha}} halos around Ly{\ensuremath{\alpha}} emitters explored by spectral energy distribution fitting and clustering analysis}",
      journal = {\pasj},
         year = 2019,
        month = jun,
       volume = {71},
       number = {3},
          eid = {55},
        pages = {55},
          doi = {10.1093/pasj/psz029},
archivePrefix = {arXiv},
       eprint = {1803.10265},
 primaryClass = {astro-ph.GA},
       adsurl = {https://ui.adsabs.harvard.edu/abs/2019PASJ...71...55K}
}

@ARTICLE{2016ApJ...831...39B,
       author = {{Borisova}, Elena and {Cantalupo}, Sebastiano and {Lilly}, Simon J. and {Marino}, Raffaella A. and {Gallego}, Sofia G. and {Bacon}, Roland and {Blaizot}, Jeremy and {Bouch{\'e}}, Nicolas and {Brinchmann}, Jarle and {Carollo}, C. Marcella and {Caruana}, Joseph and {Finley}, Hayley and {Herenz}, Edmund C. and {Richard}, Johan and {Schaye}, Joop and {Straka}, Lorrie A. and {Turner}, Monica L. and {Urrutia}, Tanya and {Verhamme}, Anne and {Wisotzki}, Lutz},
        title = "{Ubiquitous Giant Ly{\ensuremath{\alpha}} Nebulae around the Brightest Quasars at z {\ensuremath{\sim}} 3.5 Revealed with MUSE}",
      journal = {\apj},
         year = 2016,
        month = nov,
       volume = {831},
       number = {1},
          eid = {39},
        pages = {39},
          doi = {10.3847/0004-637X/831/1/39},
archivePrefix = {arXiv},
       eprint = {1605.01422},
 primaryClass = {astro-ph.GA},
       adsurl = {https://ui.adsabs.harvard.edu/abs/2016ApJ...831...39B}
}

@ARTICLE{2026A&A...708A.214P,
       author = {{Pessa}, Ismael and {Wisotzki}, Lutz and {Urrutia}, Tanya and {Bouch{\'e}}, Nicolas F. and {Leclercq}, Floriane and {Augustin}, Ramona and {Guo}, Yucheng and {Kozlova}, Daria and {Kusakabe}, Haruka and {Pharo}, John},
        title = "{First statistical constraints on galactic-scale outflow properties traced by their extended Mg II emission with MUSE}",
      journal = {\aap},
         year = 2026,
        month = apr,
       volume = {708},
          eid = {A214},
        pages = {A214},
          doi = {10.1051/0004-6361/202557875},
archivePrefix = {arXiv},
       eprint = {2602.11280},
 primaryClass = {astro-ph.GA},
       adsurl = {https://ui.adsabs.harvard.edu/abs/2026A&A...708A.214P}
}

@ARTICLE{2023AN....34420091H,
       author = {{Herenz}, Edmund Christian},
        title = "{Revisiting the emission line source detection problem in integral field spectroscopic data}",
      journal = {Astronomische Nachrichten},
         year = 2023,
        month = jun,
       volume = {344},
       number = {5},
          eid = {e20220091},
        pages = {e20220091},
          doi = {10.1002/asna.20220091},
archivePrefix = {arXiv},
       eprint = {2212.00662},
 primaryClass = {astro-ph.IM},
       adsurl = {https://ui.adsabs.harvard.edu/abs/2023AN....34420091H}
}

@ARTICLE{2020MNRAS.495.1501C,
       author = {{Cullen}, F. and {McLure}, R.~J. and {Dunlop}, J.~S. and {Carnall}, A.~C. and {McLeod}, D.~J. and {Shapley}, A.~E. and {Amor{\'\i}n}, R. and {Bolzonella}, M. and {Castellano}, M. and {Cimatti}, A. and {Cirasuolo}, M. and {Cucciati}, O. and {Fontana}, A. and {Fontanot}, F. and {Garilli}, B. and {Guaita}, L. and {Jarvis}, M.~J. and {Pentericci}, L. and {Pozzetti}, L. and {Talia}, M. and {Zamorani}, G. and {Calabr{\`o}}, A. and {Cresci}, G. and {Fynbo}, J.~P.~U. and {Hathi}, N.~P. and {Giavalisco}, M. and {Koekemoer}, A. and {Mannucci}, F. and {Saxena}, A.},
        title = "{The VANDELS survey: a strong correlation between Ly {\ensuremath{\alpha}} equivalent width and stellar metallicity at 3 {\ensuremath{\leq}} z {\ensuremath{\leq}} 5}",
      journal = {\mnras},
         year = 2020,
        month = jun,
       volume = {495},
       number = {1},
        pages = {1501-1510},
          doi = {10.1093/mnras/staa1260},
archivePrefix = {arXiv},
       eprint = {2001.11063},
 primaryClass = {astro-ph.GA},
       adsurl = {https://ui.adsabs.harvard.edu/abs/2020MNRAS.495.1501C}
}

@ARTICLE{2023A&A...678A..68S,
       author = {{Saxena}, Aayush and {Robertson}, Brant E. and {Bunker}, Andrew J. and {Endsley}, Ryan and {Cameron}, Alex J. and {Charlot}, Stephane and {Simmonds}, Charlotte and {Tacchella}, Sandro and {Witstok}, Joris and {Willott}, Chris and {Carniani}, Stefano and {Curtis-Lake}, Emma and {Ferruit}, Pierre and {Jakobsen}, Peter and {Arribas}, Santiago and {Chevallard}, Jacopo and {Curti}, Mirko and {D'Eugenio}, Francesco and {De Graaff}, Anna and {Jones}, Gareth C. and {Looser}, Tobias J. and {Maseda}, Michael V. and {Rawle}, Tim and {Rix}, Hans-Walter and {Del Pino}, Bruno Rodr{\'\i}guez and {Smit}, Renske and {{\"U}bler}, Hannah and {Eisenstein}, Daniel J. and {Hainline}, Kevin and {Hausen}, Ryan and {Johnson}, Benjamin D. and {Rieke}, Marcia and {Williams}, Christina C. and {Willmer}, Christopher N.~A. and {Baker}, William M. and {Bhatawdekar}, Rachana and {Bowler}, Rebecca and {Boyett}, Kristan and {Chen}, Zuyi and {Egami}, Eiichi and {Ji}, Zhiyuan and {Kumari}, Nimisha and {Nelson}, Erica and {Perna}, Michele and {Sandles}, Lester and {Scholtz}, Jan and {Shivaei}, Irene},
        title = "{JADES: Discovery of extremely high equivalent width Lyman-{\ensuremath{\alpha}} emission from a faint galaxy within an ionized bubble at z = 7.3}",
      journal = {\aap},
         year = 2023,
        month = oct,
       volume = {678},
          eid = {A68},
        pages = {A68},
          doi = {10.1051/0004-6361/202346245},
archivePrefix = {arXiv},
       eprint = {2302.12805},
 primaryClass = {astro-ph.GA},
       adsurl = {https://ui.adsabs.harvard.edu/abs/2023A&A...678A..68S}
}

@ARTICLE{2020ApJ...891..177Z,
       author = {{Zhang}, Haibin and {Ouchi}, Masami and {Itoh}, Ryohei and {Shibuya}, Takatoshi and {Ono}, Yoshiaki and {Harikane}, Yuichi and {Inoue}, Akio K. and {Rauch}, Michael and {Kikuchihara}, Shotaro and {Nakajima}, Kimihiko and {Yajima}, Hidenobu and {Arata}, Shohei and {Abe}, Makito and {Iwata}, Ikuru and {Kashikawa}, Nobunari and {Kawanomoto}, Satoshi and {Kikuta}, Satoshi and {Kobayashi}, Masakazu A.~R. and {Kusakabe}, Haruka and {Mawatari}, Ken and {Nagao}, Tohru and {Shimasaku}, Kazuhiro and {Taniguchi}, Yoshiaki},
        title = "{CHORUS. III. Photometric and Spectroscopic Properties of Ly{\ensuremath{\alpha}} Blobs at z = 4.9-7.0}",
      journal = {\apj},
         year = 2020,
        month = mar,
       volume = {891},
       number = {2},
          eid = {177},
        pages = {177},
          doi = {10.3847/1538-4357/ab7917},
archivePrefix = {arXiv},
       eprint = {1905.09841},
 primaryClass = {astro-ph.GA},
       adsurl = {https://ui.adsabs.harvard.edu/abs/2020ApJ...891..177Z}
}

@ARTICLE{2016MNRAS.461.1076C,
       author = {{Comparat}, Johan and {Zhu}, Guangtun and {Gonzalez-Perez}, Violeta and {Norberg}, Peder and {Newman}, Jeffrey and {Tresse}, Laurence and {Richard}, Johan and {Yepes}, Gustavo and {Kneib}, Jean-Paul and {Raichoor}, Anand and {Prada}, Francisco and {Maraston}, Claudia and {Y{\`e}che}, Christophe and {Delubac}, Timoth{\'e}e and {Jullo}, Eric},
        title = "{The evolution of the [O II], H {\ensuremath{\beta}} and [O III] emission line luminosity functions over the last nine billions years}",
      journal = {\mnras},
         year = 2016,
        month = sep,
       volume = {461},
       number = {1},
        pages = {1076-1087},
          doi = {10.1093/mnras/stw1393},
archivePrefix = {arXiv},
       eprint = {1605.02875},
 primaryClass = {astro-ph.GA},
       adsurl = {https://ui.adsabs.harvard.edu/abs/2016MNRAS.461.1076C}
}

@ARTICLE{2015A&A...575A..75B,
       author = {{Bacon}, R. and {Brinchmann}, J. and {Richard}, J. and {Contini}, T. and {Drake}, A. and {Franx}, M. and {Tacchella}, S. and {Vernet}, J. and {Wisotzki}, L. and {Blaizot}, J. and {Bouch{\'e}}, N. and {Bouwens}, R. and {Cantalupo}, S. and {Carollo}, C.~M. and {Carton}, D. and {Caruana}, J. and {Cl{\'e}ment}, B. and {Dreizler}, S. and {Epinat}, B. and {Guiderdoni}, B. and {Herenz}, C. and {Husser}, T.-O. and {Kamann}, S. and {Kerutt}, J. and {Kollatschny}, W. and {Krajnovic}, D. and {Lilly}, S. and {Martinsson}, T. and {Michel-Dansac}, L. and {Patricio}, V. and {Schaye}, J. and {Shirazi}, M. and {Soto}, K. and {Soucail}, G. and {Steinmetz}, M. and {Urrutia}, T. and {Weilbacher}, P. and {de Zeeuw}, T.},
        title = "{The MUSE 3D view of the Hubble Deep Field South}",
      journal = {\aap},
         year = 2015,
        month = mar,
       volume = {575},
          eid = {A75},
        pages = {A75},
          doi = {10.1051/0004-6361/201425419},
archivePrefix = {arXiv},
       eprint = {1411.7667},
 primaryClass = {astro-ph.GA},
       adsurl = {https://ui.adsabs.harvard.edu/abs/2015A&A...575A..75B}
}

@ARTICLE{2026ApJ..1001..150M,
       author = {{McKay}, Brianna and {Tuttle}, Sarah and {Cooper}, Erin Mentuch and {Davis}, Dustin and {Ciardullo}, Robin and {Farrow}, Daniel J. and {Finkelstein}, Steven L. and {Gebhardt}, Karl and {Liu}, Chenxu and {Zeimann}, Gregory and {Byrohl}, Chris},
        title = "{The Redshift Evolution of Extended Ly{\ensuremath{\alpha}} Emission in HETDEX. I. Imaging Methods and Emission Measurements}",
      journal = {\apj},
         year = 2026,
        month = apr,
       volume = {1001},
       number = {2},
          eid = {150},
        pages = {150},
          doi = {10.3847/1538-4357/ae4de7},
       adsurl = {https://ui.adsabs.harvard.edu/abs/2026ApJ..1001..150M}
}

@ARTICLE{2026MNRAS.tmp..886S,
       author = {{Scholtz}, J. and {Carniani}, S. and {Parlanti}, E. and {D'Eugenio}, F. and {Curtis-Lake}, E. and {Jakobsen}, P. and {Bunker}, A.~J. and {Cameron}, A.~J. and {Arribas}, S. and {Baker}, W.~M. and {Charlot}, S. and {Chevellard}, J. and {Circosta}, C. and {Curti}, M. and {Duan}, Q. and {Eisenstein}, D.~J. and {Hainline}, K. and {Ji}, Z. and {Johnson}, B.~D. and {Jones}, G.~C. and {Kumari}, N. and {Maiolino}, R. and {Maseda}, M.~V. and {Perna}, M. and {P{\'e}rez-Gonz{\'a}lez}, P.~G. and {Rawle}, T. and {Rieke}, M. and {Rinaldi}, P. and {Robertson}, B. and {Saxena}, A. and {Shivaei}, I. and {Silcock}, M.~S. and {Sun}, Y. and {Rodr{\'\i}guez Del Pino}, B. and {Tacchella}, S. and {{\"U}bler}, H. and {Venturi}, G. and {Williams}, C.~C. and {Willmer}, C.~N.~A. and {Willott}, C. and {Witstok}, J.},
        title = "{JADES Data Release 4 - Paper II: Data reduction, analysis and emission-line fluxes of the complete spectroscopic sample}",
      journal = {\mnras},
         year = 2026,
        month = may,
          doi = {10.1093/mnras/stag939},
archivePrefix = {arXiv},
       eprint = {2510.01034},
 primaryClass = {astro-ph.GA},
       adsurl = {https://ui.adsabs.harvard.edu/abs/2026MNRAS.tmp..886S}
}

@ARTICLE{2026MNRAS.tmp..935C,
       author = {{Curtis-Lake}, Emma and {Cameron}, Alex J. and {Bunker}, Andrew J. and {Scholtz}, Jan and {Carniani}, Stefano and {Parlanti}, Eleonora and {D'Eugenio}, Francesco and {Jakobsen}, Peter and {Willmer}, Christopher N.~A. and {Arribas}, Santiago and {Baker}, William M. and {Charlot}, St{\'e}phane and {Chevallard}, Jacopo and {Circosta}, Chiara and {Curti}, Mirko and {Duan}, Qiao and {Eisenstein}, Daniel J. and {Hainline}, Kevin and {Ji}, Zhiyuan and {Johnson}, Benjamin D. and {Jones}, Gareth C. and {Maiolino}, Roberto and {Maseda}, Michael V. and {Perna}, Michele and {P{\'e}rez-Gonz{\'a}lez}, Pablo G. and {Rawle}, Tim and {Rieke}, Marcia and {Rinaldi}, Pierluigi and {Robertson}, Brant and {Rodr{\'\i}guez Del Pino}, Bruno and {Saxena}, Aayush and {Shivaei}, Irene and {Smit}, Renske and {Tacchella}, Sandro and {{\"U}bler}, Hannah and {Venturi}, Giacomo and {Williams}, Christina C. and {Willott}, Chris},
        title = "{JADES Data Release 4 Paper I: Sample Selection, Observing Strategy and Redshifts of the complete spectroscopic sample}",
      journal = {\mnras},
         year = 2026,
        month = may,
          doi = {10.1093/mnras/stag836},
archivePrefix = {arXiv},
       eprint = {2510.01033},
 primaryClass = {astro-ph.GA},
       adsurl = {https://ui.adsabs.harvard.edu/abs/2026MNRAS.tmp..935C}
}

@ARTICLE{2024MNRAS.531.2701T,
       author = {{Tang}, Mengtao and {Stark}, Daniel P. and {Ellis}, Richard S. and {Sun}, Fengwu and {Topping}, Michael and {Robertson}, Brant and {Tacchella}, Sandro and {Arribas}, Santiago and {Baker}, William M. and {Bhatawdekar}, Rachana and {Boyett}, Kristan and {Bunker}, Andrew J. and {Charlot}, St{\'e}phane and {Chen}, Zuyi and {Chevallard}, Jacopo and {Jones}, Gareth C. and {Kumari}, Nimisha and {Lyu}, Jianwei and {Maiolino}, Roberto and {Maseda}, Michael V. and {Saxena}, Aayush and {Whitler}, Lily and {Williams}, Christina C. and {Willott}, Chris and {Witstok}, Joris},
        title = "{Ly{\ensuremath{\alpha}} emission in galaxies at z ≃ 5-6: new insight from JWST into the statistical distributions of Ly{\ensuremath{\alpha}} properties at the end of reionization}",
      journal = {\mnras},
         year = 2024,
        month = jun,
       volume = {531},
       number = {2},
        pages = {2701-2730},
          doi = {10.1093/mnras/stae1338},
archivePrefix = {arXiv},
       eprint = {2402.06070},
 primaryClass = {astro-ph.GA},
       adsurl = {https://ui.adsabs.harvard.edu/abs/2024MNRAS.531.2701T}
}

@ARTICLE{2024A&A...691A..66G,
       author = {{Guo}, Yucheng and {Bacon}, Roland and {Wisotzki}, Lutz and {Garel}, Thibault and {Blaizot}, J{\'e}r{\'e}my and {Schaye}, Joop and {Matthee}, Jorryt and {Leclercq}, Floriane and {Boogaard}, Leindert and {Richard}, Johan and {Verhamme}, Anne and {Brinchmann}, Jarle and {Michel-Dansac}, L{\'e}o and {Kusakabe}, Haruka},
        title = "{Spatially resolved spectroscopic analysis of Ly{\ensuremath{\alpha}} haloes: Radial evolution of the Ly{\ensuremath{\alpha}} line profile out to 60 kpc}",
      journal = {\aap},
         year = 2024,
        month = nov,
       volume = {691},
          eid = {A66},
        pages = {A66},
          doi = {10.1051/0004-6361/202347958},
archivePrefix = {arXiv},
       eprint = {2309.06311},
 primaryClass = {astro-ph.GA},
       adsurl = {https://ui.adsabs.harvard.edu/abs/2024A&A...691A..66G}
}

\begin{appendix}
\onecolumn
\section{Stacks}\label{sec:stacks_ap}

    Figure~\ref{stack_full} shows the SB profiles of stacked NB images of all galaxies in each sample, as well as those with or without a detected halo.

    \begin{figure*}[ht!]
        \centering
        \includegraphics[width=\linewidth]{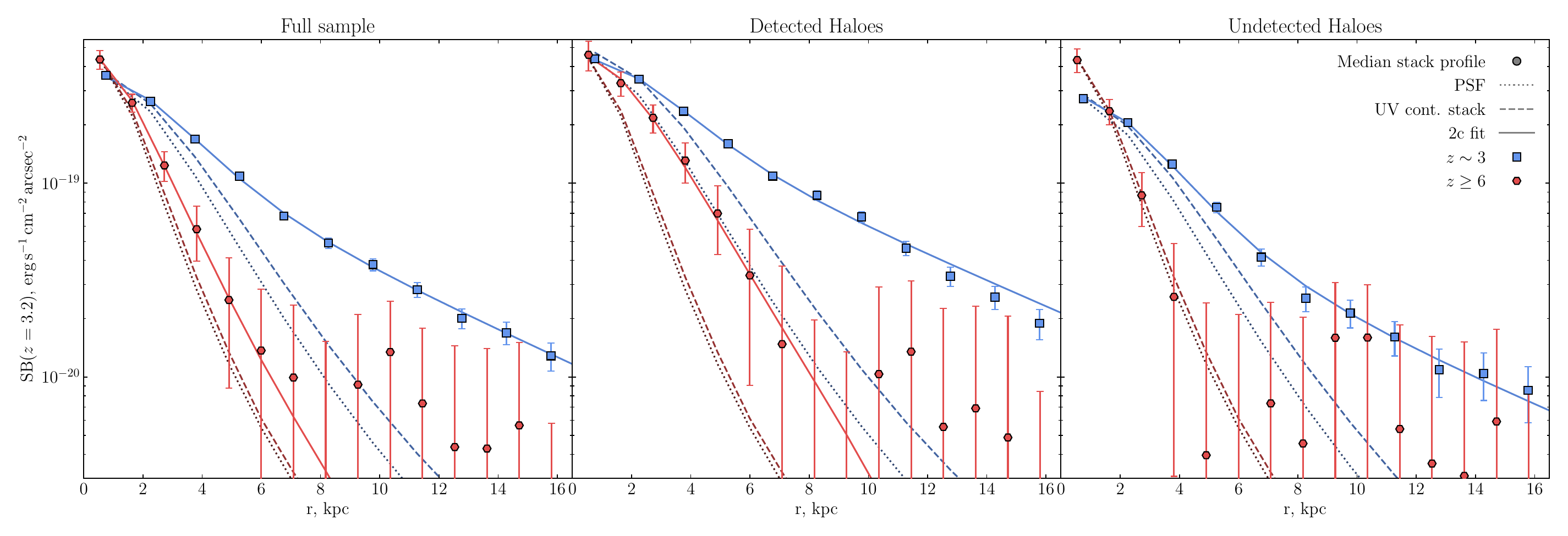}
        \caption{Median \lya surface brightness profiles of various samples of LAEs. \textit{Left panel} shows the median stacks of all LAEs in the $z\sim3$ nd $z\geq6$ samples, while \textit{central} and \textit{right panels} depicts sources with and without individually detected \lya haloes. Measured data is shown in blue squares ($z\sim3$ sample) and red hexagons ($z\geq6$ sample). For an easier comparison, the high-redshift measurement is shifted to $z=3.2$ -- median redshift of the lower-$z$ sample. PSF  UV continuum stack SB profiles are shown in dotted and dashed lines respectively. The solid line shows a SB profile of a 2 component fit to the NB stacked data.}
        \label{stack_full}
    \end{figure*}

\section{Effects of sample properties on inferred \texorpdfstring{\lya}{Lyα }line widths and halo sizes}\label{sec:matched_samples}
Here we quantify how minor differences in galaxy properties between samples affect the main results of our analysis, namely mean \lya line widths, halo scalelengths and halo detection fractions. These parameters are known to correlate with galaxy properties (\citeads{2021MNRAS.506.5129B}, \citeads{2022A&A...662A..64R}) such as stellar mass, star-formation rate, UV or \lya luminosity, so the observed differences in \lya line widths and halo scales between the $z\geq6$ and $z\sim3$ samples could be attributed to their mismatch in these parameters, rather than being a reflection of the redshift evolution.

To test this we compare $z\geq6$ and $z\sim3$ subsamples matched in M$_*$, SFR, M$_\mathrm{UV}$ or L$_\mathrm{Ly\alpha}$. To create the matched samples we first identify the galaxy parameter range occupied by both samples as $\left[\max(\min(Q_{z\geq6}), \min(Q_{z\sim3})), \min(\max(Q_{z\geq6}), \max(Q_{z\sim3})) \right]$ where $Q$ is one of four investigated parameters. The high-$z$ subsample matched in $Q$ is then all $z\geq6$ LAEs that fall within this interval. We then calculate the fractional number of objects from the $z\geq6$ subsample in 4 bins of Q and draw a random subsample from the $z\sim3$ with the same fractional number of objects in the same bins. This is the $z\sim3$ subsample matched in $Q$. We chose 4 bins to construct the distribution to minimize the effects of small number statistics from the high-$z$ sample. The examples of the matched subsamples are shown in the third column in \autoref{sample_match}. Since the drawing pool of the $z\sim3$ LAEs is much larger than at $z\geq6$ and need not have the same distribution of $Q$, one randomly drawn sample might be unrepresentative. To mitigate this we draw 100 random $z\sim3$ matched subsamples and average our results over them. For each subsample we measure 3 values: mean \lya halo scale length, \lya halo detection rate and a Kaplan-Meier estimate of the \lya line width. The summary of the experiment is presented in \autoref{sample_match}. As can be seen from the plot, while in most cases we find a $\rho=0.2-0.5$ correlation between both $r_{s, H}$ and especially FWHM$_\mathrm{Ly\alpha}$ with the studied galaxy properties, the reported difference between $z\geq6$ and $z\sim3$ remains almost unchanged.

\begin{figure*}[ht!]
    \centering
    \includegraphics[width=\linewidth]{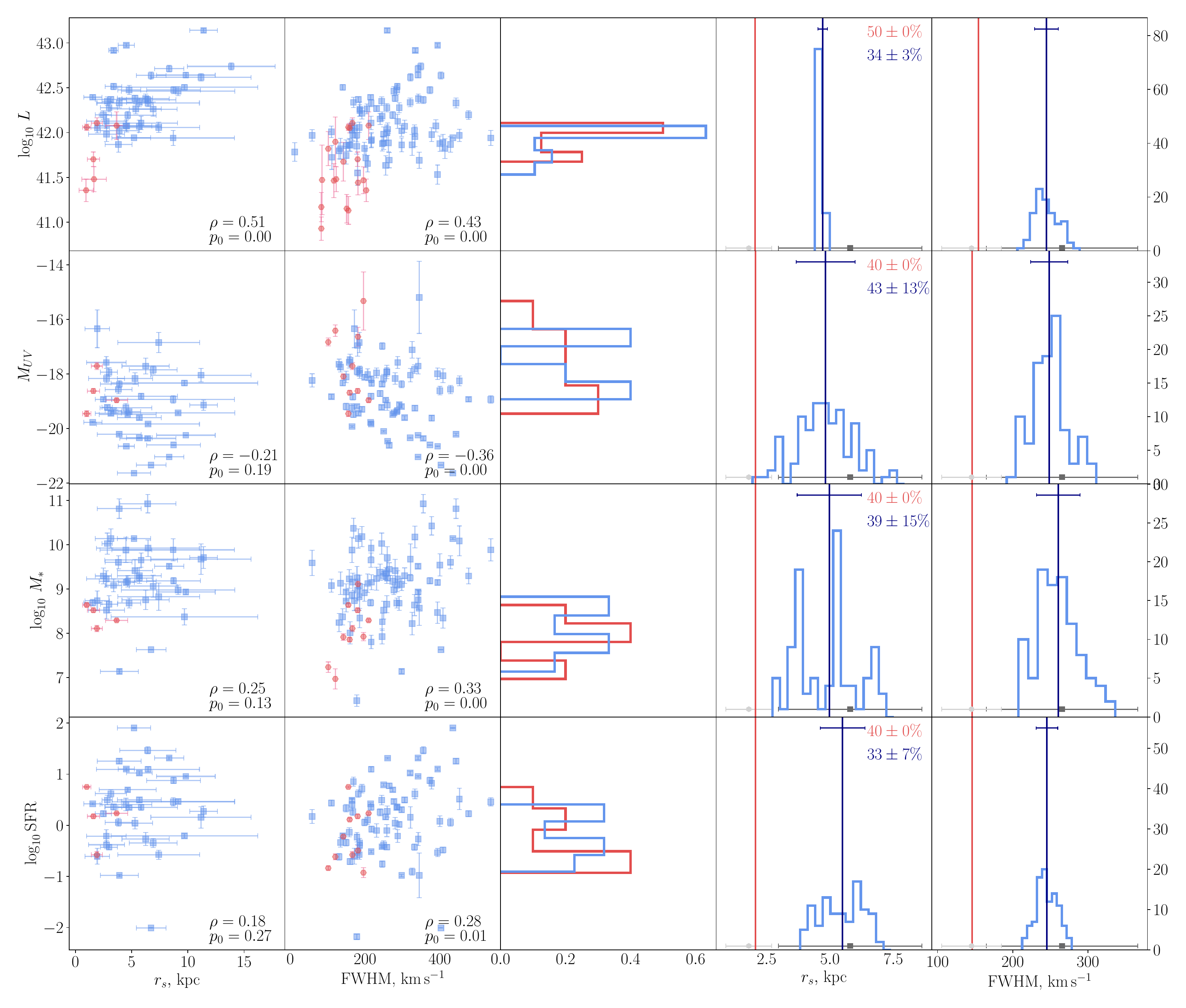}
    \caption{The summary of the sample matching exercise. The $z\sim3$ objects are shown in blue (squares), the $z\geq6$ objects are in red (hexagons). \textit{First column} shows \lya halo scale length against major galaxy parameters for LAEs with detected \lya haloes. \textit{Second column}: \lya line widths against major galaxy parameters for all LAEs. The numbers in the bottom right corners indicate the Spearman correlation coefficient for the combined sample and its associated $p_0$ value. \textit{Third column}: distributions of the example matched samples across major galaxy properties. \textit{Fourth column}: comparison of the mean values of \lya halo scale lengths of the matched samples. The $z\geq6$ matched subsample is indicated by the vertical red line, since the sample is always the same while the $z\sim3$ subsamples are shown in blue histogram with vertical blue line indicating the mean over the 100 realisations and its error. The grey errorbars at the bottom display the means of the full samples. Halo detection fractions are indicated at the top right corner with $z\geq6$ value being the top one. \textit{Fifth column:} same as the previous column, but for the \lya line widths.}
    \label{sample_match}
\end{figure*}
\twocolumn

\section{Resolution effects and the recovery of \texorpdfstring{\lya}{Lyα }halo sizes}

Another possible effects that might influence the observed scale length distributions are a change in PSF size and angular diameter distance ($D_\mathrm{A}$) evolution. As $D_\mathrm{A}$ decreases with redshift beyond $z=1.6$, objects of the same physical size become apparently more extended. In the context of this study, this constitutes a factor 1.4 increase in the apparent size of the objects at $z\approx6$ compared to $z\approx3$, which increases the detectability of compact haloes. The change of the MUSE PSF size with wavelength also introduces a dependence of the LAH detection threshold on redshift. As the detection procedure compares the UV continuum model convolved with the MUSE PSF at the observed wavelength, this dependence is linear in PSF FWHM to first order. A typical ratio between PSF FWHM at 5105\,\AA\ ($z_{\mathrm{Ly}\alpha}=3.2\approx\left<z_\mathrm{MUSCATEL}\right>$) and 8860\,\AA\ ($z_{\mathrm{Ly}\alpha}=6.29\approx\left<z_\mathrm{MXDF}\right>$) is 1.5. Both these effects combined increase the minimal size of a detectable LAH by a factor of 2 when comparing high-redshift LAEs to lower-$z$ ones. To test this, we conducted a simple experiment: using derived GALFIT models, we created mock NB images of the high-$z$ LAHs at $z=3.2$, taking into account both PSF broadening and angular size decrease. We then applied our detection algorithm to mock images under MUSCATEL-SF conditions and recovered only 3 haloes (out of 6), all of which have \rsh$>2\,$ kpc, which is very close to twice the minimal \rsh\ detected in the MXDF sample. Thus, this effect can explain the apparent lack of compact ($\rsh\lesssim2\,$kpc) haloes in the $z\sim3$ sample. The observed discrepancy could be solved statistically by assuming all LAEs in the MUSCATEL sample without a detected LAH to host a compact halo below the limit with $1\,\mathrm{kpc}\lesssim\rsh\lesssim2$\,kpc. However, studies with higher SB sensitivity and hence lower \rsh\ detection limit do not find such an overabundance of compact haloes (\citetalias{2016A&A...587A..98W}, \citetalias{2017A&A...608A...8L}, \citeads{2024A&A...690A.343P}), making this assumption unreasonable. Moreover, this argument does not explain the absence of large haloes in the high-redshift sample since such objects lie sufficiently above the detection limit.

\section{Completeness of the samples}\label{sec:fc}

Completeness of the investigated samples is another aspect that can drive the difference between the inferred sample properties. As has been shown by \citetads{2024A&A...690A.343P} LAE detection probability depends strongly on \lya halo and \lya line parameters. This means that if two samples have different selection functions, their inferred properties would reflect those of the underlying populations differently.

To understand the completeness of our samples we use the LAE selection function (SF) constructed in a same way as \citetads{2024A&A...690A.343P} for 2 specific MUSE deep fields -- MUSCATEL-Shallow and MXDF. The SF is built by inserting mock LAEs with spatial and spectral properties into empty MUSE cubes and running the standard LSDCat detection algorithm on them. Thus, each insertion constitutes one node in a multidimensional grid which tells whether a \lya emitter at a given flux, redshift, line shape and halo parameters would be detected. Marginalising this grid over the spatial and spectral properties yields the completeness $f_c$ as a function of \lya flux and redshift. The high value of $f_c$ means that an object at this flux would have been detected whatever its halo and line shape; conversely, a detected LAE with low $f_c$ must owe its detectability to a particular combination of spatial and spectral properties, or the error in the flux measurement.

We calculate the $f_c$ of each object in our sample using the two SFs constructed in the same way. The $f_c$ distributions are shown in \autoref{f_c}. It is evident that the $z\geq6$ sample has a much higher fraction of objects with extremely low completeness and is inconsistent with that at lower redshift. This is explained by the different source detection algorithm used for constructing the AMUSED database (ORIGIN, \citetads{2020A&A...635A.194M}) compared to MUSCATEL catalogues and the used selection function itself (LSDCat). To make the comparison more fair we ran LSDCat on the MXDF field and discovered that five objects with the lowest completeness are not recovered above the detection S/N threshold of 6 adopted in MUSCATEL. Dropping these five LSDCat non-detections brings the $f_c$ distribution of the $z\geq6$ sample much closer to the $z\sim3$ one. No \lya halo objects are dropped this way, so the halo size difference remains unchanged, while the halo detection fraction at $z\geq6$ rises to 46\% -- very similar to the $z\sim3$ detection fraction of 44\%. The mean \lya line width rises minimally from 145 km\,s$^{-1}$ to 156 km\,s$^{-1}$. Therefore we conclude that the observed differences between the samples can not be explained by the differences in the completeness of the catalogues.

\begin{figure}[hb!]
    \centering
    \includegraphics[width=\linewidth]{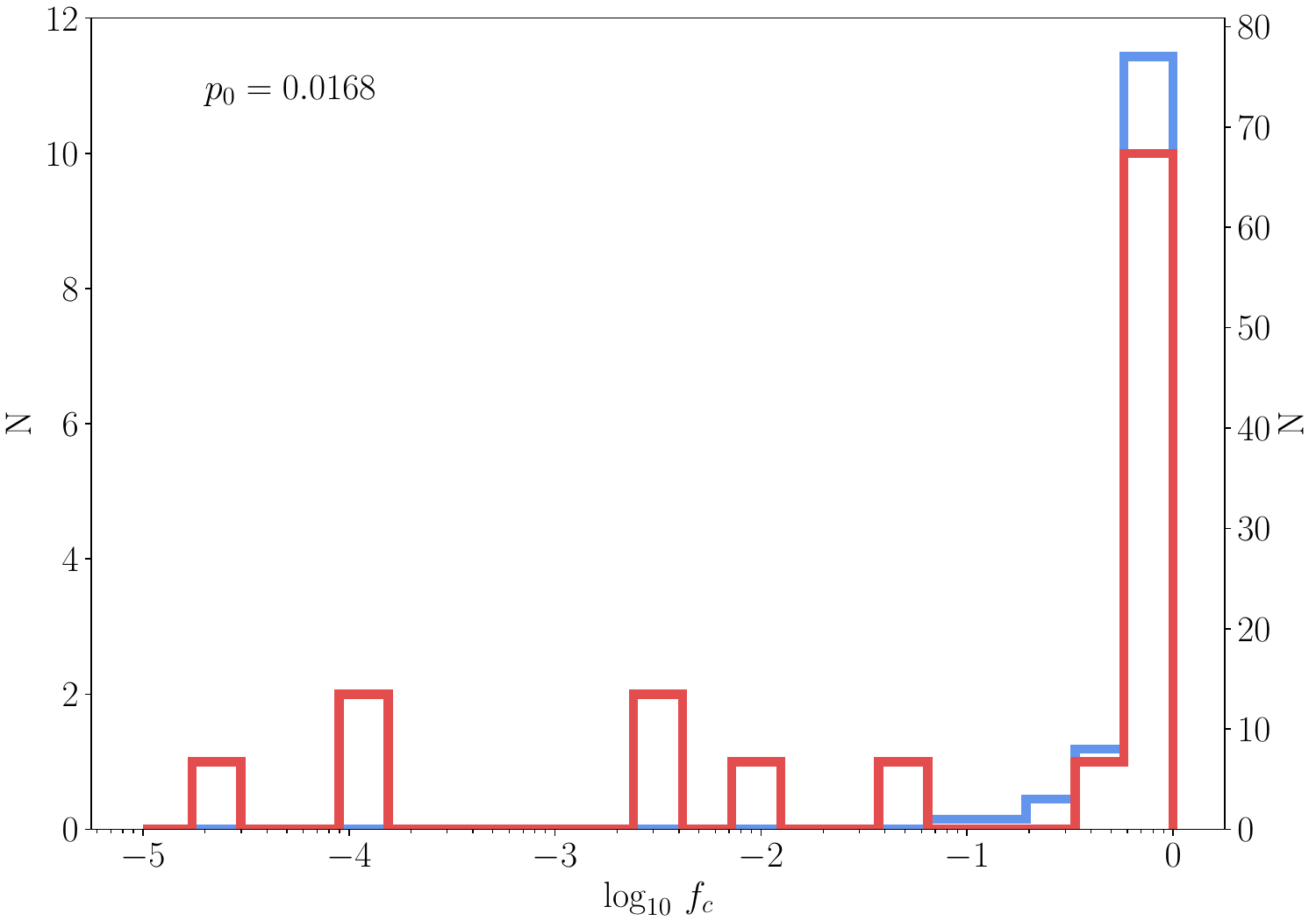}
    \caption{Distribution of the detection probabilities of objects in the high (red line) and lower (blue line) redshift samples. Due to the different sample sizes number counts for the $z\geq6$ sample are shown on the left y-axis and on right the $z\sim3$ sample. A two sample K-S test's $p_0$ value is indicated in the top left corner.}
    \label{f_c}
\end{figure}

\section{Upper limits of undetected \texorpdfstring{\lya}{Lyα }haloes}\label{sec:up_lim}

To determine the limiting parameters of haloes around LAEs that were not detected in this study, we conduct the following experiment. For each such object we generate a grid of mock haloes varying scale length and the halo flux fraction while fixing the axis ratio to 1. Each mock halo was then added to the extracted NB image of the source and fed to the detection procedure described in \autoref{sec:det}. On the resulting grid of $\chi^2$ values in the $r_{s, \mathrm{H}} - f_\mathrm{H}$ plane we determine the detection upper limit as a $\chi^2$ contour corresponding to $p_0=0.05$. \autoref{up_lims} shows derived upper limits of undetected haloes in both samples with detected objects also superposed. The emergent upper limit curves have a typical U shape as it is hard to distinguish a very compact halo from the central component and detect a very extended and hence a low SB halo. The change in the upper limits between the two samples is very similar to the change in detected halo properties seen in \autoref{par_comp}: while the $f_\mathrm{H}$ constraints span the same parameter space, upper limit curves reach minima at smaller $r_{s, \mathrm{H}}$ values at higher redshift. We also show parameters inferred from fitting a 2 component model to the stacked narrowband data. As one would expect, the undetected halo stack resides in the lower $f_\mathrm{H}$ region of the plot, compared to the detected population below most of the upper limits lines, while the detected halo stack is more similar to the individual objects. Based on behaviour of the $z\sim3$ stacks, it is possible to put the upper limits on the parameters of $z\geq6$ undetected halo stack as $f_\mathrm{H}\lesssim0.4$ and $r_\mathrm{s, H}\lesssim1.3\,$kpc.

\begin{figure}[hb!]
    \centering
    \includegraphics[width=\linewidth]{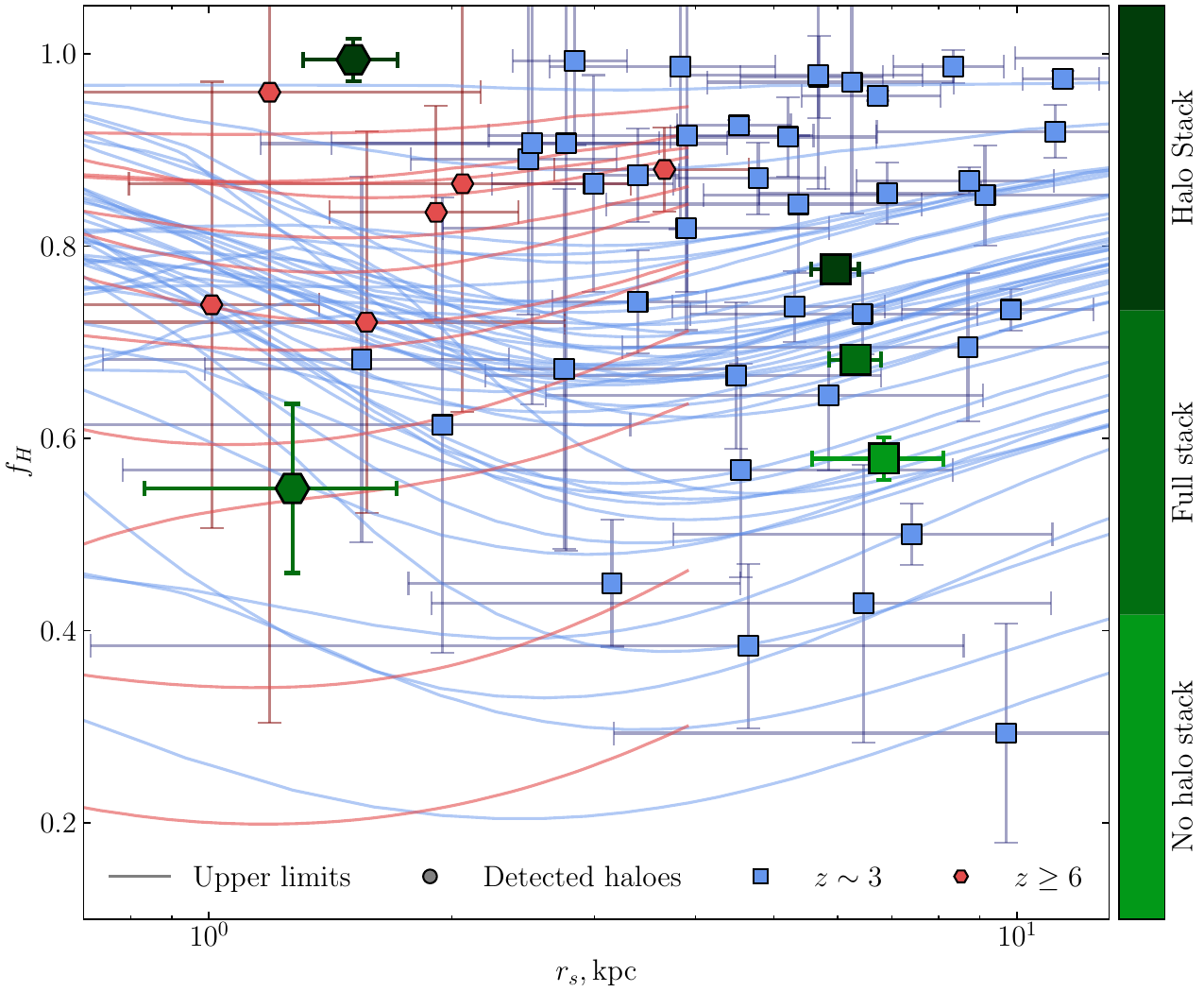}
    \caption{Upper limits of undetected objects compared to properties of individual LAHs and stacks. Curves show the upper limits for undetected haloes and markers show measurements of individual haloes, both colour coded by sample. Parameters measured from the stacked images are represented by large markers with the type of stack encoded by shade of green.}
    \label{up_lims}
\end{figure}
\end{appendix}

\end{document}